\documentclass[11pt,twoside]{article}

\usepackage{geometry}
\usepackage{amsopn}
\usepackage{lipsum}
\usepackage{amsfonts}
\usepackage{graphicx}
\usepackage{epstopdf}
\usepackage{algorithmic}
\usepackage{comment}
\usepackage{epsf}
\usepackage{graphicx}
\usepackage{bbm}
\usepackage{color}
\usepackage{nicefrac}
\usepackage{mathtools}
\usepackage{amsfonts}
\usepackage{amsthm}
\usepackage{amsmath, bm}
\usepackage{hyperref}
\usepackage{cleveref}
\usepackage{caption}
\usepackage{subcaption}
\usepackage{amssymb}
\usepackage{textcomp}
\usepackage{adjustbox}

\newenvironment{fminipage}%
  {\begin{Sbox}\begin{minipage}}%
  {\end{minipage}\end{Sbox}\fbox{\TheSbox}}

\newcommand{\algname}{\textrm{EXACT}}

\newtheorem{theorem}{Theorem}
\newtheorem{example}{Example}
\newtheorem{assumption}{Assumption}
\newtheorem{proposition}{Proposition}
\newtheorem{lemma}{Lemma}

\usepackage[square,sort,comma,numbers]{natbib}

\newlength{\widebarargwidth}
\newlength{\widebarargheight}
\newlength{\widebarargdepth}
\DeclareRobustCommand{\widebar}[1]{%
 \settowidth{\widebarargwidth}{\ensuremath{#1}}%
 \settoheight{\widebarargheight}{\ensuremath{#1}}%
 \settodepth{\widebarargdepth}{\ensuremath{#1}}%
 \addtolength{\widebarargwidth}{-0.3\widebarargheight}%
 \addtolength{\widebarargwidth}{-0.3\widebarargdepth}%
 \makebox[0pt][l]{\hspace{0.3\widebarargheight}%
   \hspace{0.3\widebarargdepth}%
   \addtolength{\widebarargheight}{0.3ex}%
   \rule[\widebarargheight]{0.95\widebarargwidth}{0.1ex}}%
 {#1}}

\makeatletter
\long\def\@makecaption#1#2{
       \vskip 0.8ex
       \setbox\@tempboxa\hbox{\small {\bf #1:} #2}
       \parindent 1.5em 
       \dimen0=\hsize
       \advance\dimen0 by -3em
       \ifdim \wd\@tempboxa >\dimen0
               \hbox to \hsize{
                       \parindent 0em
                       \hfil 
                       \parbox{\dimen0}{\def\baselinestretch{0.96}\small
                               {\bf #1.} #2
                               } 
                       \hfil}
       \else \hbox to \hsize{\hfil \box\@tempboxa \hfil}
       \fi
       }
\makeatother

\long\def\comment#1{}

\DeclareMathOperator*{\argmin}{argmin}

\newcommand{\inprod}[2]{\ensuremath{\langle #1 , \, #2 \rangle}}

\newcommand{\EE}{\ensuremath{{\mathbb{E}}}}

\newcommand{\1}{\ensuremath{{\sf (i)}}}
\newcommand{\2}{\ensuremath{{\sf (ii)}}}
\newcommand{\3}{\ensuremath{{\sf (iii)}}}
\newcommand{\4}{\ensuremath{{\sf (iv)}}}

\newcommand{\NORMAL}{\ensuremath{\mathcal{N}}}

\newcommand{\real}{\mathbb{R}}

\newcommand{\order}{\ensuremath{\mathcal{O}}}
\newcommand{\ordertil}{\ensuremath{\widetilde{\order}}}

\newcommand{\monotone}{\nu}
\newcommand{\norm}{\|x_{\star}\|_{2}}
\newcommand{\N}{W}
\newcommand{\coeff}{\mu}
\newcommand{\Photon}{\widebar{I}}

\newcommand{\Err}{\mathsf{Err}}

\newcommand{\lambdamax}{\lambda_{\mathsf{max}}}

\newcommand{\lambdamin}{\lambda_{\mathsf{min}}}

\newcommand{\bA}{\boldsymbol{A}}

\newcommand{\R}{\mathbb{R}}

\newcommand{\Constant}{16}

\begin{document}

\begin{center}

  {\bf{\LARGE{\mbox{Accurate, provable and fast polychromatic tomographic}  reconstruction: A variational inequality approach}}}

\vspace*{.2in}

{\large{
\begin{tabular}{ccc}
Mengqi Lou$^{\S}$, Kabir Aladin Verchand$^{\|}$, Sara Fridovich-Keil$^{\dagger}$, Ashwin Pananjady$^{\S, \dagger}$
\end{tabular}
}}
\vspace*{.2in}

\begin{tabular}{c}
Georgia Institute of Technology, School of Industrial and Systems Engineering$^\S$ and \\
School of Electrical and Computer Engineering$^\dagger$ \\ 
University of Southern California, Department of Data Sciences and Operations$^{\|}$ \\
\end{tabular}

\vspace*{.2in}

\today

\vspace*{.2in}

\begin{abstract}
We consider the problem of signal reconstruction for computed tomography (CT) under a nonlinear forward model that accounts for exponential signal attenuation, a polychromatic X-ray source, general measurement noise (e.g., Poisson shot noise), and observations acquired over multiple wavelength windows. We develop a simple iterative algorithm for single-material reconstruction, which we call \algname{} (EXtragradient Algorithm for Computed Tomography), based on formulating our estimate as the fixed point of a monotone variational inequality. We prove guarantees on the statistical and computational performance of \algname{} under realistic assumptions on the measurement process. We also consider a recently introduced variant of this model with Gaussian measurements and present sample and iteration complexity bounds for \algname{} that improve upon those of existing algorithms. We apply our \algname{} algorithm to a CT phantom image recovery task and show that it often requires fewer X-ray views, lower source intensity, and less computation time to achieve reconstruction quality similar to existing methods. Code is available at \url{https://github.com/voilalab/exact}.
\end{abstract}
\end{center}

\section{Introduction and  methodology}

X-ray computed tomography (CT) is a core imaging modality in modern medicine, and is also widely used in diverse nonmedical applications ranging from security inspection to nondestructive testing. CT typically works by rotating an X-ray source and detector along a prespecified trajectory around the subject, and recording the total energy or number of photons that are transmitted to the detector at each angle along the tomographic trajectory. One then solves the inverse problem of reconstructing the underlying 2D or 3D image of the subject from these measurements. Many CT reconstruction techniques have been proposed across multiple communities~\cite[see, e.g.,][]{elbakri2002statistical,geyer2015state,schirra2013statistical,sawatzky2014proximal,schmidt2017spectral, alvarez1976energy, jadue2024mace}, and several algorithms are now built into commercial CT scanners~\cite[see. e.g.,][]{fu2016comparison}. 
In this paper, for concreteness, we focus on spectral CT with photon-counting detectors~\cite{taguchi2013vision, persson2020detective, danielsson2021photon, nakamura2023introduction}. 

The starting point of any reconstruction algorithm is a forward model for CT measurements; see~\cite{tang2023spectral,fu2016comparison,gjesteby2016metal} for an overview and discussion of the importance of modeling and artifacts induced by preprocessing. The forward model for photon-counting CT (see~\cite[Eq. (2)]{mory2018comparison} or~\cite[Eq.(3)]{schmidt2023constrained}) is given by 
\begin{align}\label{eqn:referenceforwardmodel}
   y_{\omega, i}(x) &= I_{\omega, i} \sum_{j = 1}^{W} s_{\omega, i, j} \exp \left (-\sum_{m = 1}^M \sum_{k = 1}^d \mu_{m, j}A_{i, k}x_{k,m} \right ).
\end{align}
Note that our notation differs slightly from the aforementioned papers, and we use the following convention: 
\begin{itemize}
    \item $\omega$ indexes over wavelength windows in the X-ray detector;
    \item $i$ indexes over the $n$ X-ray projection measurements, where each measurement corresponds to a ray passing through the imaging target and being recorded by one of the detector cells\footnote{Note that CT data are typically represented as a sinogram; for notational convenience, we consider each angle/offset pair as a single measurement, so that each sinogram gives rise to multiple measurements.};
    \item $I_{\omega, i}$ is the (random) number of incident photons (produced by the X-ray source) in wavelength window $\omega$ at angle $i$; we denote its mean by the \emph{intensity} $\Photon_{\omega, i} := \EE[I_{\omega, i}]$; 
    \item $y_{\omega, i}$ is the (random) number of transmitted photons that arrive at the X-ray detector in wavelength window $\omega$ at angle $i$;
    \item $j$ indexes over $\N$ individual X-ray wavelengths that fall inside wavelength window $\omega$ --- in this sense, each measurement (even for a fixed wavelength window) is \emph{polychromatic};
    \item $s_{\omega, i, j}$ is the normalized spectral sensitivity of the detector; $\sum_j s_{\omega, i, j} = 1$;
    \item $m$ indexes over the $M$ basis materials that are expected to be present in the imaging target;
    \item $k$ indexes over the $d$ total pixels or voxels in the imaging target;
    \item $\mu_{m, j}$ is the mass attenuation coefficient for basis material $m$ at X-ray wavelength $j$;
    \item $A_{i, k}$ is the entry of the projection matrix (derived from the Radon transform) denoting the fractional contribution of pixel/voxel $k$ to the line integral at angle $i$ --- each such scalar is nonnegative;
    \item $\{x_{k,m}\}$ represents the unknown target image with $k$ indexing spatial location (pixel or voxel index) and $m$ indexing basis material; each scalar $x_{k,m}$ is nonnegative.
\end{itemize}
In this forward model, the strictly positive scalars  $s_{\omega, i, j}$ and $\mu_{m,j}$ can be estimated in a calibration step, as can the average intensities $\Photon_{\omega,i}$. Accordingly, we assume that these quantities are known. For notational convenience, we assume 
that the detector sensitivity $s_{\omega, i, j}$ and average incident photon count $\Photon_{\omega, i}$ depend only on wavelength (governed by $\omega$ and $j$) and not on angle $i$, to write the forward model (in expectation) as
\begin{align}\label{model:multi-material}
    \EE[y_{\omega, i}(x)] &= \Photon_{\omega} \sum_{j = 1}^W s_{\omega, j} \exp \left (-\sum_{m = 1}^M \sum_{k = 1}^d \mu_{m, j}A_{i, k}x_{k,m} \right ).
\end{align}
We focus on a special case of this forward model that corresponds to a single basis material ($M = 1$), allowing us to remove the indexing over $m$ and thereby write the model as
\begin{align*}
    \EE[y_{\omega, i}(x)] &= \Photon_{\omega} \sum_{j = 1}^W s_{\omega, j} \exp \left (-\sum_{k = 1}^d \mu_{j}A_{i, k}x_{k} \right ).
\end{align*}
The restriction to a single basis material is fundamental to our proposed optimization strategy.  
This assumption is reasonable in several CT applications~\cite{chandel2014micromechanical, sietins2021fiber}, including those in which other materials are known and the goal is to recover a single unknown material. For example, in medical CT, it approximates perfusion imaging of a single contrast agent against a known background~\cite{pourmorteza2016reconstruction}, as well as imaging of the human body without contrast agents. See our experiment in Figure~\ref{fig:comparison} and the corresponding caption for an illustration. 

\newcommand{\plotcropF}[1]{%
\adjincludegraphics[trim={{0\width} {0\height} {0.02\width} {0\height}}, clip, width=\linewidth]{#1}%
}

\newcommand{\plotcropTR}[1]{%
\adjincludegraphics[trim={{0\width} {0\height} {0.2\width} {0\height}}, clip, width=\linewidth]{#1}%
}

\newcommand{\plotinsetTR}[1]{%
\adjincludegraphics[trim={{0.3\width} {0.6\height} {0.4\width} {0.1\height}}, clip, width=\linewidth]{#1}%
}

\begin{figure}[t]
\centering
    \begin{subfigure}[b]{0.22\textwidth}
        \centering
        \plotcropF{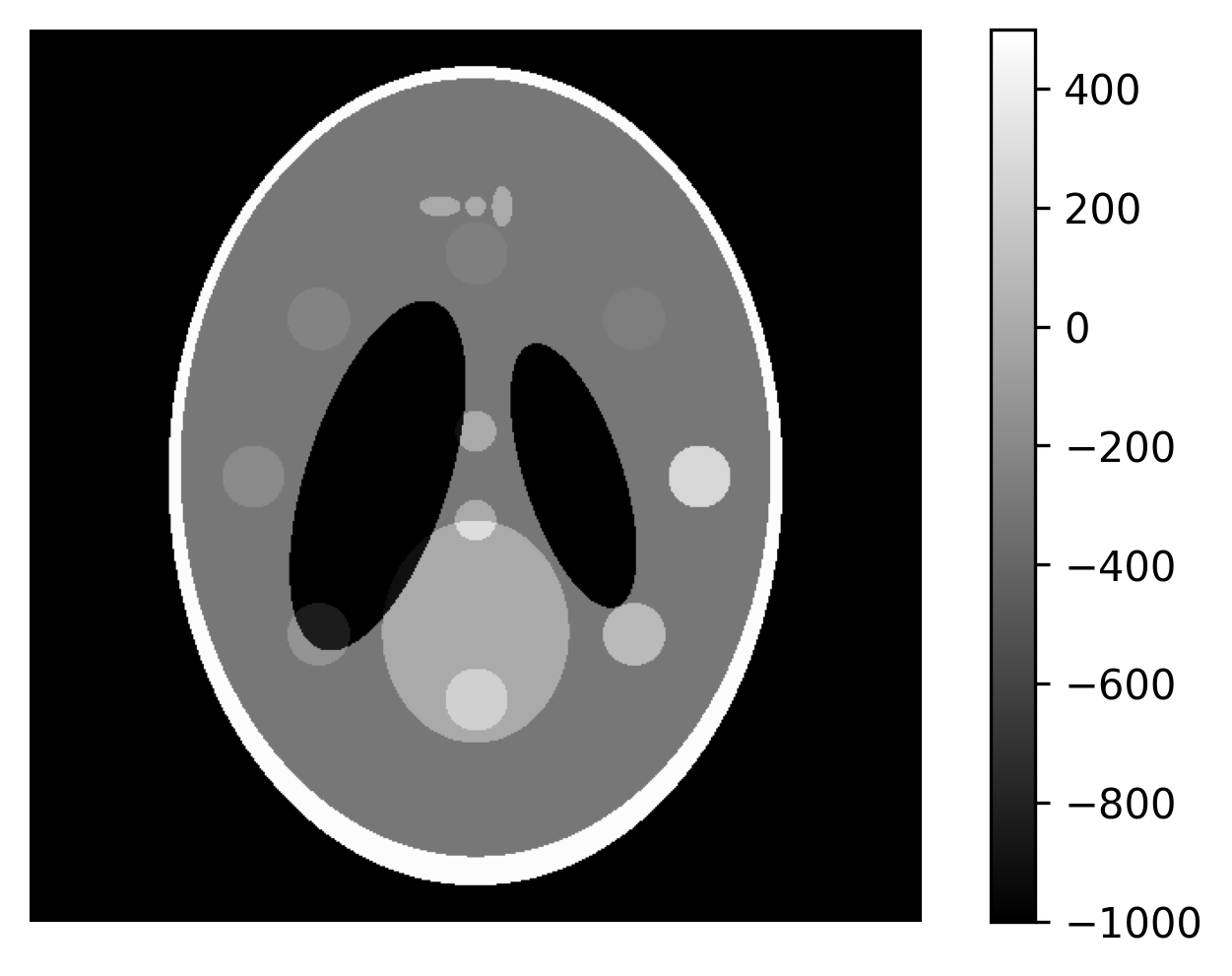}
        \plotinsetTR{figures/linear_comparison/gt.png}
        \caption{Ground truth}
        \label{fig:subfig1}
    \end{subfigure}
    \hfill
    \begin{subfigure}[b]{0.2\textwidth}
        \centering
        \plotcropTR{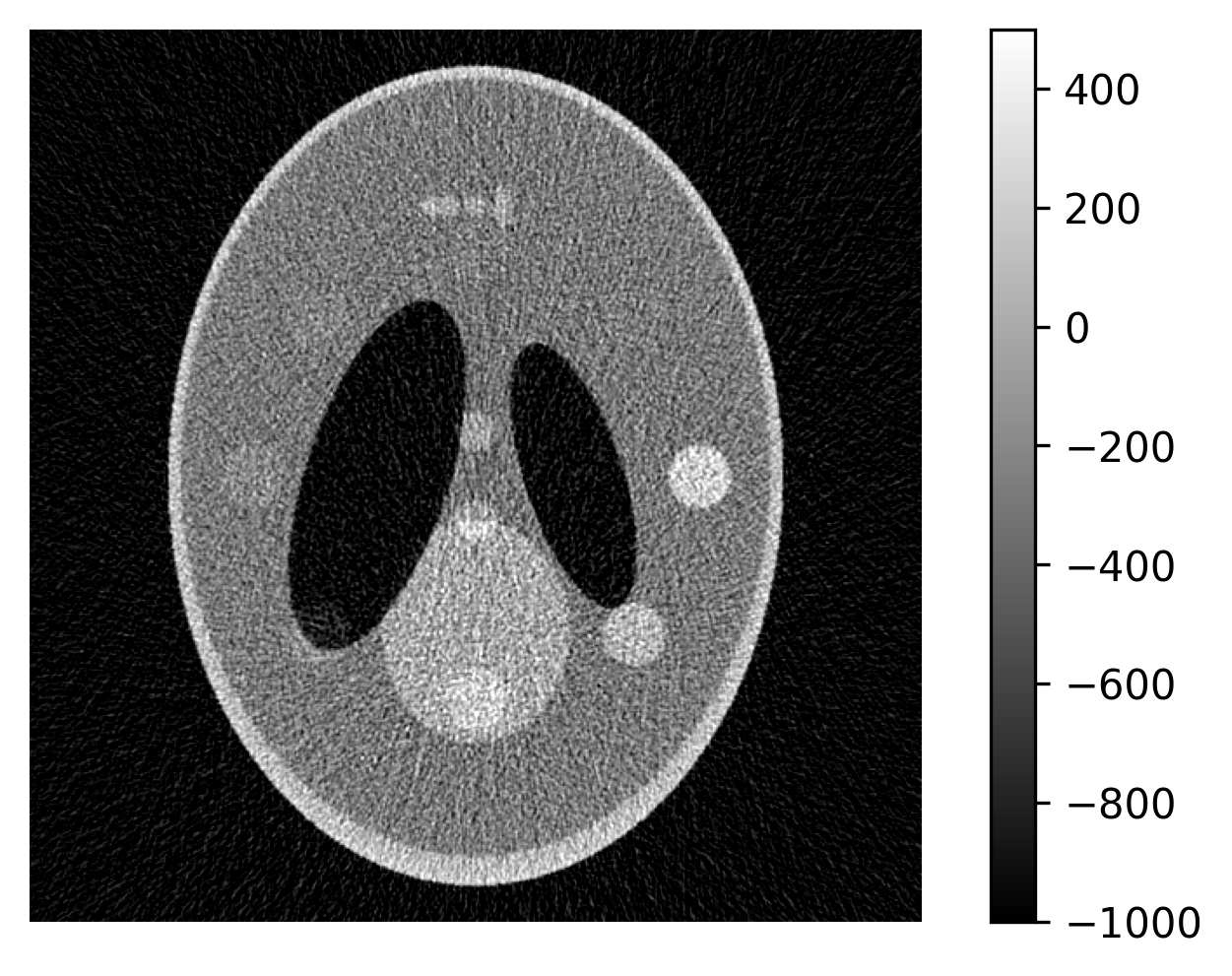}
        \plotinsetTR{figures/linear_comparison/fdk.png}
        \caption{FDK}
        \label{fig:subfig2}
    \end{subfigure}
    \hfill
    \begin{subfigure}[b]{0.2\textwidth}
        \centering
        \plotcropTR{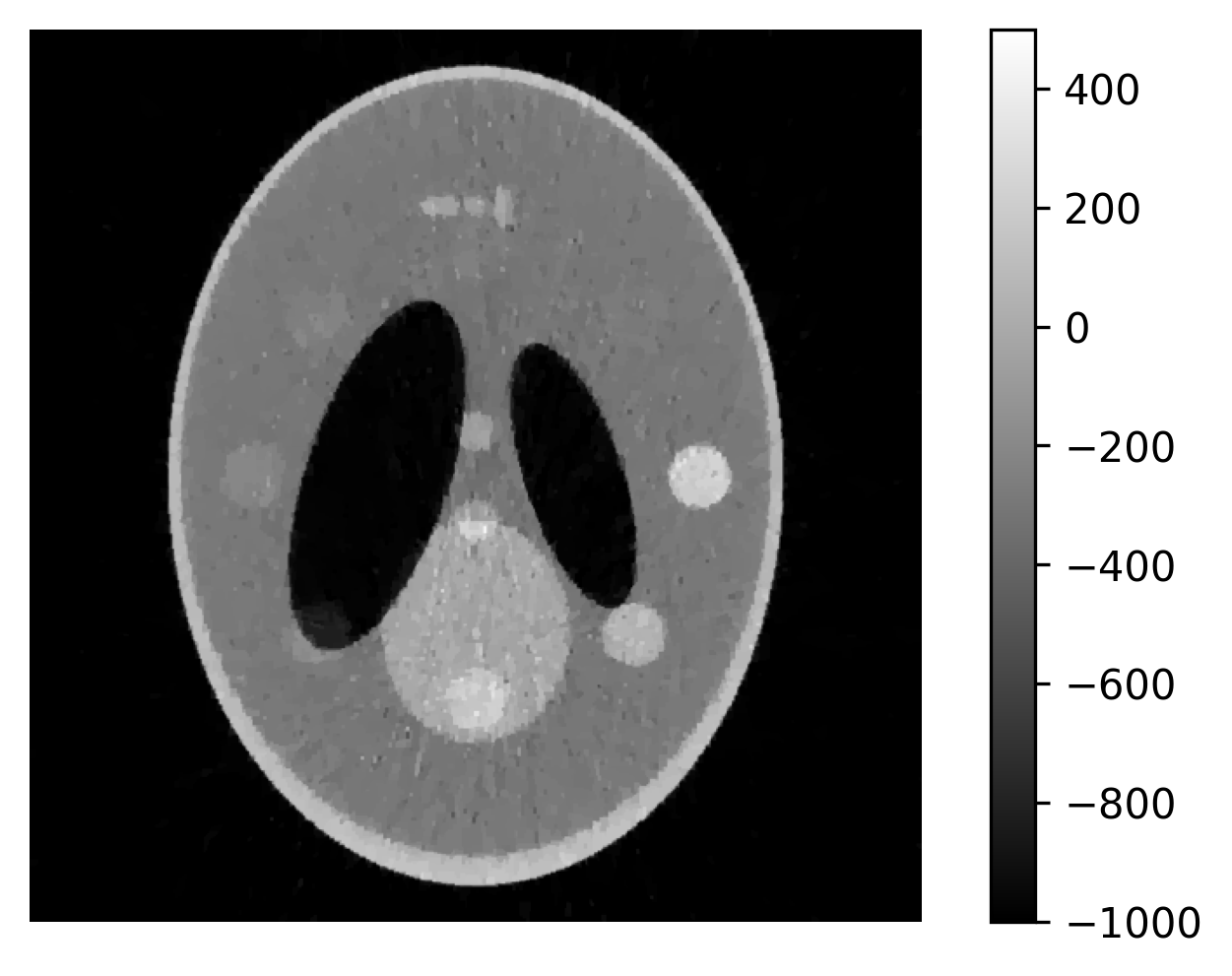}
        \plotinsetTR{figures/linear_comparison/fdk_tv.png}
        \caption{FDK with TV}
        \label{fig:subfig3}
    \end{subfigure}
    \hfill
    \begin{subfigure}[b]{0.2\textwidth}
        \centering
        \plotcropTR{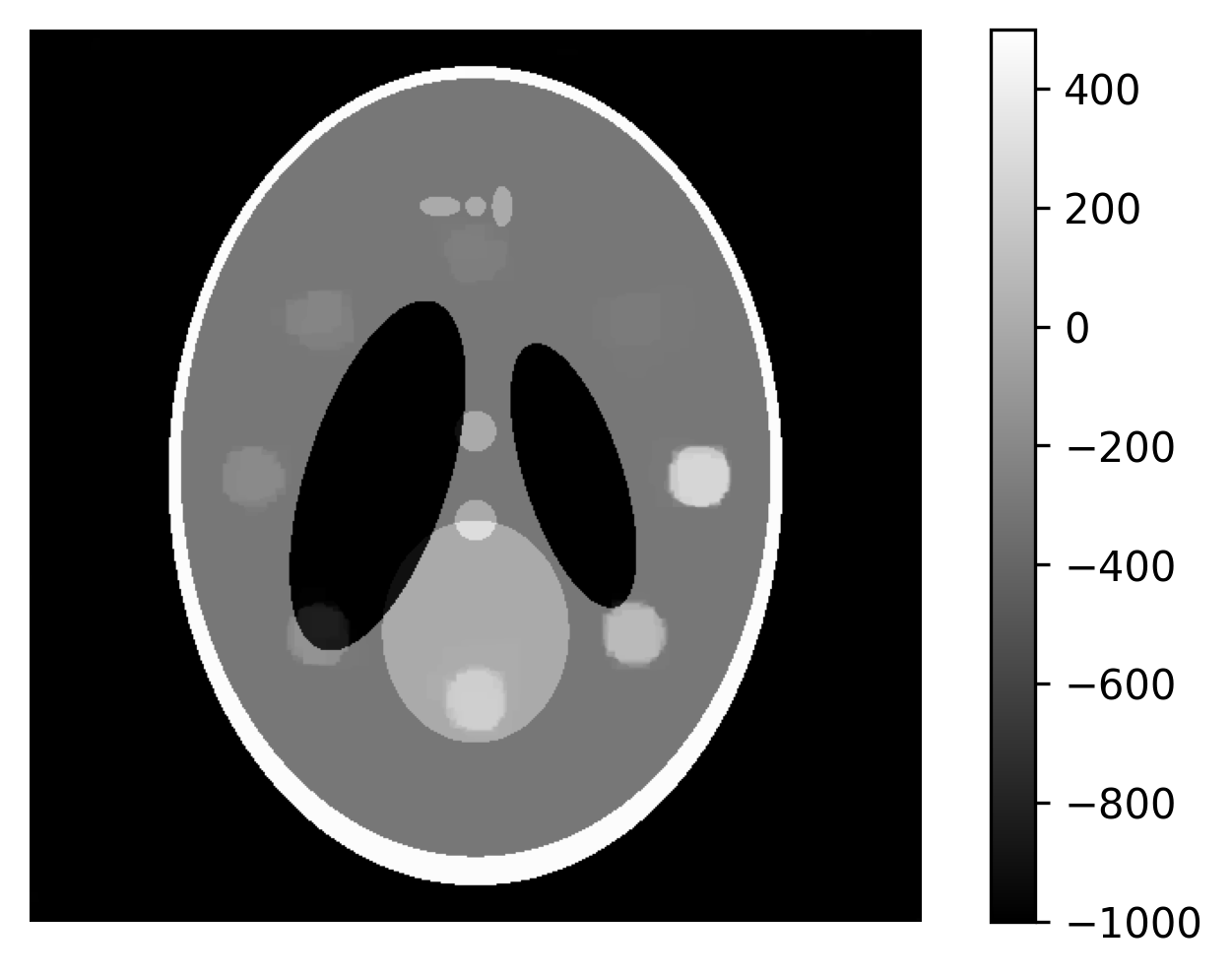}
        \plotinsetTR{figures/linear_comparison/exact.png}
        \caption{EXACT}
        \label{fig:subfig4}
    \end{subfigure}
    \caption{Panel~(a) shows the ground truth of the target phantom, which has dimension $d = 512 \times 512$ and contains three materials: water in the center, an outer ring of bone, and circular regions of iodine at varying concentrations. Panels~(b) and (c) show reconstructions using the FDK method~\cite{feldkamp1984practical} designed for the linear model via a monochromatic approximation. Panel~(c) additionally projects the FDK solution onto the oracle TV ball, whereas panel~(b) omits this projection. Panel~(d) shows the reconstruction of the EXACT method proposed in this paper, which is designed for the polychromatic model. In panel~(d), the water and bone regions are assumed to be known, and we solve the single-material reconstruction task of recovering the iodine regions; see~\ref{sec:math-model-multi-material} for the mathematical formulation of recovering a single material when other materials are assumed to be known. The EXACT method employs a projection step onto the total variation (TV) ball using the oracle TV information. In all experiments, the sample size is $n = 984 \times 1024$, the average source intensity is $10^{3}$ photons per detector cell, and the data are generated according to the Poisson noise model. All plots are in Hounsfield Units following the same colorbar as in panel~(a).}
    \label{fig:comparison}
\end{figure}

To arrive at the model we study in this paper, we restrict to a single detector window for notational convenience,\footnote{This restriction is made purely for convenience and is not fundamental to our approach. Indeed, in our CT simulations we do include multiple (distinct but overlapping) detector windows, as visualized in \Cref{fig:simulation-setup}.}
thereby removing our indexing over $\omega$ to write
\begin{align} \label{eq:model-orig}
    \EE[y_{i}(x)] &= \Photon \sum_{j = 1}^W s_{j} \exp \left (-\mu_{j} \langle a_i, x \rangle \right ),
\end{align}
where $a_i \in \R^{d}$ is the row of $A$ corresponding to projection angle $i$ and $x \in \R^d$ is the target image or volume. 
In our analysis, we consider both a general model in which the projection vectors $a_i$ are entrywise nonnegative (as they are in practice), as well as a model in which we can prove more precise results and compare directly with recent work, which assumes Gaussian measurements $a_i$. We unify these models using 
the convention of~\cite{fridovich2023gradient, charisopoulos2024nonlinear} and work with the forward model
\begin{align} \label{eq:model-pre}
    \EE[y_{i}(x)] &= \Photon \sum_{j = 1}^W s_{j} \exp \left (-\mu_{j} \langle a_i, x \rangle_+ \right ).
\end{align}
Here, we use the shorthand $t_{+} = \max\{ t,0 \}$ for any scalar $t \in \real$; the thresholding operation ensures that the ray projection values are always nonnegative even if the entries of $a_i$ may be negative, as is possible in the Gaussian model. 

In practice, we may have prior information about the signal, which we encode by assuming that $x \in X \subseteq \real^d$. For example, the set with small total variation (TV) norm~\cite{strong2003edge,osher2005iterative} is one such structured set which is particularly suited to imaging applications; we use this TV constraint along with a nonnegativity constraint in our simulations. Prior work on signal recovery under models~\eqref{eqn:referenceforwardmodel} and~\eqref{eq:model-pre} is discussed in detail in Section~\ref{sec:contributions}.

\subsection{Comparison with monochromatic approximation}
A classical approach to estimating the unknown $x$ relies on the monochromatic approximation and linearization of the nonlinear forward model~\eqref{model:multi-material}. Specifically, one applies the logarithm to both sides of Eq.~\eqref{model:multi-material} and then solves the resulting linear system. This logarithmic preprocessing is, in fact, built into commercial CT scanners~\citep{fu2016comparison}. However, the operation becomes numerically unstable when $y_{\omega, i}(x)$ approaches zero, which commonly occurs for rays traversing high-density materials (e.g., metal) or in low-dose CT scans with few X-ray photons. Such instability manifests as streak artifacts, as illustrated in~\cite[Figures~1 and~2]{fridovich2023gradient}. 

Moreover, the monochromatic approximation of the polychromatic model~\eqref{model:multi-material} introduces reconstruction errors and prevents the recovery of a specific single material when the other materials are known. In contrast, the nonlinear forward model that accommodates the polychromatic nature of measurements enables the design of algorithms that reconstruct a single material under the assumption that the others are known; see~\ref{sec:math-model-multi-material} for how to model single material reconstruction when the other materials are assumed to be known. Such a capability is especially valuable in applications such as perfusion imaging~\cite{pourmorteza2016reconstruction}. To illustrate the advantages of using the polychromatic forward model and as a preview of our results, we compare two methods: the FDK algorithm~\cite{feldkamp1984practical} --- which is based on the linear model via monochromatic approximation --- 
and the EXACT method proposed in this paper (see Eq.~\eqref{extragrad-method}), which operates directly with the polychromatic model. In our experiments, the ground-truth phantom (Figure~\ref{fig:subfig1}) contains three materials: water in the center, an outer ring of bone, and circular iodine regions with varying concentrations. We assume the regions of water and the bone are known to us, for example from a pre-scan of a patient before injection of contrast agent\footnote{In this comparison, the EXACT method relies on a strong prior: knowledge of the true regions of water and bone, while the FDK method functions without such information; see~\ref{sec:math-model-multi-material} for details.}, and our reconstruction goal is to recover the iodine regions, which are designed to reflect realistic contrast agent concentrations as they might vary over time during perfusion imaging. We simulate a low X-ray source intensity of only $10^3$ photons per detector cell, since perfusion imaging requires repeated scans of the same patient over time and thus X-ray dose is accumulated. As shown in Figures~\ref{fig:subfig2} and~\ref{fig:subfig3}, the FDK reconstructions are noticeably noisier, and several iodine regions are not accurately recovered. In contrast, Figures~\ref{fig:subfig4} demonstrates that the EXACT method accurately recovers the iodine regions and can distinguish between different iodine concentrations. This demonstrates a clear advantage of leveraging the polychromatic model together with our proposed algorithm.

\subsection{Formal model and approach}
As described above, we are interested in recovering an unknown signal from a nonlinear measurement model of the form~\eqref{eq:model-pre}. We assume throughout that our structured set $X \subseteq \mathbb{R}^d$ is a closed, convex set.
Henceforth, we denote the ground truth signal by $x_{\star}$, and our task is to recover $x_{\star} \in X$ from measurement vectors $\{a_{i}\}_{i=1}^{n}$ and the associated responses $\{y_{i}\}_{i=1}^{n}$. Formally, we rewrite the measurement model as
\begin{align}\label{model}
\EE[y_{i} | a_{i} ] = \Photon \sum_{j=1}^{\N} s_{j} \exp \big(- \coeff_{j} \langle a_{i}, x_{\star} \rangle_{+} \big) \quad \text{for } i = 1, \ldots, n.
\end{align}
The parameters $\Photon$ and $\{s_{j}, \coeff_{j}\}_{j=1}^{\N}$ are known and satisfy 
\[
    \Photon > 0, \quad s_{j} \in (0,1], \quad \sum_{j=1}^{\N} s_{j} = 1, \quad \text{and} \quad \coeff_{j} >0 \quad \text{for all } j = 1, \ldots, \N.
\]
Besides the conditional expectation relation in Eq.~\eqref{model}, we do not place any assumptions on the conditional distribution $y_i | a_i$. This flexibility allows us to cover many practical noise models. A salient such example is the Poisson shot-noise model~\cite{schmidt2023constrained} --- 
in our notation, this is given by
\begin{align}\label{Poisson-model}
    y_{i} \overset{\mathsf{i.i.d.}}{\sim} \mathsf{Poi} \bigg( \Photon \sum_{j=1}^{\N} s_{j} \exp \big(- \coeff_{j} \langle a_{i}, x_{\star} \rangle_{+} \big)  \bigg).
\end{align}
 In the CT literature, there are other noise assumptions for this problem, including those that model pulse-pileup or additional noise induced due to instrumentation errors~\cite{danielsson2021photon, persson2020detective}. With an additional calibration step, our model can in principle incorporate all these sources of noise, since the only requirement in Eq.~\eqref{model} is that the responses are unbiased. The effect of noise is then captured by an error quantity $\Err(x_{\star},X)$ that we define presently --- see Definition~\ref{def-error-term}. In~\ref{sec:math-model-multi-material}, we illustrate how Poisson noise and electronic noise can be modeled within our framework, and we derive bounds on the associated error $\Err(x_{\star},X)$.

For notational convenience in the rest of the paper, we define the function $h: \real \to \real$ via
\begin{align}\label{function-h-plus}
  h(t) := \sum_{j=1}^{\N} s_{j} \exp\big( - \coeff_{j} t_{+} \big), \quad \text{for all } t\in \real,
\end{align}
where we recall that $t_+ = \max(0, t)$ denotes the ReLU operation.
Using this shorthand, our model is given by 
\[
\EE[y_i|a_i] = \Photon \cdot h(\inprod{a_i}{x_*}).
\]

Our approach is based on defining a vector-valued map $F:X \to \real^{d}$, given by
\begin{align}\label{operator}
  F(x) = \frac{1}{n} \sum_{i=1}^{n} \big[y_{i} - \Photon \cdot h( \inprod{a_{i}}{x} )\big] \cdot a_{i}, \;\; \text{for all } x \in X.
\end{align}
By construction, the map $F$ is continuous. Two crucial observations that will motivate our methodological development are: \\
\noindent (a) The \emph{expectation} of $F$ --- call this operator $\overline{F}$ --- has a zero at $x_{\star}$, in that $\overline{F}(x_{\star}) = 0$; and \\
\noindent (b) The operator $F$ is also \emph{monotone}, in that for all $x, x' \in \real^d$, we have 
\[
\langle F(x) - F(x'), x - x' \rangle \geq 0. 
\]
Indeed, a short calculation to verify the second fact is provided below; for any $x, x' \in \real^d$, we have
\begin{align}\label{monotone}
  \langle F(x) - F(x'), x - x' \rangle = \frac{1}{n} \sum_{i=1}^{n} \Photon \cdot \big[h( \inprod{a_{i}}{x'}) - h(\inprod{a_{i}}{x})\big] \cdot \big[ \inprod{a_{i}}{x} - \inprod{a_{i}}{x'} \big] \geq 0,
\end{align}
where the inequality follows since the function $h(\cdot)$ is nonincreasing.

Given these observations, we now frame the problem as one of finding a fixed point of the variational inequality defined by the operator $F$; see the classical references~\cite{kinderlehrer2000introduction,rockafellar2009variational} for background on variational inequalities. We are interested in a point $\widehat{x} \in X$ that satisfies 
\begin{align}\label{def-strong-VI-solution}
\inprod{F(\widehat{x})}{x - \widehat{x}} \geq 0 \text{ for all } x \in X.
\end{align} 

With the goal of converging to $\widehat{x}$ --- which one should expect to be close to the zero $x_{\star}$ of $\overline{F}$ --- we study the projected extragradient method~\cite{korpelevich1976extragradient} for this problem, which proceeds as follows: With an initialization $x_{1} \in X$, execute the steps
\begin{subequations}\label{extragrad-method}
\begin{align}
\label{step1-extragrad}
x_{t+0.5} &= \mathcal{P}_{X}\big(x_{t} - \gamma_{t} F(x_{t})\big), \\
\label{step2-extragrad}
x_{t+1} &= \mathcal{P}_{X}\big(x_{t} - \gamma_{t} F(x_{t+0.5}) \big), \text{ for }t = 1,2,\dots
\end{align}
\end{subequations}
where $\gamma_{t}$ is the step size at iterate $t$, and $\mathcal{P}_{X}:\real^{d} \to X$ is the $\ell_2$-projection operator onto $X$. Ignoring the cost of these projections for the moment, the remainder of iteration~\eqref{extragrad-method} can be implemented in linear time. Furthermore, owing to the convexity of $X$, the projection step can typically be implemented via an efficient algorithm. For example, when $X$ is a ball with bounded total variation norm, achieving an $\epsilon$-optimal solution of the projection (in $\ell_2$ norm) has computational complexity $\mathcal{O}\big(d / \epsilon\big)$; see~\cite[Theorem~2]{fadili2010total}. 
In our experiments, we implement the projection operator for the case where the constraint set $X$ is the intersection of the bounded total variation set and the nonnegative orthant; see Section~\ref{sec:experiments} for details. 

Our methodology is inspired by algorithmic ideas that were proposed for other monotone single-index models~\cite{gallant1990perceptron,kalai2009isotron,kakade2011efficient} as well as the recent formalism of these methods using variational inequalities due to~\cite{juditsky2019signal, juditsky2020statistical}. 
We refer to this extragradient algorithm~\eqref{extragrad-method} applied to the specific CT forward model~\eqref{model} as \algname{}: EXtragradient Algorithm for Computed Tomography. 

Two additional comments about our algorithm are worth making. First, note that the EXACT method is not a loss-minimization-based method, whereas several popular algorithms are explicitly designed to minimize specific loss functions. For example, the gradient descent method~\cite{fridovich2023gradient} minimizes an $\ell_{2}$ loss, PolyakSGM~\cite{charisopoulos2024nonlinear} minimizes a nonsmooth $\ell_{1}$ loss, and ADMM~\cite{barber2024convergence} minimizes the negative log-likelihood of the Poisson model (or a similarly constructed posterior distribution). See Section~\ref{sec:experiments} for a detailed description of these loss-minimization-based algorithms. Second, note that the EXACT method enforces structure through projection as opposed to penalization, which is common in many iterative algorithms for CT (e.g.~\cite{barber2016algorithm,long2014multi,weidinger2016polychromatic,mechlem2017joint}). These two notions of regularization should be regarded as very similar to each other, and one can replace projection steps in Eqs.~\eqref{extragrad-method} by proximal steps. We discuss such extensions in more detail in Section~\ref{sec:discussion}.

\subsection{Additional notation} 

For a positive integer $k$, we let $[k]$ denote the set of natural numbers less than or equal to $k$. For $a,b\in \real$, we let $a\vee b = \max\{a,b\}$ and $a \wedge b = \min\{a,b\}$.
For notational convenience, we define $A \in \real^{n \times d}$ to be the data matrix whose $i$-th row is the measurement vector $a_{i}$ for each $i\in[n]$, and let $\Sigma = \frac{1}{n}A^{\top} A \in \real^{d \times d}$ denote the associated average correlation matrix. We let $I_{d}$ denote the identity matrix of dimension $d \times d$.
 For a symmetric and positive semidefinite matrix $M \in \real^{d\times d}$ and a closed set $S \subseteq \real^{d}$, we define the restricted minimum and maximum eigenvalues as 
\begin{align} \label{eq:restricted-eigenvalues}
    \lambda_{\mathsf{min}}(M,S) = \min_{v \in S} \frac{\| M v \|_{2}}{\|v\|_{2}} \quad \text{and} \quad 
    \lambda_{\mathsf{max}}(M,S) = \max_{v \in S} \frac{\| M v \|_{2}}{\|v\|_{2}}.
\end{align}
For a matrix $M \in \real^{n \times m}$, we define its Frobenius norm and operator norm as 
\[
    \|M\|_{\mathsf{F}} := \Big( \sum_{i=1}^{n} \sum_{j=1}^{m} M_{i,j}^{2} \Big)^{1/2} \quad \text{and} \quad \|M\|_{\mathsf{op}} : = \max_{v \in \real^{m},\|v\|_{2} = 1} \| Mv\|_{2}.
\] 
For two sets $S_{1},S_{2} \subseteq \real^{d}$, we define their Minkowski sum $S_{1}+S_{2} = \{s_{1} + s_{2}:s_{1} \in S_{1}, s_{2} \in S_{2}\}$, and denote their Minkowski difference by $S_1 - S_2$.
For $r>0$, we let $\mathbb{B}(r):=\{x\in \real^{d}: \|x\|_{2} \leq r\}$ denote the $\ell_2$ ball of radius $r$. For $v\in \real^{d}$ and $r>0$, we let $\mathbb{B}(r;v):= \{x\in \real^{d}: \|x-v\|_{2} \leq r\}$ denote the $\ell_2$ ball of radius $r$ centered at $v$. For $v\in \real^{d} \setminus \{0\}$, we let $P_{v} = vv^{\top}/\|v\|_{2}^{2}$ and $P_{v}^{\perp} = I_{d} - vv^{\top}/\|v\|_{2}^{2}$ denote ${d\times d}$ projection matrices that when applied to a vector, project it onto the subspace that is parallel and orthogonal to $v$, respectively. We let $\mathbbm{1}\{\cdot\}$ denote the indicator function, which takes value $1$ when its argument is true and $0$ otherwise. 
We denote by $\NORMAL(\mu, M)$ a multivariate normal distribution with mean $\mu$ and covariance matrix $M$. For two sequences of nonnegative reals $\{f_n\}_{n\geq 1}$ and $\{g_n \}_{n \geq 1}$, we use $f_n \lesssim g_n$ to indicate that
there is a universal positive constant $C$ such that $f_n \leq C g_n$ for all $n \geq 1$. The relation $f_n \gtrsim g_n$ implies that $g_n \lesssim f_n$, and we say that $f_n \asymp g_n$ if both $f_n \lesssim g_n$ and $f_n \gtrsim g_n$ hold simultaneously. We also use standard order notation $f_n = \order(g_n)$ to indicate that $f_n \lesssim g_n$ and $f_n = \ordertil(g_n)$ to indicate that $f_n \lesssim
g_n \log^c n$, for a universal constant $c>0$. We say that $f_n = \Omega(g_n)$ (resp. $f_n = \widetilde{\Omega}(g_n)$) if $g_n = \order(f_n)$ (resp. $g_n = \ordertil(f_n)$). The notation $f_n = o(g_n)$ is used when $\lim_{n \to \infty} f_n / g_n = 0$, and $f_n = \omega(g_n)$ when $g_n = o(f_n)$. Throughout, we use $c, C$ to denote universal positive constants, and their values may change from line to line. All logarithms are with respect to the natural base unless otherwise stated. 

\subsection{Contributions and comparison to related work} \label{sec:contributions}

With this setup in hand, we are now in a position to describe our contributions and to compare them with prior work. 
\begin{enumerate}
    \item \textbf{Convergence guarantees for general CT model.} We provide a rigorous convergence analysis for \algname{} (\Cref{extragrad-method}) under the assumption that the measurement vectors and true signal have nonnegative entries, which is indeed the case in CT applications. Our results accommodate an arbitrary noise model, and we show that the eventual recovery error depends explicitly on a functional of this noise process. We emphasize that our results provide, to the best of our knowledge, the first rigorous guarantees for signal reconstruction in the setting of nonnegative measurements that encompass the Radon transform, a polychromatic X-ray source, and multiple detector windows.
    In more detail, we show that the iterates $\{x_{t}\}$ generated by \algname{} converge linearly to within a small neighborhood of $x_{\star}$ when the restricted smallest eigenvalue of the matrix $\Sigma$ is positive, specifically, when $\lambda_{\min}(\Sigma, X - X) > 0$. 
    Note that such a condition is necessary even for the identifiability of the linear variant of the model, i.e., even if we observed linear functions of the structured signal $x_{\star}$ using the classical Radon transform. For our detailed guarantees and examples see Theorem~\ref{thm-positive-measurements} and the accompanying discussion.
    
For comparison, we note that several other algorithms have been proposed for signal recovery in the model~\eqref{eq:model-orig}, based on logarithmic transformations as preprocessing~\cite{maass2009image,niu2014iterative}, conjugate gradient methods~\cite{cai2013full}, separable quadratic surrogates~\cite{long2014multi, weidinger2016polychromatic, mechlem2017joint}, and the ADMM algorithm~\cite{barber2016algorithm, schmidt2017spectral}. See the paper by~\cite{mory2018comparison} for a detailed empirical comparison of these methods on some synthetic and clinical reconstruction tasks. Preprocessing-based methods are known to perform poorly with noise since they are based on heuristic inversion of nonlinearities in the model~\cite{mory2018comparison}, and all the iterative methods, save for one, do not come equipped with theoretical guarantees. The exception is the coSSCIR method by~\cite{schmidt2017spectral}, which has since been applied to a variety of CT reconstruction tasks~\cite{sidky2018three, schmidt2022addressing,rizzo2022material}. Rigorous guarantees on its performance have been proved under assumptions such as restricted strong convexity of the associated Poisson negative log-likelihood objective function~\cite{barber2024convergence}. However, to our knowledge, this assumption has not been rigorously verified when the observations are generated\footnote{The only exception to this that we are aware of is the monochromatic case, where the Poisson negative log-likelihood is a convex function~\cite{barber2024convergence}. Note that this function is nonconvex in the polychromatic models~\eqref{eq:model-orig} and~\eqref{eq:model-pre}.} by either model~\eqref{eqn:referenceforwardmodel} or model~\eqref{eq:model-orig}.

    Besides our theoretical contribution, we also implement our algorithm in practice. In Section~\ref{sec:experiments}, we present a suite of experiments for recovering a phantom image from measurements of the form~\eqref{eq:model-pre}, comparing \algname{} with algorithms that come with provable guarantees in at least some setting --- in particular, coSSCIR~\cite{schmidt2017spectral} as well as two first-order methods~\cite{fridovich2023gradient,charisopoulos2024nonlinear} that we discuss shortly. Overall, our algorithm achieves the lowest recovery error among all these methods while also having the smallest running time. In particular, our approach often yields comparable reconstruction quality, both visually and quantitatively, using lower X-ray dose compared to these baselines. 
    
    \item \textbf{Convergence guarantees for Gaussian model.} To further explore the statistical and algorithmic properties of our algorithm, we study convergence guarantees for \algname{} with Gaussian measurements. Indeed,~\cite{fridovich2023gradient} introduced a monochromatic version of the stylized model~\eqref{eq:model-pre} under i.i.d. Gaussian measurements $a_i$ in order to provide rigorous sample and iteration complexity guarantees on signal recovery. 
    They studied the gradient descent algorithm on the mean squared error objective
    in the case with a single X-ray wavelength and noiseless measurements. 
    A follow-up paper by~\cite{charisopoulos2024nonlinear} considered the same setting and proposed a different first-order method based on optimizing the nonsmooth $\ell_1$ loss, and showed improved sample and iteration complexity guarantees, again in the monochromatic, noiseless setting with Gaussian measurements. 
    
    Our guarantees in this setting are easiest to describe and compare with prior work when there is no structure and no noise.
    In the proportional regime, where $n \asymp d$, we show that it takes  \algname{} (\Cref{extragrad-method})
    \[   
    t = \mathcal{O}\bigg(\exp(24\norm) \norm^{4} \log\bigg( \frac{5\norm}{\epsilon} \bigg)  \bigg) \text{ iterations to achieve }     \|x_{t} - x_{\star}\| \leq \epsilon.
\]
    On the other hand, in the regime $n \asymp \norm^{4} d$, it takes  \algname{}
    \[
    t = \mathcal{O}\bigg( \norm^{4} \log\bigg( \frac{5\norm}{\epsilon} \bigg) \bigg)
\text{ iterations to achieve } 
    \|x_{t} - x_{\star}\| \leq \epsilon.
    \]
    This guarantee (see Theorem~\ref{thm:mono-gaussian} for details) reveals an interesting trade-off between statistical performance and computational efficiency. It indicates that practitioners can use fewer samples—specifically, $\mathcal{O}(d)$—but this will result in increased computational time. Conversely, using more samples—approximately $\mathcal{O}(\norm^{4} d)$—reduces computation time significantly.
    
    Comparing with prior work in the monochromatic, noiseless setting with Gaussian measurements, we find that \algname{} requires fewer samples and converges faster than the gradient descent method developed in~\cite{fridovich2023gradient} as well as the non-smooth optimization method introduced by~\cite{charisopoulos2024nonlinear}. Concretely, \algname{} needs $\mathcal{O}(d)$ samples for recovery (without multiplicative factors in $\norm$), whereas~\cite{fridovich2023gradient} and~\cite{charisopoulos2024nonlinear} require $\mathcal{O}(\exp(\norm)d)$ and $\mathcal{O}(\norm^{4}d)$ samples, respectively. There are also significant gains in iteration complexity. In the regime $ n \gtrsim \norm^{4} d $, achieving $\epsilon$-accurate signal recovery using \algname{} requires $ \mathcal{O}(\norm^{4} \log(5\norm/\epsilon)) $ iterations. This represents an improvement over the iteration complexity presented by \cite{fridovich2023gradient}, which has a multiplicative factor scaling as $\exp(\norm)$, and by \cite{charisopoulos2024nonlinear}, in which the iteration complexity is scaled by a polynomial factor of $\norm$. For a detailed comparison, see Table~\ref{tab:complexity}.
    
\end{enumerate}


\section{Theoretical results}
We begin by giving a general-purpose recovery guarantee for the \algname{} algorithm in the presence of noisy measurements.  
Recall the continuous and monotone map $F$ defined in Eq.~\eqref{operator}. To state our general result in Theorem~\ref{general-thm} to follow, we assume that $F$ is $L$-Lipschitz continuous, in that it satisfies
\begin{subequations}
\begin{align}\label{lipschitzness}
\| F(x) - F(x') \|_{2} \leq L \| x-x' \|_{2} \quad \text{ for all } \;x,x' \in X.
\end{align}
We also assume that $F$ is $\monotone$-strongly pseudo-monotone around the point $x_{\star}$, in that
\begin{align}\label{strong-monotone}
\langle F(x) - F(x_{\star}),\;x- x_{\star} \rangle \geq \monotone \|x- x_{\star}\|_{2}^{2} \quad \text{ for all } \;x \in X.
\end{align} 
\end{subequations}
Note that we always have $L \geq \monotone \geq 0$. Conditions~\eqref{lipschitzness} and~\eqref{strong-monotone} will be verified concretely in the results to follow under variants of the following assumption\footnote{Assumption~\ref{assump-X} can likely be relaxed by assuming a fixed initialization, say at $0$.} that $X$ is a bounded convex set, i.e., 
\begin{assumption}\label{assump-X}
    For some finite scalar $R>0$, we have 
    \[
        X \subseteq \big\{ x \in \real^{d} : \|x - x_{\star}\|_{2} \leq R \big\}.
    \] 
\end{assumption} 
Such an assumption is weak, and only requires some rough knowledge of the signal strength $\| x_{\star} \|_2$ to set a suitable value of the maximum radius $R$ and thereby define the constraint set $X$. Note that $X$ can obey further structural assumptions in addition to satisfying Assumption~\ref{assump-X}.

A crucial object that enters our guarantees is the vector
\begin{align}\label{noise-vector}
    \zeta = F(x_\star) = \frac{1}{n} \sum_{i=1}^{n} \big( y_{i} - \Photon \cdot h(\langle a_{i}, x_{\star} \rangle) \big)\cdot a_{i},
\end{align}
where the function $h$ was defined in Eq.~\eqref{function-h-plus}. Note that $\EE[\zeta] = 0$ according to model~\eqref{model} and in the noiseless case we have $\zeta = 0$. We further define the following quantity that will be a convenient measure of statistical error in the problem: 
\begin{align}\label{def-error-term}
    \mathsf{Err}(x_{\star},X) := \sup_{x,x' \in X,\, x\neq x'} \Big\langle \zeta, \frac{x-x'}{\|x-x'\|_{2}} \Big\rangle.
\end{align}
This error quantity is a random variable that provides a measure how the noise interacts with the geometry of the constraint set $X$.  For concreteness, let us explicitly evaluate $\mathsf{Err}(x_{\star},X)$ in some special cases.

\begin{example}
    In the noiseless case, we have $\zeta = 0$, and therefore, $\mathsf{Err}(x_{\star},X) = 0$ pointwise for any $x_{\star}$ and $X$. \hfill $\clubsuit$
\end{example}

\begin{example} \label{prop-error-term}
    Consider the Poisson shot-noise model~\eqref{Poisson-model}, and suppose we are in the unstructured setting with $X = \mathbb{B}(R)$. Then there exists a universal positive constant $C$ such that with probability at least $1-\delta$, we have
    \begin{align*}
        \mathsf{Err}(x_{\star},X) \leq \frac{ \Big( \Photon \cdot \sum_{i=1}^{n} \|a_{i}\|_{2}^{2} \cdot h(\langle a_{i},x_{\star}\rangle) \Big)^{1/2} }{n} + C \cdot \frac{ \Photon \big(\| A A^{\top} \|_F\big)^{1/2} }{n} \cdot \log(2/\delta).
    \end{align*}
   We provide a proof of this claim in~\ref{sec:pf-prop-error-term}. \hfill $\clubsuit$
\end{example}

We are now ready to state our first result, which is a convergence result for monotone VIs.  \begin{theorem}\label{general-thm}
    Suppose that the data $\{(a_i, y_i)\}_{i=1}^{n}$ satisfy the single-index relationship~\eqref{model}, Assumption~\ref{assump-X} holds, and that the operator $F$ in Eq.~\eqref{operator} is both $L$-Lipschitz~\eqref{lipschitzness} as well as $\nu$-strongly pseudo-monotone~\eqref{strong-monotone}. Then the iterates $\{x_{t}\}$ generated according to \algname~\eqref{extragrad-method} with step size $\gamma_{t} = 1/(4L)$ and any $x_{1} \in X$ satisfy, for all $t > 0$, 
    \begin{align*}
        \| x_{t+1} - x_{\star} \|_{2} \leq \underbrace{\Big(1-\frac{\monotone}{8L}\Big)^{t/2} \|x_{1} - x_{\star}\|_{2}}_{\mathrm{Optimization\; error}} + \underbrace{\frac{4\mathsf{Err}(x_{\star},X)}{\monotone}}_{\mathrm{Statistical\; error}}.
    \end{align*}
\end{theorem}
While the quantity $\mathsf{Err}(x_{\star},X)$ is random, the conclusion of Theorem~\ref{general-thm} is deterministic. An immediate consequence is that for each $\epsilon \in (0,1)$, if algorithm~\eqref{extragrad-method} is run for
\begin{align}
    t \geq \frac{16L}{\monotone} \log\Big(\frac{\|x_{1} - x_{\star}\|_{2}}{\epsilon} \Big) \quad \text{iterations, then } \quad \| x_{t+1} - x_{\star} \|_{2} \leq \epsilon + \frac{4\mathsf{Err}(x_{\star},X)}{\monotone}.
\end{align}
Since we can set $\epsilon$ to be arbitrarily small, the eventual estimation error is controlled by the strong pseudo-monotonicity parameter $\monotone$ and the quantity $\mathsf{Err}(x_{\star},X)$ that depends on the geometry of the set $X$. The convergence is geometric, and the rate is governed by the condition number $L/\monotone$.
Note that our convergence guarantee decomposes into a sum of two terms, the first capturing the optimization error and the second capturing the eventual statistical error. We provide the proof of Theorem~\ref{general-thm} in Section~\ref{sec:pf-general-thm}. Since we impose strong pseudo-monotonicity only around $x_{\star}$ and not the sample fixed point $\widehat{x}$, the proof is non-standard and requires careful decompositions and control of the various error terms. Finally, let us remark that while Theorem~\ref{general-thm} is stated for the CT model~\eqref{model}, it should really be thought of as a result for the extragradient algorithm if applied to an operator $F$ of the form~\eqref{operator}, if this operator happens to be Lipschitz and strongly pseudo-monotone. Accordingly, this result may find broader applications in the study of monotone single-index models.

\subsection{Positive measurements and positive signal}
In this section, we assume that the measurement vectors and the signal $x_{\star}$ all have nonnegative entries. Note that in computed tomography applications, the measurement vector $a_{i}$ is a set of known coefficients derived from the Radon transform~\cite{4307775}, whose entries are indeed nonnegative. All entries of $x_{\star}$ are also nonnegative since entry $j$ of $x_{\star}$ captures the density of the unknown material at voxel $j$.

Recall that we use $X-X := \{x-x':x,x' \in X\}$ to denote the Minkowski difference between the set $X$ and itself. Recall the definition of restricted eigenvalues from Eq.~\eqref{eq:restricted-eigenvalues}.
\begin{assumption}\label{assump-positive}
    For each $i \in [n]$, the measurement vector $a_{i}$ has nonnegative entries. Furthermore, the signal $x_{\star}$ has nonnegative entries. The structured set satisfies $X \subseteq \{x\in \real^{d}: x\geq 0\}$. The correlation matrix $\Sigma = \frac{1}{n} A^{\top} A$ satisfies $\lambda_{\mathsf{min}}(\Sigma,X-X)>0$. 
\end{assumption}
Note that it typically holds that $\lambda_{\mathsf{min}}(\Sigma, X - X) \gg \lambda_{\mathsf{min}}(\Sigma, \real^{d})$ when the constraint set $X$ has fewer degrees of freedom, such as the set of vectors with bounded TV norm. Thus, even if $\lambda_{\mathsf{min}}(\Sigma,\real^{d}) = 0$, it is still possible to have $\lambda_{\mathsf{min}}(\Sigma, X - X) > 0$ by choosing an appropriate constraint set.

Under Assumption~\ref{assump-positive}, we verify in Section~\ref{sec:pf-thm-positive-measurements} that the strong pseudo-monotonicity parameter $\monotone$ and the Lipschitz parameter $L$ satisfy
\[
    \monotone \geq \Photon \cdot \lambda_{\mathsf{min}}(\Sigma, X-X) \sum_{j=1}^{\N} s_{j}\coeff_{j} \exp\big( - \coeff_{j} \underset{x \in X, i\in [n]}{\max} \langle a_{i}, x\rangle \big) \quad \text{and} \quad L \leq \Photon \cdot \lambda_{\mathsf{max}}(\Sigma, \real^{d}) \sum_{j=1}^{\N} s_{j}\coeff_{j},
\]
respectively.
To reduce the notational burden, we define the quantity
\begin{align}\label{def-kappa}
    \kappa = \exp\Big( \max_{j \in [\N]} \coeff_{j} \cdot \max_{x \in X, i\in [n]} \inprod{a_{i}}{x} \Big) \cdot \frac{\lambdamax(\Sigma,\real^{d})}{ \lambdamin(\Sigma,X-X)},
\end{align}
which is an upper bound on the condition number $L/\nu$. Under Assumptions~\ref{assump-X} and~\ref{assump-positive}, the scalar $\kappa$ is guaranteed to be nonnegative and finite. As a consequence, applying Theorem~\ref{general-thm} yields the following result. 
\begin{theorem}\label{thm-positive-measurements}
    Suppose that the data $\{(a_i, y_i)\}_{i=1}^{n}$ satisfy the single-index relationship~\eqref{model}, and recall the model parameters $\{s_{j},\coeff_{j}\}_{j=1}^{\N}$ and $\Photon$ in Eq.~\eqref{model} and the error quantity $\mathsf{Err}(x_{\star},X)$ in Eq.~\eqref{def-error-term}. Suppose Assumption~\ref{assump-X} holds for some $R>0$, Assumption~\ref{assump-positive} holds, and the iterates $\{x_{t}\}$ are generated according to \algname~\eqref{extragrad-method} with step size
    \[
       \gamma_{t} = \frac{1}{ 4 \Photon \cdot \lambda_{\mathsf{max}}(\Sigma, \real^{d}) \cdot \sum_{j=1}^{\N} s_{j}\coeff_{j}} \quad \text{and} \quad  x_{1} \in X.
    \]
    Then, for each $\epsilon \in (0,1)$, as long as
    $
    t \geq 16 \kappa \cdot  \log\Big(\frac{\|x_{1}-x_{\star} \|_{2}}{\epsilon} \Big),
    $    
    we have
    \begin{align*}
        \|x_{t} - x_{\star}\|_{2} \leq \epsilon \;  + \; \frac{1}{ \lambda_{\mathsf{min}}(\Sigma,X-X) } \cdot \frac{4 \mathsf{Err}(x_{\star},X)}{\Photon } \cdot\frac{1}{ \sum_{j=1}^{\N} s_{j}\coeff_{j} \exp\big( - \coeff_{j} \underset{x \in X, i\in [n]}{\max} \langle a_{i}, x\rangle \big)  }.
    \end{align*}
\end{theorem}
We provide the proof of Theorem~\ref{thm-positive-measurements} in Section~\ref{sec:pf-thm-positive-measurements}. The bulk of the proof involves computing bounds on the Lipschitz parameter $L$ and the strong pseudo-monotonicity parameter $\nu$. Equipped with these estimates, our convergence guarantee is an immediate consequence of our general-purpose result in Theorem~\ref{general-thm}. 
Two additional remarks are in order.  

First, we emphasize that Theorem~\ref{thm-positive-measurements} is applicable to real spectral CT observation models in which the measurements are polychromatic, and the measurement vectors are derived from the Radon transform and consist of nonnegative entries. To the best of our knowledge, this is the first provable recovery guarantee for model~\eqref{model} under a realistic assumption that accommodates measurements arising from the Radon transform. By contrast, previous work focusing on rigorous guarantees studies non-convex formulations of the problem and provides guarantees under either a Gaussian random matrix assumption~\cite{fridovich2023gradient,charisopoulos2024nonlinear} or a restricted strong convexity assumption that has not been verified in applications to CT~\cite{barber2024convergence}.

Second, let us elaborate on our sample size condition, considering the noiseless case --- where $\zeta = 0$ and $\mathsf{Err}(x_{\star},X) = 0$ --- for simplicity. From Theorem~\ref{thm-positive-measurements}, we obtain that the iterates $\{x_{t}\}$ converge linearly to the ground truth $x_{\star}$ as long as the minimum restricted eigenvalue is positive, i.e., $\lambda_{\mathsf{min}}(\Sigma, X-X)>0$. Note that this condition would have been necessary even \emph{without} nonlinearities in the model. In the unstructured setting where $X = \mathbb{B}(R;x_{\star})$, the condition $\lambda_{\mathsf{min}}(\Sigma, X-X)>0$ typically holds once $n \geq d$, i.e., we have more observations than unknowns. When $X$ captures meaningful structure in the signal, the condition $\lambda_{\mathsf{min}}(\Sigma, X-X)>0$ can hold with $n \ll d$, thereby enabling the reconstruction of $x_{\star}$ from a number of samples smaller than the ambient dimension of the problem. The experiments in Section~\ref{sec:experiments} demonstrate the feasibility of compressed sensing (i.e., recovery with underdetermined systems): with signal dimension $d = 25^{2}$ and sample size $n = 500$, where $n < d$, the signal can still be reconstructed with high quality. Indeed, these are globally ill-posed problems where $\Sigma$ has a nontrivial null space, i.e., $\lambda_{\mathsf{min}}(\Sigma, \mathbb{R}^{d}) = 0$, but it is still possible to have $\lambda_{\mathsf{min}}(\Sigma, X - X) > 0$ if the true signal lies in a set $X$ having fewer degrees of freedom.

Third, and as a related point, note that the condition $\lambda_{\mathsf{min}}(\Sigma, X-X) > 0$ is necessary for identifiability. To see this, suppose $\lambda_{\mathsf{min}}(\Sigma, X-X) = 0$. Then there must exist two distinct vectors $x_{1}, x_{2} \in X$ such that $\Sigma(x_{1}-x_{2}) = 0$. Since $\Sigma$ and $A$ have the same null space, it follows that $Ax_{1} = Ax_{2}$. In this case, one cannot tell purely from the observed measurements $\{a_{i},y_{i}\}_{i=1}^{n}$ whether $x_{1}$ or $x_{2}$ is the true signal, leading to a lack of identifiability.

\subsection{Random matrix measurements}
In this section, we consider the model in which measurements are assumed to be Gaussian random vectors rather than discretized line integrals. While this does not correspond to a physical CT model, the main attractions of the Gaussian assumption are its analytical tractability and its facilitation of comparison to previous theoretical results~\cite{fridovich2023gradient,charisopoulos2024nonlinear}. Let us state this assumption concretely:
\begin{assumption}\label{Gaussian-assump}
    Measurements are generated at random, according to $\{a_{i}\}_{i=1}^{n} \overset{\mathsf{i.i.d.}}{\sim} \NORMAL(0,I_{d})$. 
\end{assumption}
Recalling our notation for projection matrices, we first define a few quantities that are useful in the statement of our results. Define the geometric quantity $\overline{\omega}(x_{\star},X)$, which depends on the local behavior of the set $X$ around $x_{\star}$ as
\begin{align}\label{def-Gaussian-width}
    \overline{\omega}(x_{\star},X) := \EE \bigg\{ \sup_{v \in X} \frac{\langle P_{x_{\star}}^{\perp}v , g  \rangle}{ \| P_{x_{\star}}^{\perp}v \|_{2} }  \bigg\}, \quad \text{where } g \sim \NORMAL(0,I_{d}).
\end{align}
Note that $\overline{\omega}(x_{\star},X)$ has a natural interpretation as the Gaussian width of a certain set intersected with the unit shell.
Next, we define $\gamma_{\star} \in (0,+\infty)$ as the solution of the following fixed point equation:
\begin{align}\label{fixed-point}
    \EE\bigg\{  \frac{|h'(6\norm G)|^{2} \cdot \mathbbm{1}\{G>0\} }{ \big[ |h'(6\norm G)| + \gamma_{\star}\big]^{2} } \cdot
    \int_{-G/4}^{G/4} \frac{t^2 e^{-t^2/2}}{\sqrt{2\pi}} \mathrm{d}t  \bigg\} =  \frac{ \Constant \overline{\omega}(x_{\star},X)^{2} }{n},
\end{align}
where the expectation is taking over $G \sim \NORMAL(0,1)$.
We prove in~\ref{sec:pf-fixed-point} that there exists a universal and positive constant $C$ such that as long as \mbox{$n \geq C \overline{\omega}(x_{\star},X)^{2}$}, then $\gamma_{\star}$ is unique and strictly positive. We further define the scalar quantity
\begin{align}\label{def-rho-norm}
    \rho(\norm) := \bigg( \sum_{j=1}^{\N} \frac{s_{j} \coeff_{j}}{ (\coeff_{j} \norm)^{4} \vee 1 } \bigg)^{-1},
\end{align}
where we recall the notation $a \vee b = \max(a, b)$.
Note that $\rho(\norm)$ grows at most polynomially in $\norm$; in particular, we have
\begin{align}
    \rho(\norm) = \mathcal{O}\big(\norm^{4} \vee 1 \big),
\end{align}
where in the notation $\mathcal{O}(\cdot)$ we hide dependence on the set of constants $\{s_{j},\coeff_{j}\}_{j=1}^{\N}$.
We are now ready to state our main result for Gaussian measurements, which is the first such result that incorporates measurement noise.
\begin{theorem}\label{thm:mono-gaussian}
Suppose that the data $\{(a_i, y_i)\}_{i=1}^{n}$ satisfy the single-index relationship~\eqref{model}. Recall the model parameters $\{s_{j},\coeff_{j}\}_{j=1}^{\N}$ and $\Photon$ in Eq.~\eqref{model} and the error quantity $\mathsf{Err}(x_{\star},X)$ in Eq.~\eqref{def-error-term}. Further, suppose Assumption~\ref{assump-X} holds with $R = 4 \| x_{\star} \|_2$, Assumption~\ref{Gaussian-assump} holds, and the iterates $\{x_{t}\}$ are generated according to \algname~\eqref{extragrad-method} with step size
\begin{align*}
    \gamma_{t} = \frac{(n+d)/n}{40 \Photon \sum_{j=1}^{\N}s_{j} \coeff_{j} } \quad \text{and} \quad x_{1} \in \mathbb{B}(4\| x_{\star} \|_2;x_{\star}).
\end{align*}
Recall the scalars $\overline{\omega}(x_{\star},X)$, $\rho(\norm)$ and $\gamma_{\star}$ defined in Eqs.~\eqref{def-Gaussian-width},~\eqref{def-rho-norm} and~\eqref{fixed-point}, respectively.
Then there exists a tuple of universal, positive constants $(c,C,C_{1},C_{2})$ such that if
\begin{align}\label{assump-sample-size-thm}
   n \geq C \cdot \overline{\omega}(x_{\star},X)^{2}  \quad \text{and} \quad  \sqrt{ \frac{\log(n)}{n} } \leq c\frac{ \rho(\norm)^{-1} }{\sum_{j=1}^{\N} s_{j}\coeff_{j} },
\end{align}
then the following holds with probability at least $1-16\exp\big( - \overline{\omega}(x_{\star},X)^{2}/64 \big) - 12n^{-10} - 2\exp(-d)$. For all $\epsilon \in (0,1)$, as long as
\begin{align}\label{iteration-complexity-gaussian}
\begin{split}
    t \geq C_{1} \frac{n+d}{n} \bigg( \sum_{j=1}^{\N}s_{j} \coeff_{j} \bigg) \cdot \frac{\big(\gamma_{\star} + \sum_{j=1}^{\N}s_{j}\coeff_{j} \big)^{2}}{\gamma_{\star}^{2}} \cdot  \log\Big( \frac{ \|x_{1}-x_{\star}\|_{2}}{\epsilon} \Big) \cdot \rho(\norm),
\end{split}
\end{align}
we have 
\begin{align*}
    \| x_{t} - x_{\star} \|_{2} \leq \epsilon + C_{2}  \frac{\mathsf{Err}(x_{\star},X)}{\Photon} \cdot \frac{\big(\gamma_{\star} + \sum_{j=1}^{\N}s_{j}\coeff_{j} \big)^{2}}{\gamma_{\star}^{2}} \cdot \rho(\norm).
\end{align*}
\end{theorem}
We provide the proof of Theorem~\ref{thm:mono-gaussian} in Section~\ref{sec:pf-thm-mono-gaussian}.
We now carefully unpack Theorem~\ref{thm:mono-gaussian}. Note that 
condition~\eqref{assump-sample-size-thm} necessitates that 
\[ 
    n =\Omega\big( \overline{\omega}(x_{\star},X)^{2} \big) \quad \text{and} \quad n = \Omega\big(\norm^{8}\log(\norm \vee 1) \big),
\]
where in the notation $\Omega(\cdot)$ we hide dependence on the set of constants $\{s_{j},\coeff_{j}\}_{j=1}^{\N}$. Thus, perfect recovery is possible as soon as the sample size $n$ scales linearly in $\overline{\omega}(x_{\star},X)^{2}$, provided it is also larger than $\norm^{8}\log(\norm \vee 1)$.

We next consider two regimes to elucidate the result, noting that the theorem applies more generally. In the first regime, suppose $n \asymp  \overline{\omega}(x_{\star},X)^{2}$. As we show in~\ref{app-first-ineq}, the scalar $\gamma_{\star}$, which is the solution of Eq.~\eqref{fixed-point}, satisfies
\begin{align}\label{gamma-star-lower-bound-regime1}
    \gamma_{\star} \geq \sum_{j=1}^{\N} s_{j} \coeff_{j} \exp\big( - 12\coeff_{j} \norm \big) \quad \text{if} \quad n \gtrsim \overline{\omega}(x_{\star},X)^{2}.
\end{align}
As a consequence, Theorem~\ref{thm:mono-gaussian} guarantees that for all $\epsilon \in(0,1)$, after
\[
    t = \mathcal{O}\bigg( \frac{d+n}{n} \cdot \exp\Big( 24 \max_{j\in [\N]} \coeff_{j} \norm \Big) \log\Big( \frac{\|x_{1} -x_{\star} \|_{2}}{\epsilon}\Big) \cdot \big(\norm \vee 1\big)^{4} \bigg) 
\]
iterations, we obtain 
\[
    \| x_{t} - x_{\star} \|_{2} \leq \epsilon + \frac{\mathsf{Err}(x_{\star},X)}{\Photon} \cdot \mathcal{O}\Big( \exp\Big( 24 \max_{j\in [\N]} \coeff_{j} \norm \Big) \cdot \big(\norm \vee 1 \big)^{4}  \Big).
\]
While the sample complexity in this regime scales linearly in $\overline{\omega}(x_{\star}, X)^2$, this comes at a cost: The iteration complexity and the eventual estimation error increase exponentially in $\norm$. 

Next, we consider a regime in which the sample size increases polynomially in $\norm$. In particular, we show in~\ref{app-second-ineq} that
\begin{align}\label{gamma-star-lower-bound-regime2}
    \gamma_{\star} \geq \sum_{j=1}^{\N} s_{j}\coeff_{j} \quad \text{if} \quad n \gtrsim  \frac{\big( \sum_{j=1}^{\N}s_{j} \coeff_{j} \big)^{2} }{ \sum_{j=1}^{\N}s_{j}^{2} \coeff_{j}^{2} }  \Big( \max_{j \in [\N]} (\coeff_{j} \norm)^{4} \vee 1  \Big) \cdot \overline{\omega}(x_{\star},X)^{2}.
\end{align}
In this regime, Theorem~\ref{thm:mono-gaussian} guarantees that for each $\epsilon \in (0,1)$, after 
\begin{align*}
   t = \mathcal{O}\bigg( \frac{d+n}{n} \cdot \log\Big( \frac{\|x_{1} - x_{\star} \|_{2}}{\epsilon} \Big) \big(\norm \vee 1 \big)^{4} \bigg)
\end{align*}
iterations, we obtain
\begin{align*}
 \|x_{t} - x_{\star}\|_{2} \leq \epsilon +  \frac{ \mathsf{Err}(x_{\star},X) }{\Photon} \cdot \mathcal{O}\Big( \big(\norm \vee 1 \big)^{4} \Big).
\end{align*}
Notably, this iteration complexity and the eventual estimation error only grow polynomially (and not exponentially) in $\norm$. Thus, a slight increase in the sample size can lead to significant benefits in computational complexity. 

\begin{table}[ht!]
    \centering
    \begin{tabular}{c c c}
        \hline
        \\
        \textbf{Method}  &   \textbf{Sample complexity}   &   \textbf{Iteration complexity} 
        \\\\
        \hline 
        \\
        \algname{}  & $\mathcal{O}\big(\norm^{4} \cdot d \; \vee \; \norm^{8}\log(\norm ) \big)$ &  $\mathcal{O} \Big(\norm^{4} \log\Big( \frac{5\norm}{\epsilon} \Big) \Big)$ \\\\
         & $\mathcal{O}\big(d\; \vee \;  \norm^{8}\log(\norm ) \big)$ &  $\mathcal{O} \Big( \exp( 24\norm ) \norm^{4} \log\Big( \frac{5\norm}{\epsilon} \Big) \Big)$
        \\ \\
        \hline
        \\
        PolyakSGM  &  $\mathcal{O}(\norm^{4} \cdot d)$ &  $\mathcal{O} \Big(\norm^{6} \log\Big( \frac{\norm}{\epsilon} \Big) \Big)$ \\ 
        AdPolyakSGM & & \\ \\
        \hline
        \\
        GD  & $\mathcal{O}\Big( \frac{ \exp( c_1 \norm )}{\norm}  \cdot d\Big)$  & $\mathcal{O} \Big( \exp( c_{2} \norm ) \log\Big( \frac{\norm}{\epsilon} \Big) \Big)$  \\ \\
        \hline
    \end{tabular}
    \caption{Iteration and sample complexity of iterative reconstruction methods with Gaussian measurements and unstructured signal, where we consider the noiseless, monochromatic ($\N = 1$) case in order to make a comparison.  The corresponding results for PolyakSGM and AdPolyakSGM were obtained by~\cite{charisopoulos2024nonlinear} and for GD by~\cite{fridovich2023gradient}. \algname{} can achieve $\mathcal{O}(d)$ sample complexity without additional multiplicative factors in $\norm$. It also has state-of-the-art iteration complexity for a slightly larger sample complexity $\mathcal{O}(\| x_{\star} \|_2^4 d)$. All in all, the proposed method~\eqref{extragrad-method} improves the best-known results in the Gaussian monochromatic model along both axes of sample complexity and iteration complexity.}
    \label{tab:complexity}
\end{table}

Let us conclude with a comparison of Theorem~\ref{thm:mono-gaussian} for \algname{} with the gradient descent (GD) method developed in~\cite{fridovich2023gradient} and the nonsmooth optimization methods (PolyakSGM and AdPolyakSGM) developed in~\cite{charisopoulos2024nonlinear} --- we collect this comparison of both the iteration and sample complexity of all algorithms in Table~\ref{tab:complexity}. We note that a comparison can only be made in the noiseless, unstructured setting where $X = \mathbb{B}(4\norm;x_{\star})$ --- in this case, we have $\overline{\omega}(x_{\star},X)^{2} \asymp d$ --- and the monochromatic setting where $\N = 1$ and $\coeff_{1} = 1$, since the two related works~\cite{fridovich2023gradient,charisopoulos2024nonlinear} only study this setting. 
From this table, it is clear that our method 
has the best sample and iteration complexity. In particular, our method improves both the iteration and sample complexity by a polynomial factor of $\norm$.


\section{Experimental results} \label{sec:experiments}

We present simulations under both the CT model with Radon transform measurements and a phantom image, and the Gaussian measurement model with a synthetic signal. We choose a small-scale phantom image largely for illustration purposes: our goal is to show visually that our theoretically guaranteed EXACT algorithm often has state-of-the-art practical performance when compared with other algorithms that do not come with corresponding theoretic guarantees. Note that for larger phantoms, higher-order algorithms like ADMM and its variants are known to be computationally challenging to run~\cite{mory2018comparison,barber2024convergence}.

Quantifying reconstruction quality is itself a nuanced task \cite{cheng2019validation}; for simplicity, we quantify image quality using Euclidean distance or root mean squared error (RMSE) relative to a known ground truth phantom, augmented with visual comparisons for qualitative evaluation.
\vspace{-3mm}
\begin{figure}[ht!]
    \centering
    \includegraphics[height=0.21\linewidth]{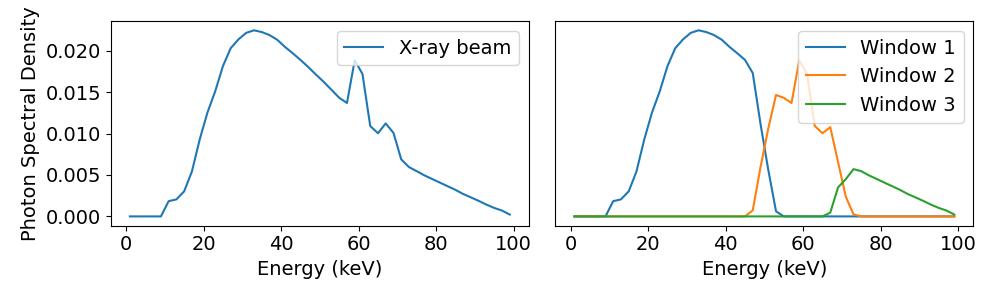}
    \includegraphics[height=0.2\linewidth]{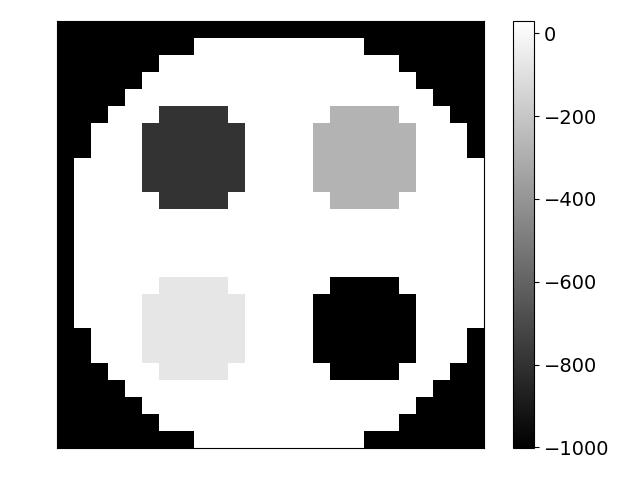}
    \caption{Spectral density of the simulated X-ray source (left), detector sensitivity for 3 simulated windows (middle), and target phantom made of simulated PMMA with 4 regions of interest (right, colorbar in Hounsfield Units). This simulation setup is based on \cite{barber2024convergence}. \vspace{-2mm}}
    \label{fig:simulation-setup}
\end{figure}

\subsection{Photon-counting CT simulation}
Our photon-counting CT simulation experiments are modeled on those from \cite{barber2024convergence}. Our imaging target is a simple simulated polymethyl methacrylate (PMMA) phantom consisting of a disc with 4 circular regions of interest (ROIs), each with different density of PMMA. The target image has $d = 25^2=625$ pixels, the simulated X-ray photon-counting detector has 3 energy windows, and the simulated X-ray source outputs $W=50$ different X-ray wavelengths. The default simulation setting follows the configuration of~\cite{barber2024convergence}, using an average source intensity of $\Photon = 10^6$ photons per detector cell. The simulated CT scanner is equipped with 50 detector cells and acquires measurements from 50 projection views evenly spaced around the unit circle. This results in a total of $n = (50$ views)$~\times~(50$ detector cells) $=2500$ rays along which measurements are collected. See~\cite[Figure 4]{barber2024convergence} for a schematic illustration of the projection geometry.
The source spectrum, detector sensitivity, and target image are shown in \Cref{fig:simulation-setup}.

\paragraph{Description of the four algorithms} In each experiment, we compare our proposed extragradient algorithm \algname{} defined by \Cref{extragrad-method} with the square loss gradient descent approach (``MSE GD'') of \cite{fridovich2023gradient}, the ADMM approach of \cite{barber2024convergence} and the Polyak subgradient descent (``PolyakSGM'') of \cite{charisopoulos2024nonlinear}, where we endow the latter with oracle knowledge of the loss incurred by the target image. 
Our simulated measurements are corrupted by Poisson noise, though only the ADMM baseline leverages specific knowledge of this noise model (by minimizing the Poisson negative log likelihood objective). More concretely, the gradient descent approach~\cite{fridovich2023gradient} minimizes the squared $\ell_{2}$ loss:
\[
\mathcal{L}_{2}(x) = \frac{1}{n} \sum_{i=1}^{n} \bigg( \Bar{I} \sum_{j=1}^{\N} s_{j} \exp \left( - \coeff_{j} \langle a_{i}, x \rangle_{+} \right) - y_{i} \bigg)^{2},
\]
while the PolyakSGM algorithm~\cite{charisopoulos2024nonlinear} minimizes the absolute $\ell_{1}$ loss:
\[
\mathcal{L}_{1}(x) = \frac{1}{n} \sum_{i=1}^{n} \bigg| \Bar{I} \sum_{j=1}^{\N} s_{j} \exp \left( - \coeff_{j} \langle a_{i}, x \rangle_{+} \right) - y_{i} \bigg|.
\]
On the other hand, the ADMM algorithm~\citep{barber2024convergence} is designed to solve the constrained optimization problem:
\[
\min_{x \in \mathbb{R}^{d},\, z \in \mathbb{R}^{n}} \; \mathsf{Loss}(z) \quad \text{subject to} \quad Ax = z,
\]
where the loss is given by the negative log-likelihood of the Poisson noise model~\eqref{Poisson-model}, i.e.,
\[
\mathsf{Loss}(z) = \sum_{i=1}^{n} \left[ \Bar{I} \sum_{j=1}^{\N} s_{j} \exp \left( - \coeff_{j} z_{i} \right) - y_{i} \cdot \log \left( \Bar{I} \sum_{j=1}^{\N} s_{j} \exp \left( - \coeff_{j} z_{i} \right) \right) \right].
\]

\paragraph{Implementation of projections onto constraint set} 
For all algorithms, reconstructions are constrained to lie in the intersection of the positive orthant and a total variation (TV) ball\footnote{Strictly speaking, this set is not bounded (as required by our theory) but defining the set via further intersection with an $\ell_2$ ball does not change the behavior of any of the algorithms.}. For a fair comparison, the TV bound for each method is set to the oracle value of the target image’s TV norm\footnote{In general, this value would have to be chosen via additional hyperparameter tuning.}. Formally, define 
\[
X \coloneqq X_{1} \cap X_{2},
\]
where
\[
X_{1} \coloneqq \left\{ x \in \mathbb{R}^{d} : \|x\|_{\mathsf{TV}} \leq \|x_{\star}\|_{\mathsf{TV}} \right\},
\quad
X_{2} \coloneqq \left\{ x \in \mathbb{R}^{d} : x_{i} \geq 0 \ \text{for all} \ i \in [d] \right\}.
\]

Let $\mathcal{P}_{X_{1}}, \mathcal{P}_{X_{2}}: \mathbb{R}^{d} \to \mathbb{R}^{d}$ denote the projection operators onto $X_{1}$ and $X_{2}$, respectively. For the projection operator $\mathcal{P}_{X_{1}}$, we leverage the fact that penalization-based approaches (i.e., proximal operators) are more readily available in software, and that the projection approach is equivalent to a suitable penalization formulation (by Lagrange duality)\footnote{Even theoretically speaking, the advantage of this approach is that the optimization problem~\eqref{eq:TV-regularization} is strongly convex, and hence enjoys quadratic growth around its minimizer. Consequently, despite its nonsmoothness~\citep{davis2024local}, an $\epsilon$-optimal solution can in principle be computed in $\mathcal{O}\big(d\,\log^{3}(\frac{1}{\epsilon})\big)$ time.}. Specifically, the projection $\mathcal{P}_{X_{1}}(z)$ is approximated by solving the strongly convex, TV-regularized problem
\begin{align}\label{eq:TV-regularization}
    \bar{x} \;=\; \argmin_{x \in \mathbb{R}^{d}} \; \|x - z\|_{2}^{2} + \lambda \,\|x\|_{\mathsf{TV}},
\end{align}
where $\lambda > 0$ is tuned via a binary search so that 
\[
\big| \| \bar{x} \|_{\mathsf{TV}} - \| x_{\star} \|_{\mathsf{TV}} \big| \leq 0.01.
\] 
Moreover, the projection $\mathcal{P}_{X_{2}}(z)$ is implemented simply by retaining the positive entries of $z$ and setting the remaining entries to zero. The above method is used for implementing the projection step across all four algorithms we compare. 

We compute the projection onto the intersection set $X$ using Dykstra’s algorithm: for any $z \in \mathbb{R}^{d}$, initialize $z_{0} = z$, $p_{0} = 0$, and $q_{0} = 0$, then iterate for $k = 0, 1, 2, \dots$:
\begin{align}\label{projection-intersection}
\begin{split}
    y_{k} &= \mathcal{P}_{X_{1}}\big(z_{k} + p_{k}\big), \\
    p_{k+1} &= z_{k} + p_{k} - y_{k}, \\
    z_{k+1} &= \mathcal{P}_{X_{2}}\big(y_{k} + q_{k}\big), \\
    q_{k+1} &= y_{k} + q_{k} - z_{k+1}.
\end{split}
\end{align}
The sequence $\{z_{k}\}$ converges to $\mathcal{P}_{X}(z)$, the projection of $z$ onto $X$. The stopping criterion is set to be $\|z_{k+1}-z_{k}\|_{2} \leq 10^{-4}$ in our experiments. This same projection routine is used for all algorithms in our comparison. 

\paragraph{Comparison of methods}
For all methods, at each iteration $t$ we consider the average $\bar x_t$ of the previous $t/2$ iterates\footnote{In their analysis of ADMM, \cite{barber2024convergence} likewise consider the average iterate.
Due to the convexity of the $\ell_2$-norm, note that our guarantees for the $t/2$-th iterate of \algname{} also apply to the average iterate at time $t$.}.
We run all algorithms until convergence, defined as the first iteration at which $\| \bar x_t - \bar x_{t-1} \|_2 \leq 10^{-5}$. We consider the average of the iterates since a stopping criterion based on this average is more stable than one that is based on a single iterate.
We quantify reconstruction quality using root mean square error (RMSE) of $\bar x_T$, where $T$ is the iteration at which the algorithm converged.

\begin{figure}[ht!]
    \centering
    \includegraphics[width=0.49\linewidth]{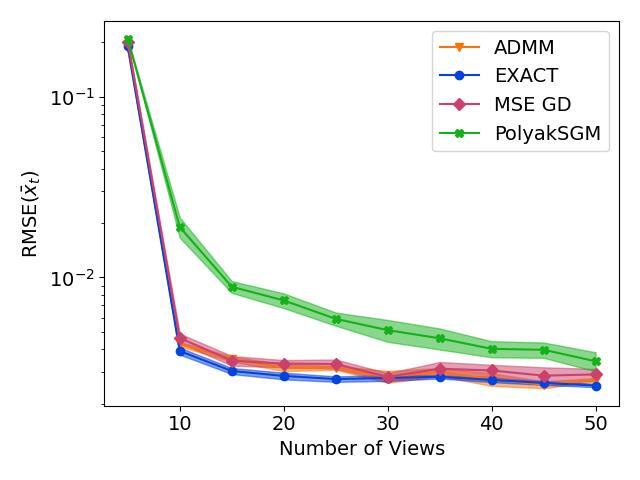}
    \includegraphics[width=0.49\linewidth]{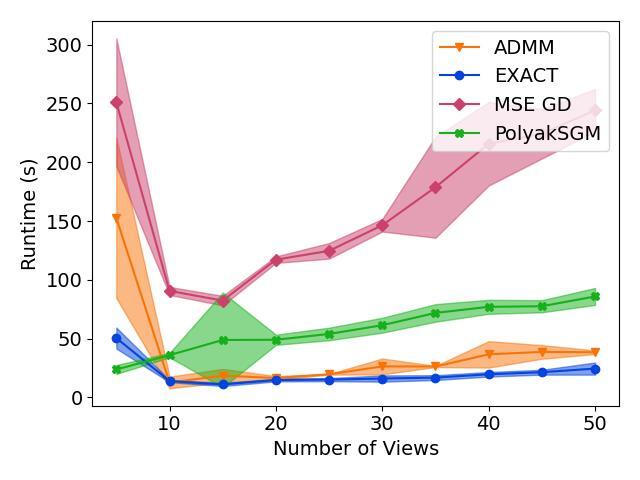}
    \caption{RMSE (left) and runtime (right), varying the number of views while the X-ray source outputs on average $10^6$ photons per detector cell. Until the number of views drops below 10, at which point all algorithms fail, our proposed method \algname{} (defined in \Cref{extragrad-method}) converges fastest and to lowest error.}
    \label{fig:1e6photons-varymeasurements}
\end{figure}

\newcommand{\plotcrop}[1]{%
\adjincludegraphics[trim={{0.08\width} {0.05\height} {0.05\width} {0.04\height}}, clip, width=\linewidth]{#1}%
}

\newcommand{\plotinset}[1]{%
\adjincludegraphics[trim={{0.1\width} {0.2\height} {0.6\width} {0.45\height}}, clip, width=\linewidth]{#1}%
}

\begin{figure}[ht!]
    \centering
    \begin{minipage}{0.24\textwidth}
        \centering
        \plotcrop{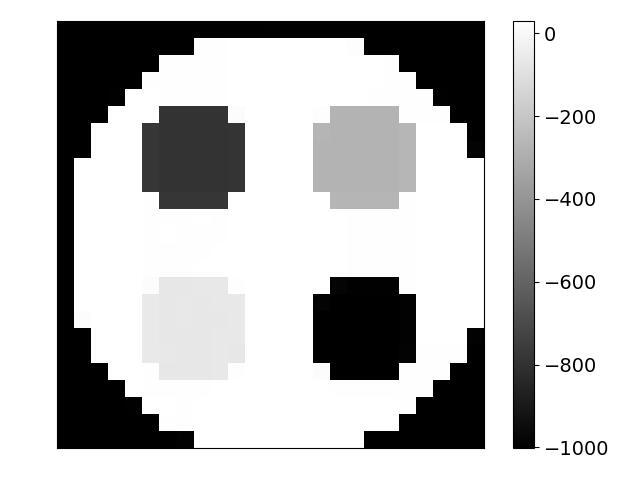}
        \plotinset{figures/1e6photons-10projections-HU/recon_extra_stepsize2e-08_TVTrue_noiseTrue.jpg}
        \par\small EXACT
    \end{minipage}
    \hfill
    \begin{minipage}{0.24\textwidth}
        \centering
        \plotcrop{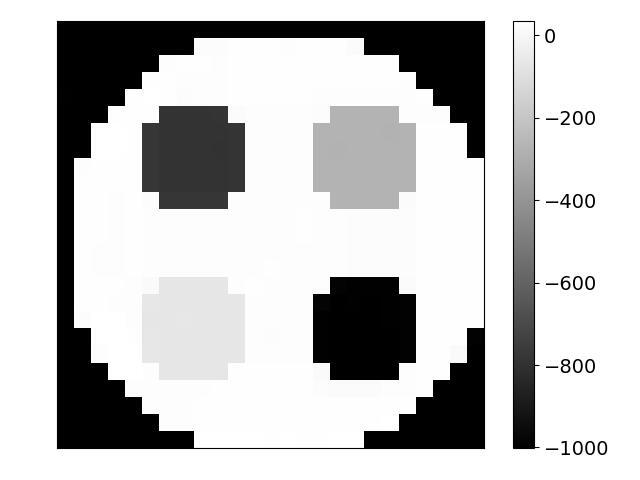}
        \plotinset{figures/1e6photons-10projections-HU/recon_msegd_stepsize5e-09_TVTrue_noiseTrue.jpg}
        \par\small MSE GD
    \end{minipage}
    \hfill
    \begin{minipage}{0.24\textwidth}
        \centering
        \plotcrop{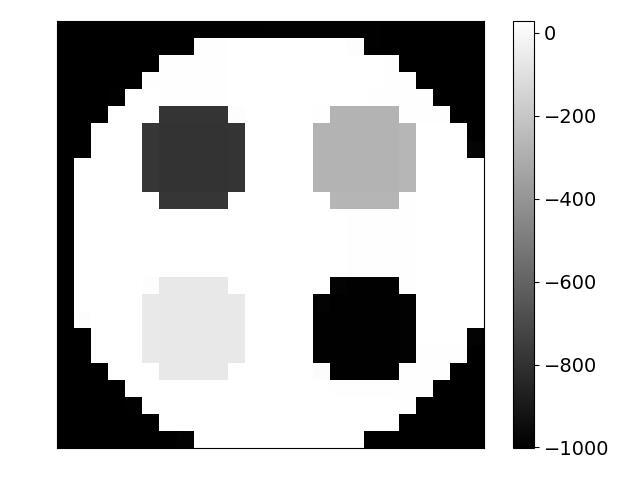}
        \plotinset{figures/1e6photons-10projections-HU/recon_sigma100_TVTrue_noiseTrue.jpg}
        \par\small ADMM
    \end{minipage}
    \hfill
    \begin{minipage}{0.24\textwidth}
        \centering
        \plotcrop{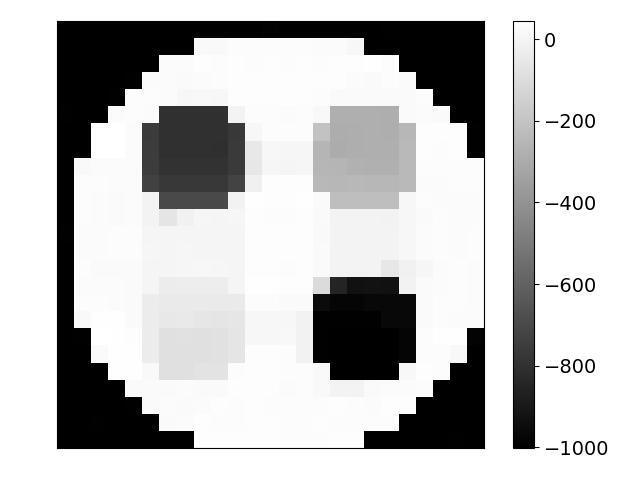}
        \plotinset{figures/1e6photons-10projections-HU/recon_polyak1_TVTrue_noiseTrue.jpg}
        \par\small PolyakSGM
    \end{minipage}
    \caption{Full reconstructions (in Hounsfield Units, HU) and zoom-ins (bottom row) after convergence of each algorithm using 10 views and an average source intensity of $10^6$ photons per detector cell, a total of 5$\times$ reduction in X-ray dose compared to the 50-views default setting in which all algorithms perform well. 
    The \algname{}, MSE GD and ADMM reconstructions are visually faithful to the target in \Cref{fig:simulation-setup}, and the PolyakSGM reconstruction exhibits more severe artifacts that obscure part of the lowest-contrast ROI.}
    \label{fig:1e6photons-10projections}
\end{figure}

We begin with a sufficiently high source intensity, on average $10^6$ photons per detector cell, to mimic the setup of \cite{barber2024convergence}. 
Results are reported in
\Cref{fig:1e6photons-varymeasurements}, including variation over 10 random seeds controlling the Poisson noise. 
For all algorithms, we tune the hyperparameters in the algorithm for the experiment with $10$ views, and then simulate the algorithm for the other view settings with that hyperparameter setting.
\algname{} (\Cref{extragrad-method}) yields the lowest RMSE and fastest convergence time even down to 10 views, while all methods fail with 5 views. 
Visual comparisons using 10 views are shown in \Cref{fig:1e6photons-10projections}. Interestingly, while EXACT, MSE GD and ADMM produce visually flawless reconstruction even with this dose reduction, Polyak SGM suffers significant artifacts.

\newcommand{\plotcroplines}[1]{%
\adjincludegraphics[trim={{0.03\width} {0.0\height} {0.03\width} {0.0\height}}, clip, width=0.32\linewidth]{#1}%
}

\begin{figure}[ht!]
    \centering
    \includegraphics[width=0.49\linewidth]{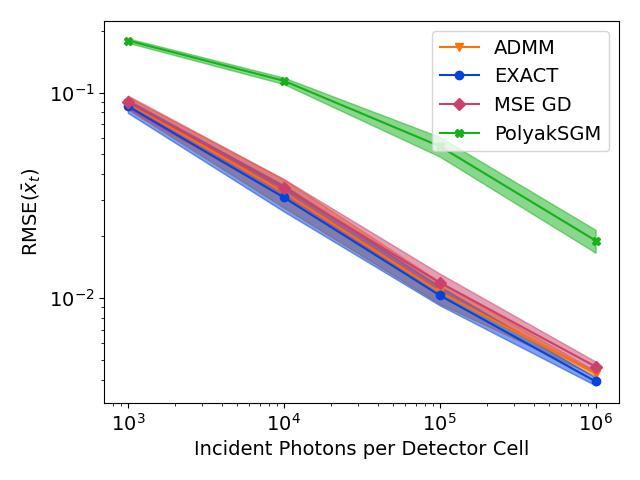}
    \includegraphics[width=0.49\linewidth]{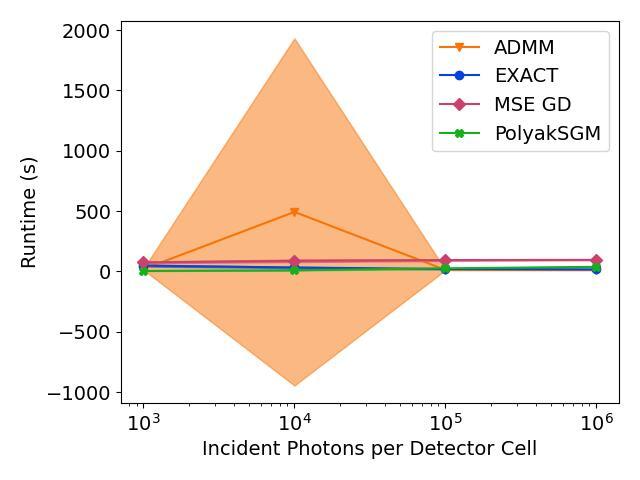}
    \caption{RMSE (left) and runtime (right), varying the source intensity while the number of views is fixed at 10, the smallest value for which recovery is possible in \Cref{fig:1e6photons-varymeasurements}. All algorithms suffer higher reconstruction error as source intensity and thus SNR is decreased. Also note that ADMM exhibits slow convergence behavior on some of the random seeds, leading to a high standard deviation in runtime.}
    \label{fig:10measurements-varyphotons}
\end{figure}

For our next experiment, we hold the number of views fixed at 10 and gradually decrease the intensity of the X-ray source. This results in further dose reduction. In this set of experiments, we scale the stepsize of our EXACT algorithm by the source intensity (as suggested by the theory) while the hyperparameters in the other algorithms are tuned for the specific source intensity employed. In that sense, we give the other algorithms access to more hyperparameter tuning resources.
RMSE and runtime results are summarized in \Cref{fig:10measurements-varyphotons}. Here again, it appears that the RMSE of EXACT, MSE GD and ADMM are comparable for most source intensities, but there are random seeds (corresponding to random draws of the noise) for which ADMM can exhibit slow convergence behavior. 

\begin{figure}[ht!]
  \centering
  \includegraphics[scale = 0.35]{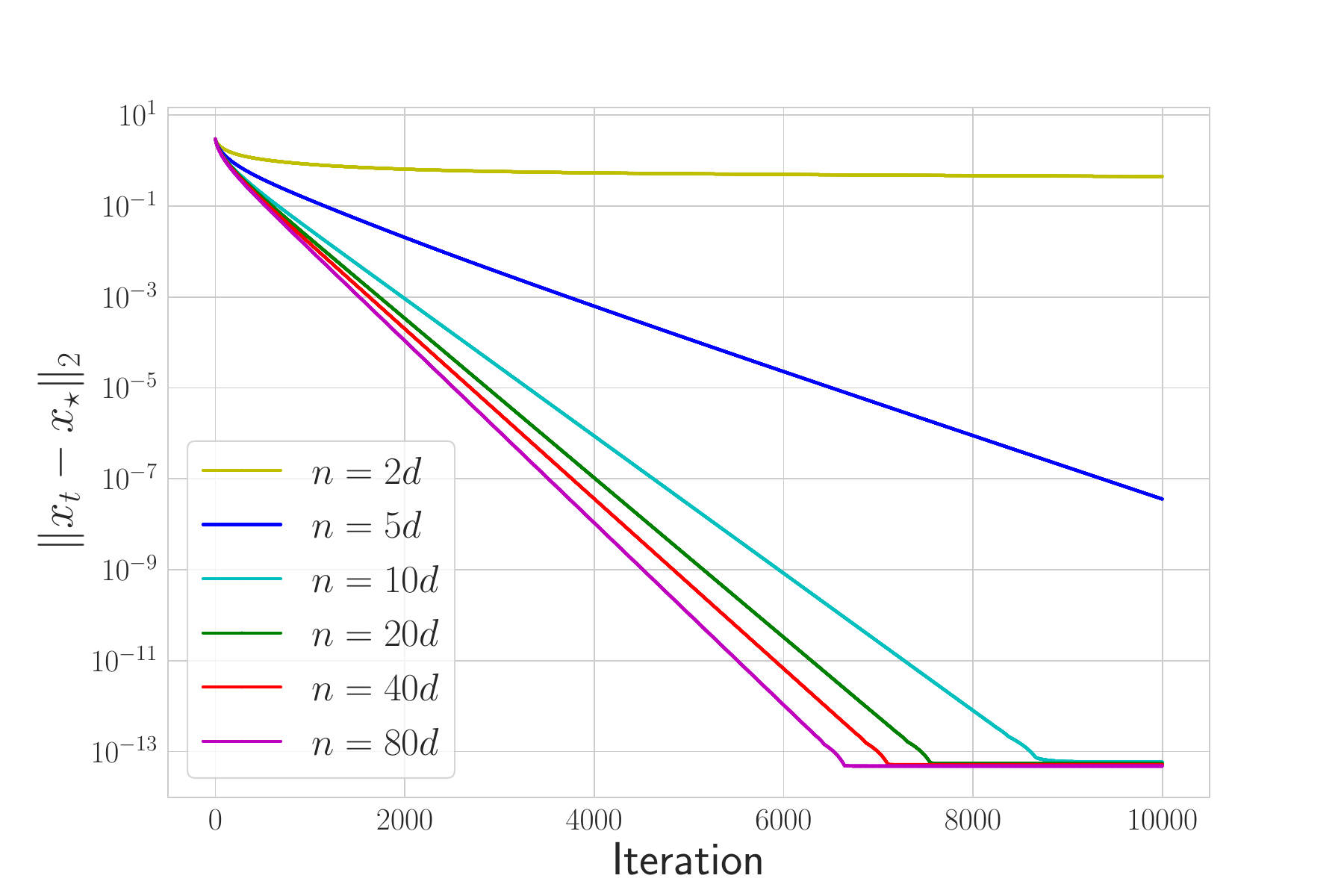}
  \caption{Estimation error over iterations for different sample size $n$. The experiment is conducted in the setting of Gaussian measurements, noiseless observations, and monochromatic model where $\N = 1,\; \Photon = 1$ and $\mu_{1} = 1$. In this simulation, we have $d = 100$ and $\norm = 3$, and we set the step size $\gamma_{t} = 0.25$ in Eq.~\eqref{extragrad-method}. As predicted theoretically, we observe that increasing the number of measurements can dramatically improve runtime up to a threshold.}    
  \label{fig:gaussian_noiseless}
\end{figure}

\subsection{Gaussian model}
In this section, we evaluate \algname{} (Algorithm~\eqref{extragrad-method}) under Gaussian measurements (see Assumption~\ref{Gaussian-assump}) and in the monochromatic setting where $\N = 1$ and $\coeff_{1} = 1$. We consider noiseless observations, i.e., $y_{i} = \exp\{- \langle a_{i}, x_{\star} \rangle \}$, set $d = 100$, use step size $\gamma_{t} = 0.25$, and vary the number of samples $n \in \{ 5d, 10d, 20d, 40d, 80d\}$. We set $\norm = 3$.

From Figure~\ref{fig:gaussian_noiseless}, we observe that as the sample size increases, it takes fewer iterations to achieve machine precision. This finding aligns with the theoretical results presented in Theorem~\ref{thm:mono-gaussian} and its associated comments, which state that a larger sample size can significantly reduce iteration complexity. Moreover, we also note that once the sample size surpasses a certain threshold, the improvement in iteration complexity from further increasing the sample size becomes less pronounced.
Showing a sharp characterization of these convergence rates --- with accompanying lower bounds --- in a fine-grained, sample size dependent manner is an interesting open question that we leave to future work.


\section{Proofs} \label{sec:proofs}
In this section, we provide proofs of all our main results. Auxiliary technical lemmas can be found in the supplementary material.
 
\subsection{Proof of Theorem~\ref{general-thm}} \label{sec:pf-general-thm}
Recall that $F$ is $\monotone$-strongly pseudo-monotone around the ground truth $x_{\star}$, i.e., 
\[
    \langle F(x) - F(x_{\star}) , x-x_{\star} \rangle \geq \monotone \cdot \| x - x_{\star} \|_{2}^{2} \quad \text{for all } x\in X,
\]
and also Lipschitz on the set $X$, i.e.,
\[
    \| F(x) - F(x') \|_{2} \leq L \cdot \| x - x'\|_{2} \quad  \text{for all } x,x' \in X.
\]
By the definition of substep~\eqref{step2-extragrad}, we have
\begin{align}\label{ineq-convergence-0}
    \| x_{t+1} - x_{\star} \|_{2}^{2} &= \big\| \mathcal{P}_{X}\big( x_{t} - \gamma_{t} F(x_{t+0.5}) \big) - \mathcal{P}_{X}(x_{\star}) \big\|_{2}^{2}  \nonumber \\
    & \overset{\1}{\leq} \big\langle x_{t+1} - x_{\star}, x_{t} - \gamma_{t} F(x_{t+0.5}) - x_{\star}  \big\rangle \nonumber \\
    & = \big \langle  x_{t+1} - x_{\star} , x_{t} - x_{\star} \big \rangle - \gamma_{t} \big\langle  F(x_{t+0.5}) , x_{t+1} - x_{\star}  \big \rangle \nonumber \\
    & = \frac{1}{2} \| x_{t+1} - x_{\star} \|_{2}^{2} + \frac{1}{2} \| x_{t} - x_{\star} \|_{2}^{2} - \frac{1}{2} \| x_{t+1} - x_{t} \|_{2}^{2} 
   \nonumber \\ 
    & \quad - \gamma_{t} \big\langle  F(x_{t+0.5}) , x_{t+1} - x_{t+0.5}  \big \rangle - \gamma_{t} \big\langle  F(x_{t+0.5}) , x_{t+0.5} - x_{\star}  \big \rangle
\end{align}
where in step $\1$ we use $1$-cocoercivity of the projection operator onto a convex set, whereby
\[
    \| \mathcal{P}_{X}(x) - \mathcal{P}_{X}(x') \|_{2}^{2} \leq \big\langle \mathcal{P}_{X}(x) - \mathcal{P}_{X}(x') , x-x' \big\rangle \quad \text{for all } x,x' \in X.
\]
Next, we bound the latter two inner product terms on the RHS. By strong monotonicity of $F$ around $x_{\star}$, we have
$
    \big\langle F(x_{t+0.5}) - F(x_{\star}), x_{t+0.5} - x_{\star} \big \rangle \geq \monotone \cdot \| x_{t+0.5} - x_{\star} \|_{2}^{2},
$
and rearranging terms yields
\begin{align}\label{ineq-convergence-1}
    &-\gamma_{t} \big \langle F(x_{t+0.5}), x_{t+0.5} - x_{\star}  \big \rangle \nonumber \\
    &\qquad \qquad \leq -\gamma_{t} \big \langle F(x_{\star}) , x_{t+0.5} - x_{\star}  \big \rangle - \monotone \gamma_{t}  \|x_{t+0.5} - x_{\star} \|_{2}^{2} \nonumber \\
    &\qquad \qquad\leq \gamma_{t} \|x_{t+0.5} - x_{\star}\|_{2} \cdot \mathsf{Err}(x_{\star},X) - \frac{1}{2}\monotone \gamma_{t}  \|x_{t} - x_{\star} \|_{2}^{2} + \monotone \gamma_{t}  \| x_{t+0.5} - x_{t} \|_{2}^{2},
\end{align}
where in the last step we use $\zeta = F(x_{\star})$ and recall the definition of $\mathsf{Err}(x_{\star},X)$~\eqref{def-error-term}. In addition, we also use the elementary inequality
$    \| x_{t} - x_{\star} \|_{2}^{2} \leq 2\| x_{t} - x_{t+0.5} \|_{2}^{2} + 2\| x_{t+0.5} - x_{\star} \|_{2}^{2}.
$

We next turn to bound the inner product term $-\gamma_{t} \langle F(x_{t+0.5}), x_{t+1} - x_{t+0.5}  \rangle$. By using the first-order optimality conditions for the projection step used in computing $x_{t+0.5}$, we have 
\[
    \langle x_{t+0.5} - x_{t} + \gamma_{t} F(x_{t}) , x - x_{t+0.5}\rangle \geq 0 \quad \text{for all } x \in X.
\]
Substituting $x = x_{t+1}$ above and rearranging the terms yields
\begin{align}
    -\gamma_{t} \langle F(x_{t}) , x_{t+1} - x_{t+0.5}  \rangle &\leq \langle x_{t+0.5} - x_{t},x_{t+1} - x_{t+0.5}   \rangle \nonumber \\
    &= \frac{1}{2} \| x_{t} - x_{t+1} \|_{2}^{2} -\frac{1}{2} \| x_{t+0.5} - x_{t} \|_{2}^{2} - \frac{1}{2} \| x_{t+1} - x_{t+0.5} \|_{2}^{2}. \label{eq:intermediate-EG}
\end{align} 
Proceeding now to the term of interest and performing some algebraic manipulation, we have
\begin{align}
    & -\gamma_{t} \langle F(x_{t+0.5}), x_{t+1} - x_{t+0.5}  \rangle \nonumber \\
    &\qquad \qquad = 
    -\gamma_{t} \langle F(x_{t+0.5}) - F(x_{t}), x_{t+1} - x_{t+0.5}  \rangle - \gamma_{t}  \langle F(x_{t}), x_{t+1} - x_{t+0.5}  \rangle \nonumber \\
    &\qquad \qquad \overset{\1}{\leq} \gamma_{t} \| F(x_{t+0.5}) - F(x_{t}) \|_{2} \cdot \| x_{t+1} - x_{t+0.5} \|_{2} - \gamma_{t}  \langle F(x_{t}), x_{t+1} - x_{t+0.5}  \rangle \nonumber \\
    &\qquad \qquad \overset{\2}{\leq} \gamma_{t} L \| x_{t+0.5} - x_{t} \|_{2} \cdot \| x_{t+1} - x_{t+0.5} \|_{2} - \gamma_{t}  \langle F(x_{t}), x_{t+1} - x_{t+0.5}  \rangle \nonumber \\
    &\qquad \qquad \overset{\3}{\leq} \frac{\gamma_{t}^{2} L^{2}}{2} \| x_{t+0.5} - x_{t} \|_{2}^{2} + \frac{1}{2}\| x_{t+1} - x_{t+0.5} \|_{2}^{2} - \gamma_{t}  \langle F(x_{t}), x_{t+1} - x_{t+0.5}  \rangle \nonumber \\
    &\qquad \qquad \overset{\4}{\leq} \frac{\gamma_{t}^{2} L^{2}}{2} \| x_{t+0.5} - x_{t} \|_{2}^{2} + \frac{1}{2} \| x_{t} - x_{t+1} \|_{2}^{2} -\frac{1}{2} \| x_{t+0.5} - x_{t} \|_{2}^{2}, \label{eq:term2-EG}
\end{align}
where step $\1$ follows from the Cauchy--Schwarz inequality, step $\2$ from the fact that $F$ is $L$-Lipschitz, step $\3$ from the AM-GM inequality, and step $\4$ from Eq.~\eqref{eq:intermediate-EG}.

Substituting Ineqs.~\eqref{eq:term2-EG} and~\eqref{ineq-convergence-1} into inequality~\eqref{ineq-convergence-0}, we obtain 
\begin{align*}
    \| x_{t+1} - x_{\star} \|_{2}^{2} &\leq  \big( 1 - \monotone \gamma_{t} \big) \cdot \|x_{t} - x_{\star}\|_{2}^{2} + \big( \gamma_{t}^{2} L^{2} + 2\monotone \gamma_{t} - 1\big) \cdot \| x_{t+0.5} - x_{t} \|_{2}^{2} \\
    & \qquad \qquad + 2\gamma_{t}\| x_{t+0.5} -x_{\star} \|_{2} \cdot \mathsf{Err}(x_{\star},X).
\end{align*}
To bound the last term on the RHS, we apply the inequality $2ab \leq a^2 + b^2$ with $a = \gamma_t \| x_{t+0.5} -x_{\star} \|_{2} \cdot \sqrt{\frac{\monotone L}{2}}$ and $b = \mathsf{Err}(x_{\star},X) \cdot \sqrt{\frac{2}{\monotone L}}$ to obtain
\begin{align*}
    2\gamma_{t}\| x_{t+0.5} -x_{\star} \|_{2} \cdot \mathsf{Err}(x_{\star},X) 
    &\leq \frac{\gamma_{t}^{2} \monotone L}{2}  \| x_{t+0.5} - x_{\star}  \|_{2}^{2} + \frac{2}{\monotone L} \cdot \mathsf{Err}(x_{\star},X)^{2} \\
    &\leq \gamma_{t}^{2} \monotone L  \Big( \| x_{t+0.5} - x_{t} \|_2^2 + \| x_t - x_{\star}  \|_{2}^{2} \Big) + \frac{2}{\monotone L} \cdot \mathsf{Err}(x_{\star},X)^{2}.
\end{align*}
Combining with the other two terms, we have
\begin{align*}
    \| x_{t+1} - x_{\star} \|_{2}^{2} &\leq  \big( 1 - \monotone \gamma_{t} + \monotone L \gamma_{t}^{2} \big) \cdot \|x_{t} - x_{\star}\|_{2}^{2} \\
    & \quad + \big( \gamma_{t}^{2} L^{2} + 2\monotone \gamma_{t} - 1 + \monotone L \gamma_{t}^{2} \big) \cdot \| x_{t+0.5} - x_{t} \|_{2}^{2} + \frac{2\mathsf{Err}(x_{\star},X)^{2}}{\monotone L}.
\end{align*}
By setting $\gamma_{t} = \frac{1}{4L}$, we obtain 
\begin{align*}
    1 - \monotone \gamma_{t} + \monotone L \gamma_{t}^{2} = 1-\frac{\monotone}{4L} + \frac{\monotone}{16L} \leq 1 - \frac{\monotone}{8L},
\end{align*}
and 
\begin{align*}
    \gamma_{t}^{2} L^{2} + 2\monotone \gamma_{t} - 1 + \monotone L \gamma_{t}^{2} = \frac{1}{16} + \frac{\monotone}{2L} - 1 + \frac{\monotone}{16L} < 0,
\end{align*}
where we use the fact that $\monotone \leq L$. Putting the pieces together yields
\begin{align*}
    \| x_{t+1} - x_{\star} \|_{2}^{2} \leq \Big( 1 - \frac{\monotone}{8L} \Big) \cdot \|x_{t} - x_{\star}\|_{2}^{2} + \frac{2 \mathsf{Err}(x_{\star},X)^{2}}{\monotone L}.
\end{align*}
Unrolling the recursion over time thus yields
\begin{align*}
    \| x_{t+1} - x_{\star} \|_{2}^{2} \leq \Big( 1 - \frac{\monotone}{8L} \Big)^{t} \cdot \|x_{1} - x_{\star}\|_{2}^{2} + \frac{16 \mathsf{Err}(x_{\star},X)^{2}}{\monotone^{2}},
\end{align*}
and taking square-roots of both sides and using the inequality $\sqrt{a + b} \leq \sqrt{a} + \sqrt{b}$ completes the proof.
\qed

\subsection{Proof of Theorem~\ref{thm-positive-measurements}} \label{sec:pf-thm-positive-measurements}

Our proof proceeds by first bounding the strong pseudo-monotonicity parameter $\monotone$ and Lipschitz parameter $L$, and then we apply Theorem~\ref{general-thm}.

\paragraph{Strong pseudo-monotonicity} By definition, the strong pseudo-monotonicity constant \newline around the point $x_{\star}$ is bounded below by the global strong monotonicity constant. For any pair $x,x' \in X$, we have
\begin{align*}
    \langle F(x)-F(x'), x-x' \rangle = \Photon \cdot  \frac{1}{n} \sum_{i=1}^{n} \big[h(\langle a_{i},x' \rangle) - h(\langle a_{i},x\rangle)\big] \cdot \big[ \langle a_{i},x \rangle - \langle a_{i}, x' \rangle \big].
\end{align*}
Note that by Assumption~\ref{assump-positive}, we have $\langle a_{i},x \rangle \geq 0$ for all $x\in X$ and $i \in [n]$. Thus, applying the mean value theorem yields that for any $x,x' \in X$, there exists some scalar $t_i$ lying in the range $ 0 \leq \min\{ \langle a_{i}, x \rangle, \langle a_{i}, x' \rangle \} \leq t_{i} \leq \max\{ \langle a_{i}, x \rangle, \langle a_{i}, x' \rangle \}$ such that 
\[
    h( \langle a_{i}, x' \rangle ) - h( \langle a_{i}, x \rangle) = h'(t_{i}) \cdot \big[ \langle a_{i} , x' \rangle - \langle a_{i}, x \rangle \big] = | h'(t_{i}) | \cdot \big[ \langle a_{i}, x \rangle - \langle a_{i}, x' \rangle \big],
\]
where $h'(t_{i}) \leq 0$ is the derivative of $h$ at $t_{i}$ (recall that $h$ is nonincreasing). 
Consequently,
\begin{align*}
  \langle F(x) - F(x'), x - x' \rangle &\geq 
   \Photon \cdot \frac{1}{n} \sum_{i=1}^{n} | h'(t_{i}) | \cdot \big[\langle a_{i}, x'\rangle - \langle a_{i}, x \rangle \big]^{2} \\
  & \geq \frac{\Photon}{n}  \min_{1\leq i \leq n} |h'(t_{i})| \cdot \sum_{i=1}^{n} \big[\langle a_{i}, x'\rangle - \langle a_{i}, x \rangle \big]^{2} \\
  & \geq \Photon \cdot \min_{1\leq i \leq n} |h'(t_{i})| \cdot \lambdamin(\Sigma,X-X) \cdot \|x - x'\|_{2}^{2},
\end{align*}
By noting that
\begin{align*}
  |h'(t_{i})| = \sum_{j=1}^{\N} s_{j} \coeff_{j} \exp\big( - \coeff_{j} \cdot t_{i} \big) \quad \text{and} \quad 0 \leq t_{i} \leq \max_{\widetilde{x} \in X} \; \langle a_{i}, \widetilde{x} \rangle,
\end{align*}
we obtain the bound
\[
  \min_{1\leq i \leq n} |h'(t_{i})| \geq \sum_{j=1}^{\N} s_{j} \coeff_{j} \cdot \exp\big(-\coeff_{j} \max_{\widetilde{x} \in X,\; 1\leq i\leq n} \langle a_{i}, \widetilde{x} \rangle \big).
\]
Putting the pieces together, we obtain 
\begin{align*}
  \langle F(x) - F(x'), x - x' \rangle \geq \Photon \cdot \lambdamin(\Sigma, X-X)  \sum_{j=1}^{\N} s_{j} \coeff_{j}  \exp\big( -\coeff_{j} \max_{\widetilde{x} \in X,\; 1\leq i\leq n} \langle a_{i}, \widetilde{x} \rangle \big)   \cdot \|x - x' \|_{2}^{2}.
\end{align*}
Since the above bound holds for any $x,x' \in X$, we conclude that
\begin{align}\label{strong-monotonicity-positive}
    \monotone \geq \Photon \cdot \lambdamin(\Sigma, X-X) \cdot \sum_{j=1}^{\N} s_{j} \coeff_{j} \cdot \exp\big( -\coeff_{j} \max_{\widetilde{x} \in X,\; 1\leq i\leq n} \langle a_{i}, \widetilde{x} \rangle \big).
\end{align}

\paragraph{Lipschitz constant} Let us now show that $F$ obeys the Lipschitz property, noting that Assumption~\ref{assump-positive} is not required for this claim. Define a vector $w \in \real^{n}$ with $w_i = h( \langle a_{i}, x \rangle) - h(\langle a_{i}, x' \rangle)$. We have
\begin{align*}
  \| F(x) - F(x') \|_{2}^{2} &= \frac{\Photon^{2}}{n^2} \Big \| \sum_{i=1}^{n} \big( h( \langle a_{i}, x \rangle) - h(\langle a_{i}, x' \rangle) \big) a_{i} \Big\|_{2}^{2} \\
  &= \frac{\Photon^{2}}{n^2} \| \bA^{\top} w \|_{2}^{2} \leq \frac{\Photon^{2}\lambdamax(\Sigma,\real^{d})}{n} \cdot \|w\|_{2}^2
\end{align*}
We may now write the squared norm of $w$ as
\begin{align*}
  \frac{1}{n}\|w\|_{2}^{2} & = \frac{1}{n} \sum_{i=1}^{n} \big[  h( \langle a_{i}, x \rangle) - h(\langle a_{i}, x' \rangle)  \big]^{2} \\
   & \overset{\1}{\leq} \Big( \sum_{j=1}^{\N}s_j \coeff_j \Big)^{2} \cdot \frac{1}{n} \sum_{i=1}^{n}  \big[ \langle a_{i}, x \rangle - \langle a_{i}, x' \rangle \big]^{2} \\
   &\leq \Big( \sum_{j=1}^{\N}s_j \coeff_j \Big)^{2} \cdot \lambdamax(\Sigma,\real^{d}) \cdot \|x- x'\|_{2}^{2}
\end{align*}
where in step $\1$ we use the Lipschitzness of function $h$, i.e., for all $t,t' \in \real$, 
\begin{align*}
    |h(t) - h(t') | \leq  \sum_{j=1}^{\N}s_j \big|  \exp\big(-\coeff_{j} \max\{t,0\}\big) - \exp\big(-\coeff_{j} \max\{t',0\}\big) \big| \leq \sum_{j=1}^{\N}s_j \coeff_{j} \cdot |t-t'|.
\end{align*}
Putting the two pieces together yields for all $x,x' \in \real^{d}$,
\begin{align}\label{ineq-Lipschitz-parameter}
  \| F(x) - F(x') \|_{2} \leq \Photon \cdot \lambdamax(\Sigma,\real^{d}) \cdot \Big( \sum_{j=1}^{\N}s_j \coeff_j \Big) \cdot \|x - x'\|_{2}.
\end{align}
In particular, we have shown that $F$ is Lipschitz with parameter $L = \Photon  \frac{\lambdamax(\Sigma,\real^{d})}{n}  \Big( \sum_{j=1}^{\N}s_j \coeff_j \Big)$. Note that in the above proof of the Lipschitzness, we do not require the measurement vectors $\{a_{i}\}_{i=1}^{n}$ to have only nonnegative entries, nor do we require $X \subseteq \{ x \in \mathbb{R}^{d} : x \geq 0 \}$ to hold.

\paragraph{Putting together the pieces}
To conclude, we apply Theorem~\ref{general-thm}, using the lower bound on $\nu$ given by Eq.~\eqref{strong-monotonicity-positive} and the the upper bound on $L$ given by Eq.~\eqref{ineq-Lipschitz-parameter}.
\qed

\subsection{Proof of Theorem~\ref{thm:mono-gaussian}}\label{sec:pf-thm-mono-gaussian}
As in the previous proof, the bulk of the argument relies on bounding the parameters $(\monotone, L)$ but now in the Gaussian measurement model. 
\begin{proposition}\label{prop-mono-gaussian}
Let $\gamma_{\star}$ be the solution of the fixed point equation~\eqref{fixed-point} and let $\overline{\omega}(x_{\star},X)$ be defined in Eq.~\eqref{def-Gaussian-width}. Suppose Assumption~\ref{assump-X} holds with $R \leq 4\norm$ and Assumption~\ref{Gaussian-assump} holds. 
Then for all sample size $n$ satisfying condition~\eqref{assump-sample-size-thm}, there exists a universal positive constant $c$ such that with probability at least $1-16\exp\big( - \overline{\omega}(x_{\star},X)^{2}/64 \big) - 12n^{-10} - 2\exp(-d)$, the following statements hold:

\noindent (a) The Lipschitz parameter $L$~\eqref{lipschitzness} satisfies
\begin{align*}
    L \leq 10  \Photon \cdot \frac{d+n}{n} \cdot \sum_{j=1}^{\N} s_{j} \coeff_{j}.
\end{align*}

\noindent (b) The strong pseudo-monotonicity parameter $\monotone$~\eqref{strong-monotone} satisfies
\begin{align*}
    \monotone \geq c \Photon \cdot \bigg( \frac{\gamma_{\star}}{ \gamma_{\star} + \sum_{j=1}^{\N}s_{j}\coeff_{j} } \bigg)^{2}  \cdot  \bigg(  \sum_{j=1}^{\N} \frac{s_{j} \coeff_{j}}{ \big(\coeff_{j} \norm\big)^{4} \vee 1 }  \bigg).
\end{align*}
\end{proposition}
We provide the proof of Proposition~\ref{prop-mono-gaussian} in Section~\ref{sec:pf-prop-mono-gaussian}.
Applying Theorem~\ref{general-thm} in conjunction with the lower bound on $\monotone$ and the upper bound on $L$ given by Proposition~\ref{prop-mono-gaussian} completes the proof of Theorem~\ref{thm:mono-gaussian}.
\qed

\subsubsection{Proof of Proposition~\ref{prop-mono-gaussian}(a)}\label{sec:pf-prop-mono-gaussian}

Under Assumption~\ref{Gaussian-assump}, applying~\cite[Theorem 4.4.5]{vershynin2018high} yields
\[
    \lambda_{\mathsf{max}}(\Sigma) = \| A \|_{\mathsf{op}}^{2} \leq 10 (d+n), \quad \text{with probability} \geq 1-2e^{-d}.
\]
Combining this inequality with inequality~\eqref{ineq-Lipschitz-parameter} yields the desired result.  

\subsubsection{Proof of Proposition~\ref{prop-mono-gaussian}(b)}

Define the shorthand
\begin{align}\label{def-high-dim-inf}
  \mathcal{P}(A) = \inf_{\substack{ x \in X,\; x \neq x_{\star}}} \; \frac{ \frac{1}{n} \sum_{i=1}^{n} \big[ h(a_{i}^{\top}x_{\star}) - h(a_{i}^{\top}x) \big] \cdot \big[ a_{i}^{\top}x - a_{i}^{\top}x_{\star} \big] }{\| x - x_{\star}\|_{2}^{2} },
\end{align}
and note that 
\[
\inf_{ \substack{x \in X,\; x \neq x_{\star}}} \; \frac{ \langle F(x) - F(x_{\star}), x - x_{\star} \rangle}{\|x - x_{\star}\|_{2}^{2}} = \Photon \cdot \mathcal{P}(A).
\]
To prove the claimed bound, it suffices to show that the following holds with probability at least $1-16\exp\big( - \overline{\omega}(x_{\star},X)^{2}/64 \big) - 12n^{-10}$: 
\[
\mathcal{P}(A) \geq \monotone.
\]
We first reduce the high-dimensional infimum in Eq.~\eqref{def-high-dim-inf} to a low-dimensional infimum by employing the Gaussian minmax theorem; we will use a variant of the result in~\cite{gordon1985some,gordon1988milman,thrampoulidis2018precise}. Let $g,\widetilde{g} \overset{\mathsf{i.i.d.}}{\sim} \NORMAL(0,I_{d})$ be Gaussian random vectors chosen independently of the data $\{a_{i}\}_{i=1}^{n}$. We define the random quantity
\begin{align}\label{Random-Gaussian-width}
  \omega(x_{\star},X) = \sup_{v \in X}\; \frac{ \langle P^{\perp}_{x_{\star}} v, g  \rangle }{\| P^{\perp}_{x_{\star}} v \|_{2}},
\end{align}
whose expectation is equal to the previously defined scalar $\overline{\omega}(x_{\star}, X)$ (Eq.~\eqref{def-Gaussian-width}).
We also define the random variable
\begin{align}\label{def-low-dim-inf}
  \mathcal{D}(Ax_{\star},g,\widetilde{g}) &: = \inf_{ \substack{\alpha \in \real,\; \beta\geq 0, \\ 0<(\alpha - \|x_{\star}\|_{2})^{2} + \beta^{2} \leq (4\norm)^{2} }} \; \frac{ \sup_{\gamma \geq 0} \; \bigg[ \frac{1}{n} \sum_{i=1}^{n} \mathcal{T}_{i}(\alpha,\beta,\gamma)  - \frac{\gamma \cdot \beta^{2}}{7n} \cdot \omega(x_{\star},X)^{2} \bigg]}{(\alpha - \|x_{\star}\|_{2})^{2} + \beta^{2}},
\end{align}
where for each $i\in[n]$, the random variable $\mathcal{T}_{i}(\alpha,\beta,\gamma)$  is defined as
\begin{align}
  \mathcal{T}_{i}(\alpha,\beta,\gamma) &:= \frac{1}{7}\frac{\gamma \cdot |h'(6a_{i}^{\top}x_{\star})|}{\gamma + |h'(6a_{i}^{\top}x_{\star})|} \cdot \Big( \Big(\frac{\alpha}{\norm} - 1\Big) a_{i}^{\top}x_{\star} + \beta\widetilde{g}_{i} \Big)^{2} \nonumber\\
   & \qquad \qquad \cdot \mathbbm{1}\Big\{ -6a_{i}^{\top}x_{\star} \leq \beta\widetilde{g}_{i} + \frac{\alpha}{\norm} a_{i}^{\top}x_{\star} \leq 6a_{i}^{\top}x_{\star}, a_{i}^{\top}x_{\star}>0\Big\}. \label{def-T-alpha-beta-gamma}
\end{align}
The notation $\mathcal{D}(Ax_{\star},g,\widetilde{g})$ reflects that $\mathcal{D}$ depends on the matrix $A$ only through the one-dimensional projection $Ax_{\star}$.
We next state an auxiliary lemma that relates quantity $\mathcal{P}(A)$~\eqref{def-high-dim-inf} and quantity $\mathcal{D}(Ax_{\star},g,\widetilde{g})$~\eqref{def-low-dim-inf}.
\begin{lemma}\label{lemma:gordon} 
Let $\mathcal{P}(A)$ and $\mathcal{D}(Ax_{\star},g,\widetilde{g})$ be defined as in Eq.~\eqref{def-high-dim-inf} and Eq.~\eqref{def-low-dim-inf}, respectively. Then
\[
  \Pr\big\{ \mathcal{P}(A)  \leq \tau \big\} \leq 2\Pr\big\{ \mathcal{D}(Ax_{\star},g,\widetilde{g}) \leq \tau \} \quad \text{for all } \tau \in \real .
\]
\end{lemma}
We provide the proof of Lemma~\ref{lemma:gordon} in~\ref{sec:proof-lemma-gordon}. 
Finally, we claim that there exists a universal and positive constant $c$ such that with probability at least $1-8\exp\big( - \overline{\omega}(x_{\star},X)^{2}/64 \big) - 6n^{-10}$,
\begin{align}\label{claim:low-dim-inf}
    \mathcal{D}(Ax_{\star},g,\widetilde{g}) \geq c\bigg( \frac{\gamma_{\star}}{ \gamma_{\star} + \sum_{j=1}^{\N}s_{j}\coeff_{j} } \bigg)^{2}  \cdot  \bigg(  \sum_{j=1}^{\N} \frac{s_{j} \coeff_{j}}{ \big(\coeff_{j} \norm\big)^{4} \vee 1 }  \bigg).
\end{align}
Combining this inequality with Lemma~\ref{lemma:gordon}, we obtain that with probability at least $1-16\exp\big( - \overline{\omega}(x_{\star},X)^{2}/64 \big) - 12n^{-10}$,
\[
    \mathcal{P}(A) \geq c\bigg( \frac{\gamma_{\star}}{ \gamma_{\star} + \sum_{j=1}^{\N}s_{j}\coeff_{j} } \bigg)^{2}  \cdot  \bigg( \sum_{j=1}^{\N} \frac{s_{j} \coeff_{j}}{ \big(\coeff_{j} \norm\big)^{4} \vee 1 }  \bigg).
\]
This concludes the proof of strong pseudo-monotonicity. It remains to verify claim~\eqref{claim:low-dim-inf}.

\paragraph{Proof of Claim~\eqref{claim:low-dim-inf}} 
To reduce the notational burden, we let 
\begin{align}\label{loss-gamma}
    \Gamma(\alpha,\beta): = \sup_{\gamma \geq 0} \; \bigg[ \frac{1}{n} \sum_{i=1}^{n} \mathcal{T}_{i}(\alpha,\beta,\gamma)  - \frac{\gamma \cdot \beta^{2}}{7n} \cdot \omega(x_{\star},X)^{2}  \bigg].
\end{align}
We claim that there exists a universal and positive constant $c$ such that uniformly for all $(\alpha,\beta)$ satisfying $\beta \geq 0$ and $0<(\alpha - \norm)^{2} + \beta^{2}\leq (4\norm)^{2}$, the following holds with probability at least $1-16\exp\big( - \overline{\omega}(x_{\star},X)^{2}/64 \big) - 12n^{-10}$.
\begin{subequations}
   \begin{align}
   \label{subclaim-1}
     &\text{If } \beta>0, \text{then }\frac{\Gamma(\alpha,\beta)}{ (\alpha-\norm)^{2} + \beta^{2} } \geq c \bigg( \frac{\gamma_{\star}}{ \gamma_{\star} + \sum_{j=1}^{\N}s_{j}\coeff_{j} } \bigg)^{2}  \bigg( \sum_{j=1}^{\N} \frac{s_{j} \coeff_{j}}{ \big(\coeff_{j} \norm\big)^{4} \vee 1 }  \bigg);\\
    \label{subclaim-2}
     &\text{If }\beta = 0, \text{then }  \frac{ \Gamma(\alpha,0) }{(\alpha - \norm)^{2}} \geq c  \bigg( \sum_{j=1}^{\N} \frac{s_{j} \coeff_{j}}{ \big(\coeff_{j} \norm\big)^{4} \vee 1 }  \bigg).
    \end{align} 
\end{subequations}
Note that Claim~\eqref{claim:low-dim-inf} immediately follows by the definition of $\mathcal{D}(Ax_{\star},g,\widetilde{g})$ in Eq.~\eqref{def-low-dim-inf}. We now turn to verify inequality~\eqref{subclaim-1} and inequality~\eqref{subclaim-2}.

\vspace{1mm}

\noindent \underline{Verifying inequality~\eqref{subclaim-1}.}
We first focus on computing the inner supremum over $\gamma>0$ in Eq.~\eqref{def-low-dim-inf}. For each $\alpha \in \real$ and $\beta> 0$, let $\overline{\gamma}(\alpha,\beta)$ be the solution of the inner $\sup_{\gamma\geq 0}$ in Eq.~\eqref{def-low-dim-inf}, i.e.,
\begin{align}\label{def-gamma-argsup}
  \overline{\gamma}(\alpha,\beta) : = \arg \sup_{\gamma \geq 0}\; \bigg[  \frac{1}{n} \sum_{i=1}^{n} \mathcal{T}_{i}(\alpha,\beta,\gamma)  - \frac{\gamma \cdot \beta^{2}}{7n} \cdot \omega(x_{\star},X)^{2} \bigg],
\end{align} 
where by convention we choose the smallest element of the $\arg\sup$ if the set is not a singleton. The following auxiliary lemma states some useful properties of $\overline{\gamma}(\alpha,\beta)$.
\begin{lemma}\label{lemma-gamma-property}
Let $\overline{\gamma}(\alpha,\beta)$ be defined in Eq.~\eqref{def-gamma-argsup} and let $\gamma_{\star}$ be defined in Eq.~\eqref{fixed-point}. Let $\overline{\omega}(x_{\star},X)$ and $\omega(x_{\star},X)$ be defined in Eq.~\eqref{def-Gaussian-width} and Eq.~\eqref{Random-Gaussian-width}, respectively. There exists a universal and positive constant $C$ such that if $n \geq C \cdot \overline{\omega}(x_{\star},X)$, then uniformly for all 
\begin{align}\label{assump-of-alpha-beta}
    \alpha \in \real,\quad \beta> 0,\quad \text{and} \quad   0< (\alpha - \norm)^{2} + \beta^{2} \leq (4\norm)^{2},
\end{align}
we have with probability at least $1-8\exp\big( - \overline{\omega}(x_{\star},X)^{2}/64 \big)$, the scalar $\overline{\gamma}(\alpha,\beta)$ is finite, and furthermore
\begin{align}\label{event-lower-bound-gamma-bar}
  \overline{\gamma}(\alpha,\beta) \geq \gamma_{\star}>0.
\end{align}
\end{lemma}
We provide the proof of Lemma~\ref{lemma-gamma-property} in~\ref{sec:proof-lemma-gamma-property}. 

Note that $\mathcal{T}_{i}(\alpha,\beta,\gamma)$ defined in Eq.~\eqref{def-T-alpha-beta-gamma} is a concave function of $\gamma$ for each $i \in [n]$. Then we have either $\overline{\gamma}(\alpha,\beta) = 0$ or $\overline{\gamma}(\alpha,\beta)$ must satisfy the first-order optimality condition, i.e.,
\begin{align*}
    \frac{1}{n} \sum_{i=1}^{n}  \frac{ \partial\mathcal{T}_{i}(\alpha,\beta,\gamma) }{ \partial \gamma} - \frac{\beta^{2}}{7n} \cdot \omega(x_{\star},X)^{2} = 0.
\end{align*}
But on the event guaranteed by Lemma~\ref{lemma-gamma-property}, we have $\overline{\gamma}(\alpha,\beta) \geq \gamma_{\star}>0$ with probability at least $1-8\exp\big( - \overline{\omega}(x_{\star},X)^{2}/64 \big)$. 
Thus, the first-order optimality condition must hold, and evaluating the partial derivative $\partial\mathcal{T}_{i}(\alpha,\beta,\gamma) / \partial \gamma$, this is equivalent to the condition
\begin{align}\label{gamma-bar-kkt}
  \frac{1}{n} \sum_{i=1}^{n} \frac{|h'(6a_{i}^{\top}x_{\star})|^{2} \cdot \big( (\frac{\alpha}{\norm}-1)a_{i}^{\top}x_{\star} + \beta \widetilde{g}_{i} \big)^{2} }{(\overline{\gamma}(\alpha,\beta) + |h'(6a_{i}^{\top}x_{\star})| )^{2}} \cdot &\mathbbm{1}\Big\{  |\beta \widetilde{g}_{i} + \frac{\alpha}{\norm} a_{i}^{\top}x_{\star}| \leq 6a_{i}^{\top}x_{\star}, a_{i}^{\top}x_{\star}>0 \Big\} \nonumber \\ &= \frac{\beta^2 \omega(x_{\star},X)^{2}}{n}.
\end{align}
Thus, recalling the definition of $\Gamma(\alpha,\beta)$ in Eq.~\eqref{loss-gamma}, we have
\begin{align}\label{ineq1-pf-claim-low-inf}
    \Gamma(\alpha,\beta) &=  \frac{1}{n} \sum_{i=1}^{n} \mathcal{T}_{i}(\alpha,\beta,\overline{\gamma}(\alpha,\beta) )  - \frac{\overline{\gamma}(\alpha,\beta) \cdot \beta^{2}}{7n} \cdot \omega(x_{\star},X)^{2} \nonumber \\
    & \overset{\1}{=} \frac{1}{7n} \sum_{i=1}^{n} \bigg( \frac{\overline{\gamma}(\alpha,\beta)  }{ \overline{\gamma}(\alpha,\beta) + |h'(6a_{i}^{\top}x_{\star})| } \bigg)^{2} \cdot |h'(6a_{i}^{\top}x_{\star})| \cdot \big[ (\alpha/\norm-1) a_{i}^{\top}x_{\star} + \beta \widetilde{g}_{i} \big]^{2} \nonumber \\ &  \qquad \qquad \qquad \qquad \qquad   \cdot \mathbbm{1}\Big\{ -6a_{i}^{\top}x_{\star} \leq \beta \widetilde{g}_{i} + \frac{\alpha}{\norm} a_{i}^{\top}x_{\star} \leq 6a_{i}^{\top}x_{\star}, a_{i}^{\top}x_{\star}>0 \Big\}.
\end{align}
where in step $\1$, we have used the definition~\eqref{def-T-alpha-beta-gamma} of $\mathcal{T}_i$ and the first-order optimality condition~\eqref{gamma-bar-kkt}.

Next, note that since $\overline{\gamma}(\alpha,\beta) \geq \gamma_{\star}$ and $|h'(t)| \leq \sum_{j=1}^{\N}s_{j} \coeff_{j}$ for all $t>0$, we have
\begin{align}\label{ineq2-pf-claim-low-inf}
    \frac{\overline{\gamma}(\alpha,\beta)  }{ \overline{\gamma}(\alpha,\beta) + |h'(6a_{i}^{\top}x_{\star})| } \geq \frac{\gamma_{\star}}{ \gamma_{\star} + \sum_{j=1}^{\N}s_{j}\coeff_{j} }  \quad \text{for} \quad a_{i}^{\top}x_{\star} > 0.
\end{align}
Also note that for $(\alpha,\beta)$ satisfying condition~\eqref{assump-of-alpha-beta}, we have
\begin{align}\label{ineq-indicator-lower-bounf}
    \mathbbm{1}\Big\{ -6a_{i}^{\top}x_{\star} \leq \beta \widetilde{g}_{i} &+ \frac{\alpha}{\norm} a_{i}^{\top}x_{\star} \leq 6a_{i}^{\top}x_{\star}, a_{i}^{\top}x_{\star}>0 \Big\} \geq \mathbbm{1}\Big\{ -a_{i}^{\top}x_{\star} \leq \beta \widetilde{g}_{i}  \leq a_{i}^{\top}x_{\star}, a_{i}^{\top}x_{\star}>0 \Big\} \nonumber \\
    &\geq \mathbbm{1}\Big\{ -\frac{a_{i}^{\top}x_{\star}}{4\norm} \leq  \widetilde{g}_{i}  \leq \frac{a_{i}^{\top}x_{\star}}{4\norm}, a_{i}^{\top}x_{\star}>0 \Big\}.
\end{align}
Substituting the inequality in the display above and inequality~\eqref{ineq2-pf-claim-low-inf} into inequality~\eqref{ineq1-pf-claim-low-inf}, we obtain that with probability at least $1-8\exp\big( - \overline{\omega}(x_{\star},X)^{2}/64 \big)$ the following holds for all $(\alpha,\beta)$ satisfying condition~\eqref{assump-of-alpha-beta}:
\begin{align}\label{ineq:lower-bound-inner-sup}
    &\Gamma(\alpha,\beta) \nonumber \\
    &\geq 
    \bigg( \frac{\gamma_{\star}}{ \gamma_{\star} + \sum_{j=1}^{\N}s_{j}\coeff_{j} } \bigg)^{2} \times  \nonumber \\
    & \qquad \qquad \frac{1}{7n} \sum_{i=1}^{n} |h'(6a_{i}^{\top}x_{\star})| \cdot \big[ (\alpha/\norm-1) a_{i}^{\top}x_{\star} + \beta \widetilde{g}_{i} \big]^{2} \cdot \mathbbm{1}\Big\{ |\widetilde{g}_{i}|  \leq \frac{a_{i}^{\top}x_{\star}}{4\norm}, a_{i}^{\top}x_{\star}>0 \Big\} \nonumber \\
    &   =:  \bigg( \frac{\gamma_{\star}}{ \gamma_{\star} + \sum_{j=1}^{\N}s_{j}\coeff_{j} } \bigg)^{2} \cdot \bigg( (\alpha - \norm)^{2} \cdot \frac{1}{7n} \sum_{i=1}^{n} \theta_{i} + \frac{\beta^{2}}{7n} \sum_{i=1}^{n} \vartheta_{i} + 2(\alpha-\norm) \beta \cdot \frac{1}{7n} \sum_{i=1}^{n} \xi_{i} \bigg),
\end{align}
where in the last step we have expanded the square and used the following shorthand: 
\begin{subequations}\label{def-theta-xi}
\begin{align}
\label{def-theta-i}
\theta_{i} &= |h'(6a_{i}^{\top} x_{\star})| \Big(\frac{a_{i}^{\top}x_{\star}}{\norm}\Big)^{2} \cdot \mathbbm{1}\Big\{ -\frac{a_{i}^{\top}x_{\star}}{4\norm} \leq  \widetilde{g}_{i}  \leq \frac{a_{i}^{\top}x_{\star}}{4\norm}, a_{i}^{\top}x_{\star}>0 \Big\}, \\
\vartheta_{i} & = |h'(6a_{i}^{\top} x_{\star})| \widetilde{g}_{i}^{2} \cdot \mathbbm{1}\Big\{ -\frac{a_{i}^{\top}x_{\star}}{4\norm} \leq  \widetilde{g}_{i}  \leq \frac{a_{i}^{\top}x_{\star}}{4\norm}, a_{i}^{\top}x_{\star}>0 \Big\}, \\
\xi_{i} &= |h'(6a_{i}^{\top} x_{\star})| \frac{a_{i}^{\top}x_{\star}}{\norm} \widetilde{g}_{i} \cdot \mathbbm{1}\Big\{ -\frac{a_{i}^{\top}x_{\star}}{4\norm} \leq  \widetilde{g}_{i}  \leq \frac{a_{i}^{\top}x_{\star}}{4\norm}, a_{i}^{\top}x_{\star}>0 \Big\}.
\end{align}
\end{subequations}
Note that all random variables defined in Eq.~\eqref{def-theta-xi} are independent of the tuple $(\alpha, \beta, \gamma)$.
The following auxiliary lemma controls these random variables with high probability.
\begin{lemma}\label{lemma:concentration-theta-xi}
Let $\{ \theta_{i}, \vartheta_{i}, \xi_{i} \}_{i=1}^{n}$ be defined in Eq.~\eqref{def-theta-xi}. Suppose the sample size $n$ satisfies condition~\eqref{assump-sample-size-thm}. Then there exists a pair of universal and positive constants $(c_{1},C_{1})$ such that with probability at least $1-6n^{-10}$, we have
\begin{align*}
    \frac{1}{n} \sum_{i=1}^{n} \theta_{i} \; \wedge \; \frac{1}{n} \sum_{i=1}^{n} \vartheta_{i}  \geq c_{1} \sum_{j=1}^{\N} \frac{s_{j} \coeff_{j}}{ \big(\coeff_{j} \norm\big)^{4} \vee 1 } \quad \text{and} \quad \bigg|  \frac{1}{n} \sum_{i=1}^{n} \xi_{i} \bigg| \leq C_{1} \sum_{j=1}s_{j} \coeff_{j} \sqrt{\frac{\log(n)}{ n}}.
\end{align*}
\end{lemma}
We provide the proof of Lemma~\ref{lemma:concentration-theta-xi} in~\ref{sec:pf-concentration-theta-xi}. Substituting the bound of Lemma~\ref{lemma:concentration-theta-xi} into inequality~\eqref{ineq:lower-bound-inner-sup} yields that with probability at least $1-8\exp\big( - \overline{\omega}(x_{\star},X)^{2}/64 \big) - 6n^{-10}$,
\begin{align}\label{ineq:lower-bound-inner-sup-1}
    \Gamma(\alpha,\beta) &\geq \frac{1}{7}\bigg( \frac{\gamma_{\star}}{ \gamma_{\star} + \sum_{j=1}^{\N}s_{j}\coeff_{j} } \bigg)^{2} \times \nonumber \\
    &\bigg( \big[ (\alpha-\norm)^{2} + \beta^{2} \big] \cdot  c_{1} \sum_{j=1}^{\N} \frac{s_{j} \coeff_{j}}{ \big(\coeff_{j} \norm\big)^{4} \vee 1 }  - 2C_{1}|\alpha - \norm| \beta \sum_{j=1}^{\N} s_{j} \coeff_{j} \sqrt{\frac{\log(n)}{n}}  \bigg) \nonumber \\
    & \overset{\1}{\geq} \frac{1}{7}\bigg( \frac{\gamma_{\star}}{ \gamma_{\star} + \sum_{j=1}^{\N}s_{j}\coeff_{j} } \bigg)^{2} \bigg(   \frac{c_{1}}{2} \sum_{j=1}^{\N} \frac{s_{j} \coeff_{j}}{ \big(\coeff_{j} \norm\big)^{4} \vee 1 }  \bigg) \Big( (\alpha-\norm)^{2} + \beta^{2} \Big),
\end{align}
where in step $\1$ we use the AM-GM inequality and condition~\eqref{assump-sample-size-thm} so that
\[
    2C_{1}|\alpha - \norm| \beta \sum_{j=1}^{\N}s_{j} \coeff_{j} \sqrt{\frac{\log(n)}{n}} \leq  \frac{c_1}{2} \big[ (\alpha-\norm)^{2} +\beta^{2} \big] \cdot \sum_{j=1}^{\N} \frac{s_{j} \coeff_{j}}{ \big(\coeff_{j} \norm\big)^{4} \vee 1 },
\]
This concludes the proof of inequality~\eqref{subclaim-1}

\vspace{1mm}

\noindent \underline{Verifying inequality~\eqref{subclaim-2}.} By the definitions of $\Gamma(\alpha,\beta)$~\eqref{loss-gamma} and $\mathcal{T}_{i}(\alpha,\beta,\gamma)$~\eqref{def-T-alpha-beta-gamma}, we obtain
\begin{align*}
    \Gamma(\alpha,0) &= \sup_{\gamma \geq 0} \bigg[ \frac{1}{7n} \sum_{i=1}^{n} \frac{\gamma |h'(6a_{i}^{\top}x_{\star})|}{\gamma +|h'(6a_{i}^{\top}x_{\star})| } \frac{(a_{i}^{\top}x_{\star})^{2}}{\norm^{2}} \cdot (\alpha -\norm)^{2} \cdot \mathbbm{1}\Big\{ |\alpha|  \leq 6\norm, a_{i}^{\top}x_{\star}>0 \Big\} \bigg] 
     \\ & \overset{\1}{=} \frac{1}{7n} \sum_{i=1}^{n} |h'(6a_{i}^{\top}x_{\star})| \frac{(a_{i}^{\top}x_{\star})^{2}}{\norm^{2}} \cdot (\alpha -\norm)^{2} \cdot \mathbbm{1}\Big\{a_{i}^{\top}x_{\star}>0 \Big\} \\
     &\overset{\2}{\geq} \frac{1}{7n} \sum_{i=1}^{n} \theta_{i}  \cdot (\alpha -\norm)^{2},
\end{align*}
where in step $\1$, the supremum is achieved when $\gamma \uparrow \infty$, since the condition $-6\norm \leq \alpha \leq 6\norm$ always holds due to the assumption
$(\alpha - \norm)^{2} \leq (4\norm)^{2}$. Step $\2$ follows from definition of $\theta_{i}$ in Eq.~\eqref{def-theta-xi}. To conclude, we use the high-probability lower bound on $\frac{1}{n} \sum_{i=1}^{n} \theta_{i}$ guaranteed by Lemma~\ref{lemma:concentration-theta-xi} to
obtain the desired inequality~\eqref{subclaim-2}.\qed


\section{Discussion} \label{sec:discussion}

We presented the first provable and efficient algorithm for single-material, polychromatic reconstruction from spectral CT measurements, which was based on carefully constructing a variational inequality whose fixed point approximated the desired signal. We presented explicit bounds on the sample and iteration complexity of our EXACT algorithm both for general nonnegative measurements, including the discretized Radon transform, and random matrix measurements, handling noise in the measurements and structure in the signal. Several downstream questions are interesting from both the theoretical and methodological points of view --- we point out a few such questions below. 

Let us start with some interesting theoretical questions. First, understanding the fundamental (i.e., information-theoretic) limits of signal recovery in this problem --- say under the Poisson noise model~\eqref{Poisson-model} --- is an important direction, since it would provide guidance on whether the performance of our estimator can be improved.  A second family of theoretical questions for the algorithm when run on Gaussian data is to characterize sharp rates of convergence and sample complexity (i.e., with exact constant factors). Note that our algorithm itself is a \emph{general first-order method} in the nomenclature of~\cite{celentano2020estimation}, so we may be able to prove a so-called ``state evolution" that characterizes its iterate-by-iterate performance as well as its eventual error. On a complementary note, information-theoretic lower bounds for the problem should be derivable via the techniques in~\cite{barbier2019optimal}. When does the eventual error of our algorithm exactly match the information-theoretic lower bound in the Gaussian model? To our knowledge, in spite of considerable research on monotone single-index models with Gaussian data~\cite[e.g.,][]{balabdaoui2019least,pananjady2021single} and although there has been some recent interest in proving state evolutions for algorithms that go beyond first-order methods~\cite[see, e.g.,][]{chandrasekher2023sharp, chandrasekher2024alternating, lou2024hyperparameter, kaushik2024precise,celentano2025state}, such questions have not received attention for algorithms that find fixed-points of (strongly) monotone VIs. 

From a methodological viewpoint also, there are several interesting open questions. The first and most important open question is how to design a variant of our methodology for the multi-material recovery problem in Eq.~\eqref{eqn:referenceforwardmodel}. Note that in that case, the natural analog of the operator $F$ that we use is no longer monotone, so we would need to take a different approach to designing a provably convergent algorithm. Another possible extension of the algorithm is to replace the projection operation onto the constraint set $X$ with other structure-enforcing operations. For instance, one can introduce a regularization function $\ell_{X} : \mathbb{R}^d \rightarrow \mathbb{R}_+$ associated with $X$. When $X$ is a total variation (TV) ball, a natural choice is $\ell_{X}(x) = \|x\|_{\mathsf{TV}}$. More generally, for a compact convex set $X$ containing the origin, a typical choice is the gauge function $\ell_{X}(x) = \inf \{ \lambda > 0 : x \in \lambda X \}$. In this setting, the iterates~\eqref{extragrad-method} can be replaced by
\begin{align*}
x_{t+0.5} &= \arg\min_{x \in \mathbb{R}^d} \left\| x - \left( x_t - \gamma_t F(x_t) \right) \right\|_2^2 + \lambda_t \,\ell_{X}(x), \\
x_{t+1} &= \arg\min_{x \in \mathbb{R}^d} \left\| x - \left( x_t - \gamma_t F(x_{t+0.5}) \right) \right\|_2^2 + \lambda_t \,\ell_{X}(x),
\end{align*}
for $t = 1, 2, \dots$, where $\lambda_t > 0$ is a regularization parameter. Such a modification may offer computational advantages over the projection step. Establishing theoretical guarantees for these extensions is an interesting direction for future work.
Moreover, our analysis also opens up interesting measurement design questions, such as how to optimally allocate a limited photon budget by trading off source intensity $\Photon$ and number of measurements $n$ to yield the highest reconstruction accuracy. Future analysis may also consider detector nonidealities such as \emph{pulse pileup} and \emph{charge sharing}, in which some incident photons are not recorded or are recorded as multiple photons with incorrect energies. For example, reducing the source intensity $\Photon$ increases shot noise but also mitigates pulse pileup by reducing the likelihood that a photon arrives during the sensor's dead time. Incorporating these effects into algorithm design and analysis may inform refinements to the reconstruction process and measurement design setup.

\subsection*{Acknowledgments}
KAV was at Georgia Tech and SFK was at Stanford when part of this work was performed. ML, KAV and AP were supported in part by the NSF through grants CCF-2107455 and DMS-2210734, and by research awards from Adobe, Amazon, Mathworks and Google. SFK was supported in part by the NSF Mathematical Sciences Postdoctoral Research Fellowship under award number 2303178. We thank Amir Pourmorteza for helpful discussions and for suggesting the experimental idea. We also thank Namhoon Kim for assistance in debugging the experiment used to produce Figure 1. Finally, we thank the anonymous reviewers for their suggestions, which improved the scope and presentation of the paper.

\bibliographystyle{abbrvnat}
\bibliography{main}

\appendix
\begin{center}
\Large{\textbf{Appendix}}
\end{center}

\section{Proof of statements in the main text}\label{sec:pf-statements}

We first present calculations for several statements that were made in the main text.

\subsection{Proof of statements in the introduction}
\label{sec:math-model-multi-material}
\paragraph{Mathematical model underlying the experiment in Figure~\ref{fig:comparison}} The task is to recover a single material when the other two materials are assumed to be known. We consider the Poisson noise model:
    \begin{align}\label{model:multiple-material}
        y_{i} \overset{\mathsf{i.i.d.}}{\sim} \mathsf{Poi}\!\left( \Photon \sum_{j=1}^{\N} s_{j} \exp\!\big( - \coeff_{j} \langle a_{i},x_{\star} \rangle  \big) \right). 
    \end{align}
    Here, $x_{\star} = \big[ x_{\star}^{1} \mid x_{\star}^{2} \mid x_{\star}^{3} \big] \in \mathbb{R}^{d \times 3}$ and $a_{i} = \big[ a_{i}^{1} \mid a_{i}^{2} \mid a_{i}^{3} \big] \in \mathbb{R}^{d \times 3}$ for all $i \in [n]$, since in our experiments there are three materials: water, bone, and iodine, modeled respectively by $x_{\star}^{1}$, $x_{\star}^{2}$, and $x_{\star}^{3}$. We focus on the setting where $x_{\star}^{1}$ and $x_{\star}^{2}$ are known, and our goal is to reconstruct $x_{\star}^{3}$. By reparameterizing the problem with
    \begin{align*}
        \Photon' &= \Photon \cdot \sum_{j=1}^{\N} s_{j} \exp\!\Big( -\coeff_{j} \big( \langle a_{i}^{1},x_{\star}^{1} \rangle + \langle a_{i}^{2},x_{\star}^{2} \rangle \big) \Big), \\
        s_{j}' &= \frac{s_{j} \exp\!\Big( -\coeff_{j} \big( \langle a_{i}^{1},x_{\star}^{1} \rangle + \langle a_{i}^{2},x_{\star}^{2} \rangle \big) \Big) }{\sum_{j=1}^{\N} s_{j} \exp\!\Big( -\coeff_{j} \big( \langle a_{i}^{1},x_{\star}^{1} \rangle + \langle a_{i}^{2},x_{\star}^{2} \rangle \big) \Big)} \quad \text{for all } j \in [\N],
    \end{align*}
    we obtain a single-material model with the new parameters, namely
    \begin{align*}
        y_{i} \overset{\mathsf{i.i.d.}}{\sim} \mathsf{Poi}\!\left( \Photon' \sum_{j=1}^{\N} s_{j}' \exp\!\big( - \coeff_{j} \langle a_{i}^{3},x_{\star}^{3} \rangle  \big) \right). 
    \end{align*}
    Our algorithmic framework and guarantees can now be applied to the task of recovering the unknown $x_{\star}^{3} \in \mathbb{R}^{d}$. By contrast, the linear model is based on a monochromatic approximation, obtained by directly applying a logarithm to both sides of model~\eqref{model:multiple-material}. Such a simplification cannot capture the setting where one seeks to recover a single material while other materials are known.

\paragraph{Mathematical model of both Poisson noise and electronic noise} We model the electronic noise as additive Gaussian noise. Specifically, we consider
\[
    y_{i} \overset{\mathsf{i.i.d.}}{\sim} \mathsf{Poi} \left( \Photon \sum_{j=1}^{\N} s_{j} \exp \left( - \coeff_{j} \langle a_{i}, x_{\star} \rangle \right) \right) + \mathcal{N}(0, \sigma^{2}),
\]
then the total error can be bounded using the triangle inequality:
\begin{align*}
    \Err(x_{\star}, X) \leq \sup_{x, x' \in X} \left\langle \xi_{1}, \frac{x - x'}{\|x - x'\|_{2}} \right\rangle + \sup_{x, x' \in X} \left\langle \xi_{2}, \frac{x - x'}{\|x - x'\|_{2}} \right\rangle,
\end{align*}
where $\xi_{1}$ and $\xi_{2}$ represent the contributions from Poisson and Gaussian noise, respectively. These are defined as:
\[
\xi_{1} = \frac{1}{n} \sum_{i=1}^{n} \left( \mathsf{Poi}(\lambda_{i}) - \lambda_{i} \right) a_{i}, \quad \text{with } \lambda_{i} = \Photon \sum_{j=1}^{\N} s_{j} \exp\left( - \coeff_{j} \langle a_{i}, x_{\star} \rangle \right),
\]
and
\[
\xi_{2} = \frac{1}{n} \sum_{i=1}^{n} \epsilon_{i} a_{i}, \quad \text{where } \{\epsilon_{i}\}_{i=1}^{n} \overset{\mathsf{i.i.d.}}{\sim} \mathcal{N}(0, \sigma^{2}).
\]

\subsection{Proof of claim in Example~\ref{prop-error-term}}\label{sec:pf-prop-error-term}

Since $X = \mathbb{B}(R)$ for some $R>0$, we obtain
\[
    \sup_{x,x' \in X,\, x \neq x'} \Big \langle \zeta, \frac{x-x'}{\|x - x'\|_{2}}  \Big \rangle = \|\zeta\|_{2}.
\]
For convenience, define the noise variables
\[
    R_{i} = y_{i} - \Photon \cdot h(\langle a_{i}, x_{\star}\rangle) \quad \text{for } i = 1,2,\dots,n,
\]
which are all independent and zero-mean, and drawn according to model~\eqref{Poisson-model}.

To proceed, note that
\begin{align*}
    \| \zeta \|_{2}^{2} = \frac{1}{n^{2}} \Big\| A^{\top} R  \Big\|_{2}^{2} = \frac{1}{n^{2}} R^{\top} A A^{\top} R,
\end{align*}
and that
\[
\EE\Big[ R^{\top} A A^{\top} R \Big] = \sum_{i=1}^{n} \|a_{i}\|_{2}^{2} \EE[R_{i}^{2}] = \Photon \sum_{i=1}^{n} \|a_{i}\|_{2}^{2} \cdot h(\langle a_{i}, x_{\star}\rangle).
\]
Next, note that $\{R_{i}\}_{i=1}^{n}$ are $\mathsf{i.i.d.}$ zero-mean random variables according to model~\eqref{Poisson-model}. Moreover, $\{ R_{i}\}_{i=1}^{n}$ are sub-exponential random variables, with bounded Orlicz $\psi_1$-norm~\cite{vershynin2018high}. Indeed, we have for all $0<t \leq 1$,  
\begin{align*}
    \EE\big[ \exp( t |R_{i}|) \big] &\leq \EE \big[ \exp(t |y_{i}|) \big] \cdot \exp\big( t \Photon h(\langle a_{i}, x_{\star} \rangle) \big) \\
    & = \exp\big( \Photon h(\langle a_{i},x_{\star} \rangle) (e^{t} - 1) \big) \cdot \exp \big(t \Photon h(\langle a_{i}, x_{\star} \rangle) \big) \\
    & \overset{\1}{\leq} \exp\big( \Photon (e^{t} - 1 + t) \big)  \\
    & \leq \exp\big( \Photon (e+1)t \big),
\end{align*}
where in step $\1$ we use $h(t) \leq 1$ for all $t \in \real$ and in the last step we use $e^{t} \leq 1 + e\cdot t$ for $t\in(0,1)$. Substituting $t = \big(10(e + 1) \Photon\big)^{-1}$, we have $\EE\Big[ \exp \Big( |R_{i}|/ \big( 10(e+1)\Photon\big) \Big) \Big] \leq e^{1/10} \leq 2$, thereby proving that $\| R_{i} \|_{\psi_{1}} \lesssim \Photon$.

We now apply the Hanson-Wright inequality for $1$-sub-exponential random variables~\cite[Proposition 1.1]{10.1214/21-EJP606}, which yields that for all $t>0$, we have
\begin{align*}
    \Pr \bigg\{ \Big| R^{\top} A A^{\top} R - \EE\Big[ R^{\top} A A^{\top} R \Big] \Big| \geq t  \bigg\} 
    \leq 2\exp\bigg( -c \min \bigg( \frac{t^{2}}{(C\Photon)^{4} \| AA^{\top} \|_{\mathsf{F}}^{2} }, \Big( \frac{t}{(C\Photon)^{2} \| AA^{\top} \|_{\mathsf{op}} } \Big)^{1/2} \bigg) \bigg),
\end{align*}
where $c$ and $C$ are two universal and positive constants. Some algebra yields that 
\begin{align*}
    R^{\top} A A^{\top} R - \EE\Big[ R^{\top} A A^{\top} R \Big] \lesssim (\Photon)^{2} \cdot \Big( \| AA^{\top} \|_{\mathsf{F}} \sqrt{\log(2/\delta)} + \| AA^{\top} \|_{\mathsf{op}} \log^2(2/\delta) \Big)
\end{align*}
with probability at least $1-\delta$. Substituting for the expectation and performing some algebraic simplifications proves the claim.
\qed

\subsection{Proof of uniqueness of fixed-point Eq.~\eqref{fixed-point}}\label{sec:pf-fixed-point}
We define a function $\psi:[0,+\infty) \mapsto \real$ as
\[
    \psi(\gamma) = \EE\bigg\{  \frac{|h'(6\norm G)|^{2} \cdot \mathbbm{1}\{G>0\} }{ \big[ |h'(6\norm G)| + \gamma \big]^{2} } \cdot
    \int_{-G/4}^{G/4} \frac{t^2 e^{-t^2/2}}{\sqrt{2\pi}} \mathrm{d}t  \bigg\}.
\]
Note that $\psi(\gamma)$ is a strictly decreasing function of $\gamma$,  
\[
     \lim_{\gamma \rightarrow +\infty} \psi(\gamma) = 0 \quad \text{and} \quad \psi(0) = \int_{0}^{+\infty} \int_{-\tau/4}^{\tau/4} \frac{t^2 e^{-t^2/2}}{\sqrt{2\pi}} \mathrm{d}t  \cdot \frac{e^{-\tau^{2}/2}} {\sqrt{2\pi}} \mathrm{d}\tau > 0,
\]
and $\psi(0)$ is a universal and positive constant. Consequently, for each $n \geq \big( \Constant/ \psi(0)\big) \cdot \overline{\omega}(x_{\star},X)^{2}$, the equation $\psi(\gamma) = \Constant \overline{\omega}(x_{\star},X)^{2}/n$ has a unique and strictly positive solution. 
\qed

\subsection{Proof of inequality~\eqref{gamma-star-lower-bound-regime1}} \label{app-first-ineq}

We obtain
\begin{align*}
     \EE\bigg\{  \frac{|h'(6\norm G)|^{2}  \mathbbm{1}\{G>0\} }{ \big[ |h'(6\norm G)| + \gamma_{\star}\big]^{2} } 
    \int_{-G/4}^{G/4} \frac{t^2 e^{-t^2/2}}{\sqrt{2\pi}} \mathrm{d}t  \bigg\} &\geq \int_{1}^{2} \frac{|h'(6\norm \tau)|^{2} }{ \big[ |h'(6\norm \tau)| + \gamma_{\star}\big]^{2} } \cdot \int_{-\tau/4}^{\tau/4} \frac{t^2 e^{-t^2/2}}{\sqrt{2\pi}} \mathrm{d}t \cdot \frac{e^{-\tau^2/2}}{\sqrt{2\pi}} \mathrm{d}\tau \\
    & \geq  \frac{|h'(12\norm)|^{2} }{ \big[ |h'(12\norm)| + \gamma_{\star}\big]^{2} } \cdot \int_{-1/4}^{1/4} \frac{t^2 e^{-t^2/2}}{\sqrt{2\pi}} \mathrm{d}t \cdot \frac{e^{-2}}{\sqrt{2\pi}}.
\end{align*}
Let $c_{0} = \int_{-1/4}^{1/4} \frac{t^2 e^{-t^2/2}}{\sqrt{2\pi}} \mathrm{d}t \cdot \frac{e^{-2}}{\sqrt{2\pi}}$, and note that it is a universal constant. Combining the inequality in the display above with definition~\eqref{fixed-point}, we obtain 
\[
    \frac{\Constant \overline{\omega}(x_{\star},X)^{2} }{n} \geq c_{0} \cdot \frac{|h'(12\norm)|^{2} }{ \big[ |h'(12\norm)| + \gamma_{\star}\big]^{2} }.
\]
Thus, as long as $n \geq (4\times \Constant/c_{0}) \cdot \overline{\omega}(x_{\star},X)^{2}$, we obtain
\[
    \gamma_{\star} \geq |h'(12\norm)| = \sum_{j=1}^{\N} s_{j} \coeff_{j} \exp\big( - 12\coeff_{j} \norm \big).
\]
\qed

\subsection{Proof of inequality~\eqref{gamma-star-lower-bound-regime2}}  \label{app-second-ineq}

Note that
$|h'(t)| \leq \sum_{j=1}^{\N}s_{j}\coeff_{j}$ for all  $t>0$. 
Thus,
\begin{align*}
    \frac{\Constant \overline{\omega}(x_{\star},X)^{2} }{n} &= \EE\bigg\{  \frac{|h'(6\norm G)|^{2}  \mathbbm{1}\{G>0\} }{ \big[ |h'(6\norm G)| + \gamma_{\star}\big]^{2} } 
    \int_{-G/4}^{G/4} \frac{t^2 e^{-t^2/2}}{\sqrt{2\pi}} \mathrm{d}t  \bigg\} \\
    &\overset{\1}{\gtrsim} \frac{ \EE\bigg\{ |h'(6\norm G)|^{2}  \mathbbm{1}\{G>0\} 
     e^{-G^{2}/16}G^{3}\bigg\} }{\big[ \sum_{j=1}^{\N}s_{j}\coeff_{j} + \gamma_{\star}  \big]^{2} }.
\end{align*}
where step $\1$ follows because $\int_{-G/4}^{G/4} \frac{t^2 e^{-t^2/2}}{\sqrt{2\pi}} \mathrm{d}t \geq \frac{e^{-G^{2}/16} }{\sqrt{2\pi}} \frac{(G/4)^{3}}{3}$ for all $G>0$.
Continuing, note that for $G>0$
\[
    |h'(6\norm G)|^{2} \geq \sum_{j=1}^{\N} s_{j}^{2}\coeff_{j}^{2} \exp\big( - 12 \coeff_{j} \norm G \big).
\]
We thus obtain
\begin{align*}
     \EE\bigg\{ |h'(6\norm G)|^{2}  \mathbbm{1}\{G>0\} 
     e^{-G^{2}/16}G^{3}\bigg\} &\geq \frac{1}{\sqrt{2\pi}} \sum_{j=1}^{\N} s_{j}^{2}\coeff_{j}^{2} \int_{0}^{+\infty} \exp\big( - 12 \coeff_{j} \norm t - t^{2}/16 - t^{2}/2\big) t^{3} \mathrm{d}t \\
     &\gtrsim \sum_{j=1}^{\N} s_{j}^{2}\coeff_{j}^{2} \int_{0}^{\frac{1}{6\coeff_{j} \norm}  \wedge \frac{1}{6}} \exp\big( - 12 \coeff_{j} \norm t - t^{2} \big) t^{3} \mathrm{d}t \\
     &\overset{\1}{\gtrsim} \sum_{j=1}^{\N} s_{j}^{2}\coeff_{j}^{2} \int_{0}^{\frac{1}{6\coeff_{j} \norm}  \wedge \frac{1}{6}} t^{3} \mathrm{d}t \\
     &\gtrsim \sum_{j=1}^{\N} s_{j}^{2}\coeff_{j}^{2} \frac{1}{ (\coeff_{j} \norm)^{4} \vee 1 },
\end{align*}
where in step $\1$ we use for $0<t \leq \frac{1}{6\coeff_{j} \norm}  \wedge \frac{1}{6}$,
\[
    \exp\big( - 12 \coeff_{j} \norm t - t^{2} \big) \geq e^{-2 - 1/36} \gtrsim 1.
\]
Putting all pieces together yields that there exists a universal and positive constant $C$ such that 
\begin{align*}
    C \frac{\overline{\omega}(x_{\star},X)^{2}}{n} \geq  \frac{ \sum_{j=1}^{\N}  \frac{s_{j}^{2}\coeff_{j}^{2}}{ (\coeff_{j} \norm)^{4} \vee 1 } }{ \big[ \sum_{j=1}^{\N}s_{j}\coeff_{j} + \gamma_{\star}  \big]^{2} }.
\end{align*}
Consequently, if $n$ satisfies
\[
        C \frac{\overline{\omega}(x_{\star},X)^{2}}{n} \leq \sum_{j=1}^{\N}  \frac{s_{j}^{2}\coeff_{j}^{2}}{ (\coeff_{j} \norm)^{4} \vee 1 }  \cdot \frac{1}{\big(2\sum_{j=1}^{\N} s_{j} \coeff_{j} \big)^{2}},
\]
then $\sum_{j=1}^{\N}s_{j}\coeff_{j} + \gamma_{\star} \geq 2\sum_{j=1}^{\N} s_{j} \coeff_{j}$ and thus $\gamma_{\star} \geq \sum_{j=1}^{\N} s_{j} \coeff_{j}$.
\qed

\section{Proof of auxiliary lemmas}
In this section, we provide proofs of all the auxiliary lemmas that were stated in Section~\ref{sec:proofs}.

\subsection{Proof of Lemma~\ref{lemma:gordon}} \label{sec:proof-lemma-gordon}
Introducing an auxiliary variable $u \in \real^n$, we may write
\begin{align*}
\mathcal{P}(A) &\overset{\1}{=} \inf_{ \substack{x \in X,\; u \in \real^{n}, \\ x \neq x_{\star} }} \; \sup_{\lambda \in \real^{n}} \; 
\frac{ \frac{1}{n} \sum_{i=1}^{n} \big[h(a_{i}^{\top}x_{\star}) - h(u_{i} ) \big] \cdot \big[ u_{i} - a_{i}^{\top}x_{\star} \big]  }{\|x - x_{\star}\|_{2}^{2}} + \frac{1}{n} \langle \lambda, Ax - u \rangle
\\& \overset{\2}{=} \inf_{ \substack{x \in X,\; u \in \real^{n}, \\ x \neq x_{\star} }} \; \sup_{\lambda \in \real^{n}} \; \frac{ \frac{1}{n} \sum_{i=1}^{n} \big[h(a_{i}^{\top}x_{\star}) - h(u_{i} ) \big] \cdot \big[ u_{i} - a_{i}^{\top}x_{\star} \big] }{ \|x - x_{\star}\|_{2}^{2} } + \frac{1}{n} \langle \lambda, A P_{x_{\star}}x - u \rangle + \frac{1}{n}\langle \lambda,A P_{x_{\star}}^{\perp} x \rangle
\end{align*}
where in step $\1$, the scalar $u_{i}$ is the $i$-th entry of $u\in \real^{n}$, and the inner sup is imposing the constraints $a_{i}^{\top}x = u_{i}$ for all $i \in [n]$. In step $\2$, we use the decomposition $x = P_{x_{\star}}x + P_{x_{\star}}^{\perp}x$, where we recall that $P_{x_\star}$ is the projection matrix onto the span of $x_{\star}$, and $P^{\perp}_{x_{\star}}$ is the projection matrix onto the orthogonal complement of this subspace. Note that $AP_{x_{\star}}^{\perp}$ and $AP_{x_{\star}}$ are independent of each other since entries of $A$ are $\mathsf{i.i.d.}$ standard Gaussian random variables. 

The decomposition above allows us to apply the CGMT (see, e.g.,~\cite{thrampoulidis2018precise} although we will use the version in Proposition 1 of \cite{chandrasekher2023sharp}) to the last term $ \frac{1}{n} \langle \lambda, A P_{x_{\star}}^{\perp} x  \rangle $ by condition on the other terms of $\mathcal{P}(A)$. As a consequence, we derive a surrogate for $\mathcal{P}(A)$. Specifically, let $g,\widetilde{g} \overset{\mathsf{i.i.d.}}{\sim} \NORMAL(0,I_{d})$ and be independent of $\{a_{i}\}_{i=1}^{n}$ and define 
\begin{align}\label{def-Q}
  \mathcal{Q}(Ax_{\star},g,\widetilde{g}) &= \inf_{ \substack{x \in X,\; u \in \real^{n}, \\ x \neq x_{\star} }} \; \sup_{\lambda \in \real^{n}} \; \frac{ \frac{1}{n} \sum_{i=1}^{n} \big[h(a_{i}^{\top}x_{\star}) - h(u_{i} ) \big] \cdot \big[ u_{i} - a_{i}^{\top}x_{\star} \big]}{\|x - x_{\star}\|_{2}^{2}} + \frac{1}{n} \langle \lambda, A P_{x_{\star}}x - u \rangle \nonumber \\
 &\qquad \qquad \qquad \qquad \qquad \qquad \qquad + \frac{1}{n} \|\lambda\|_{2} \cdot \langle P_{x_{\star}}^{\perp} x, g \rangle + \frac{1}{n} \|P_{x_{\star}}^{\perp} x\|_{2} \cdot \langle \lambda , \widetilde{g} \rangle.
\end{align}
By employing part (a) of Proposition 1 in~\cite{chandrasekher2023sharp}, we obtain 
\begin{align*}
  \Pr\big\{ \mathcal{P}(A) \leq \tau \big\} \leq 2\Pr\big\{ \mathcal{Q}(Ax_{\star}, g, \widetilde{g}) \leq \tau \big\} \quad \text{for all } \tau \in \real.
\end{align*}
Note that the statement of part (a) of Proposition~1 in~\cite{chandrasekher2023sharp} requires that the infimum and supremum be taken over compact sets.  At the end of the proof, we show that there is no loss in restricting the variational problem to compact sets.

In view of the above, it suffices to prove the following lower bound on $\mathcal{Q}$:
\[
  \mathcal{Q}(Ax_{\star}, g, \widetilde{g}) \geq \mathcal{D}(Ax_{\star},g,\widetilde{g}).
\]
To prove this lower bound, let us carefully manipulate the minmax problem~\eqref{def-Q} defining $\mathcal{Q}$. First, note that we may take the inner supremum in~\eqref{def-Q} by separately optimizing over the norm and direction of $\lambda$.
Taking the supremum over the direction of $\lambda \in \real^{n}$, we obtain
\begin{align*}
  \mathcal{Q}(Ax_{\star},g,\widetilde{g}) = \inf_{ \substack{x \in X,\; u \in \real^{n}, \\ x\neq x_{\star} }} \; \sup_{\lambda \in \real^{n}} \; & \frac{ \frac{1}{n} \sum_{i=1}^{n} \big[h(a_{i}^{\top}x_{\star}) - h(u_{i} ) \big] \cdot \big[ u_{i} - a_{i}^{\top}x_{\star} \big] }{ \|x - x_{\star}\|_{2}^{2} } \\
  &+ \frac{\|\lambda\|_{2}}{n} \big\| A P_{x_{\star}}x - u + \|P_{x_{\star}}^{\perp} x\|_{2} \cdot \widetilde{g}  \big\|_{2} + \frac{\|\lambda\|_{2}}{n} \langle P_{x_{\star}}^{\perp} x, g \rangle.
\end{align*}
Continuing, note that by definition of $\omega(x_{\star},X)$~\eqref{Random-Gaussian-width}, we obtain 
\[
  \langle P_{x_{\star}}^{\perp} x, g \rangle \geq - \| P_{x_{\star}}^{\perp} x \|_{2} \cdot \omega(x_{\star},X).
\]
Putting the pieces together yields and letting $\gamma = \|\lambda\|_{2}$, we obtain
\begin{align*}
  \mathcal{Q}(Ax_{\star},g,\widetilde{g}) \geq \inf_{ \substack{x \in X,\; u \in \real^{n}, \\ x\neq x_{\star} }} \; \sup_{\gamma \geq 0} \; & \frac{ \frac{1}{n} \sum_{i=1}^{n} \big[h(a_{i}^{\top}x_{\star}) - h(u_{i} ) \big] \cdot \big[ u_{i} - a_{i}^{\top}x_{\star} \big] }{\|x - x_{\star}\|_{x}^{2}} \\
  &+ \frac{\gamma}{n} \Big( \big\| A P_{x_{\star}}x - u + \|P_{x_{\star}}^{\perp} x\|_{2} \cdot \widetilde{g}  \big\|_{2} - \| P_{x_{\star}}^{\perp} x \|_{2} \cdot \omega(x_{\star},X) \Big).
\end{align*}
In the above program, the decision variable $x$ only enters through two low dimensional functionals $\inprod{x}{x_{\star}}$ and $\| P^{\perp}_{x_{\star}} x \|_2$. This suggests that we should define the scalar decision variables
\[
  \alpha = \langle x, x_{\star} \rangle / \norm \quad \text{and} \quad \beta = \| P_{x_{\star}}^{\perp} x\|_{2}.
\]
Note that by Assumption~\ref{assump-X}, we have 
\begin{align*}
    &\big\{ (\alpha,\beta) : \alpha = \langle x, x_{\star} \rangle / \norm, \beta = \| P_{x_{\star}}^{\perp} x\|_{2}, x \in X, x\neq x_{\star} \big\} \\
    & \quad \quad \subseteq \big\{ (\alpha,\beta): \alpha \in \real, \beta\geq 0, 0<(\alpha - \norm)^{2} + \beta^{2} \leq (4\norm)^{2} \big\}.
\end{align*}
The above infimum over $x$ can therefore be lower bounded using a low-dimensional infimum, thereby obtaining the bound
\begin{align}
  \mathcal{Q}(Ax_{\star},g,\widetilde{g}) \geq \inf_{ \substack{\alpha \in \real,\; \beta\geq 0, \; u\in \real^{n}, \\0< (\alpha - \|x_{\star}\|_{2})^{2} + \beta^{2} \leq (4\norm)^{2} }} \; \sup_{\gamma \geq 0} \;& \frac{ \frac{1}{n} \sum_{i=1}^{n} \big[h(a_{i}^{\top}x_{\star}) - h(u_{i} ) \big] \cdot \big[ u_{i} - a_{i}^{\top}x_{\star} \big] }{ (\alpha - \norm)^{2} + \beta^{2} } \nonumber \\
  &+ \frac{\gamma}{n} \Big( \big\|  A x_{\star} \alpha / \norm - u + \beta  \widetilde{g}  \big\|_{2} - \beta  \omega(x_{\star},X) \Big). \label{eq:intermediate-lb}
\end{align}
The rest of our proof is dedicated to lower-bounding the objective function in Eq.~\eqref{eq:intermediate-lb} in a particular form.
To do so, it is convenient to define a function $f:\real \times \real \mapsto \real$ as
\begin{align}\label{lower-bound-first-part}
  f(a_{i}^{\top}x_{\star},u_{i}) := \frac{1}{7} |h'(6a_{i}^{\top}x_{\star})| \cdot \big[ u_{i} - a_{i}^{\top}x_{\star} \big]^{2} \cdot \mathbbm{1}\big\{ a_{i}^{\top}x_{\star}>0, - 6a_{i}^{\top}x_{\star} \leq u_{i} \leq 6a_{i}^{\top}x_{\star} \big\}.
\end{align}
We will make use of the following relation: 
\begin{align}\label{claim:lower-bound-first-part}
  \big[h(a_{i}^{\top}x_{\star}) - h(u_{i} ) \big] \cdot \big[ u_{i} - a_{i}^{\top}x_{\star} \big] \geq f(a_{i}^{\top}x_{\star},u_{i}) \quad \text{for all } u_{i} \in \real \text{ and } a_{i}^{\top}x_{\star} \in \real.
\end{align}
We defer the proof of Claim~\eqref{claim:lower-bound-first-part} to the end of the proof, and use it to manipulate our lower bound for $\mathcal{Q}$.
Applying Claim~\eqref{claim:lower-bound-first-part} to Eq.~\eqref{eq:intermediate-lb}, we obtain
\begin{align*}
  &\mathcal{Q}(Ax_{\star},g,\widetilde{g}) \\
    &\geq \inf_{ \substack{\alpha \in \real,\; \beta\geq 0, \; u\in \real^{n}, \\ 0<(\alpha - \|x_{\star}\|_{2})^{2} + \beta^{2} \leq (4\norm)^{2} }} \; \sup_{\gamma \geq 0} \; \frac{ \frac{1}{n} \sum_{i=1}^{n} f(a_{i}^{\top}x_{\star}, u_i)}{(\alpha - \norm)^{2} + \beta^{2}} + \frac{\gamma}{n} \Big( \big\|  A x_{\star} \alpha / \norm - u + \beta  \widetilde{g}  \big\|_{2} - \beta  \omega(x_{\star},X) \Big) \\
  & \overset{\1}{=} \inf_{ \substack{\alpha \in \real,\; \beta\geq 0, \; u\in \real^{n}, \\ 0<(\alpha - \|x_{\star}\|_{2})^{2} + \beta^{2} \leq (4\norm)^{2} }} \; \sup_{\gamma \geq 0} \; \frac{ \frac{1}{n} \sum_{i=1}^{n} f(a_{i}^{\top}x_{\star}, u_i) }{(\alpha - \norm)^{2} + \beta^{2}} + \frac{\gamma}{7n} \bigg( \frac{ \big\|  A x_{\star} \alpha / \norm - u + \beta  \widetilde{g}  \big\|_{2}^{2} - \beta^{2}  \omega(x_{\star},X)^{2} }{(\alpha - \norm)^{2} + \beta^{2}} \bigg).
\end{align*}
To obtain step $\1$, we use the following reasoning: By virtue of taking the supremum over $\gamma \geq 0$, we are imposing the constraint $ \big\|  A x_{\star} \alpha / \norm - u + \beta  \widetilde{g}  \big\|_{2} = \beta  \omega(x_{\star},X)$ in the first inequality. On the other hand, and by the same logic, the  inner supremum in the second inequality is imposing the constraint
\[
     \frac{ \big\|  A x_{\star} \alpha / \norm - u + \beta  \widetilde{g}  \big\|_{2}^{2}  }{(\alpha - \norm)^{2} + \beta^{2}} = \frac{  \beta^{2}  \omega(x_{\star},X)^{2}}{(\alpha - \norm)^{2} + \beta^{2}},
\]
where we note that the denominator in the expression above is always nonzero by assumption.

Continuing, by the max-min inequality, we can switch $\inf_{u\in \real^{n}}$ and $\sup_{\gamma\geq 0}$, and obtain
\begin{align*}
&\mathcal{Q}(Ax_{\star},g,\widetilde{g}) \\
&\quad \geq  \inf_{ \substack{\alpha \in \real,\; \beta\geq 0, \\ (\alpha - \|x_{\star}\|_{2})^{2} + \beta^{2} = R^{2} }} \; \sup_{\gamma \geq 0} \; \inf_{u \in \real^{n}} \; \frac{ \frac{1}{n} \sum_{i=1}^{n} f(a_{i}^{\top}x_{\star}, u_i) + \frac{\gamma}{7n} \Big( \big\|  A x_{\star} \alpha / \norm - u + \beta  \widetilde{g}  \big\|_{2}^{2} - \beta^{2}  \omega(x_{\star},X)^{2} \Big) }{ (\alpha - \norm)^{2} + \beta^{2} }.
\end{align*}
Note that the $\inf_{u \in \real^{n}}$ is separable over $i\in [n]$, and each $\inf_{u_{i} \in \real}$ has an explicit solution given by
\begin{align}\label{explicit-inf-u}
  &\inf_{u_{i} \in \real} \; f(a_{i}^{\top}x_{\star},u_{i}) + \frac{\gamma}{7} \cdot \Big( \alpha a_{i}^{\top}x_{\star}/\norm + \beta\widetilde{g}_{i} - u_{i} \Big)^{2} \nonumber \\
  & \geq \begin{cases} \frac{1}{7} \frac{\gamma \cdot |h'(6a_{i}^{\top}x_{\star})|}{\gamma + |h'(6a_{i}^{\top}x_{\star})| }  \cdot \Big( (\alpha/\norm - 1) a_{i}^{\top} x_{\star} +\beta\widetilde{g}_{i} \Big)^{2}, &\text{if }
  -6a_{i}^{\top}x_{\star} \leq
  \beta \widetilde{g}_{i} + \frac{\alpha}{\norm}a_{i}^{\top}x_{\star} \leq 6a_{i}^{\top}x_{\star}, a_{i}^{\top}x_{\star}>0, \\
    0, &\text{otherwise}.  \end{cases}  \\
  & = \mathcal{T}_{i}(\alpha,\beta,\gamma), \nonumber
\end{align}
where in the last step we use definition~\eqref{def-T-alpha-beta-gamma}. 
We defer the explicit calculations of inequality~\eqref{explicit-inf-u} to the end of this section. Now, combining the two pieces in the display above, we obtain
\begin{align*}
  \mathcal{Q}(Ax_{\star},g,\widetilde{g}) \geq  \inf_{ \substack{\alpha \in \real,\; \beta\geq 0, \\ (\alpha - \|x_{\star}\|_{2})^{2} + \beta^{2} = R^{2} }} \; \sup_{\gamma \geq 0} \; \frac{  \frac{1}{n} \sum_{i=1}^{n} \mathcal{T}_{i}(\alpha,\beta,\gamma) - \frac{\gamma}{7n} \beta^{2}  \omega(x_{\star},X)^{2} }{ (\alpha - \norm)^{2} + \beta^{2} } = \mathcal{D}(Ax_{\star}, \widetilde{g}),
\end{align*} 
where in the last step we recall definition~\eqref{def-low-dim-inf}. This concludes the proof. 

It remains to prove Claim~\eqref{claim:lower-bound-first-part}
and to verify Eq.~\eqref{explicit-inf-u}.

\paragraph{Proof of Claim~\eqref{claim:lower-bound-first-part}} We consider three cases.\\
\textbf{Case 1: $a_{i}^{\top}x_{\star}\leq 0$ or $u_i > 6a_{i}^{\top}x_{\star}$ or $u_{i} < -6a_{i}^{\top}x_{\star}$.}
Note that in this case, $f(a_{i}^{\top}x_{\star},u_{i}) = 0$. Moreover, since $h(\cdot)$ is non-increasing, we obtain 
\[
  \big[h(a_{i}^{\top}x_{\star}) - h(u_{i} ) \big] \cdot \big[ u_{i} - a_{i}^{\top}x_{\star} \big] \geq 0 \geq f(a_{i}^{\top}x_{\star},u_{i}).
\]

\noindent \textbf{Case 2: $a_{i}^{\top}x_{\star}>0$ and $ 0 \leq u_i \leq 6a_{i}^{\top}x_{\star}$.} By the mean value theorem, we obtain 
\[
  h(a_{i}^{\top}x_{\star}) - h(u_{i}) = h'(t_{i}) ( a_{i}^{\top}x_{\star} - u_{i}) \text{ for some }\min\big( a_{i}^{\top}x_{\star},u_{i}\big) \leq t_{i} \leq \max\big(a_{i}^{\top}x_{\star},u_{i}\big).
\]
Note that $h'(t_{i})\leq 0$ and thus
\[
  \big[ h(a_{i}^{\top}x_{\star}) - h(u_{i}) \big] \cdot \big[ u_{i} - a_{i}^{\top}x_{\star} \big] = |h'(t_{i})| \cdot  \big[ u_{i} - a_{i}^{\top}x_{\star} \big]^{2} \geq |h'(6a_{i}^{\top}x_{\star})| \cdot  \big[ u_{i} - a_{i}^{\top}x_{\star} \big]^{2},
\]
where in the last step we use the fact that $0 \leq t_{i} \leq 6a_{i}^{\top}x_{\star}$ and that $|h'(\cdot)|$ is strictly decreasing when its argument is positive. \\
\textbf{Case 3: $a_{i}^{\top}x_{\star}>0$ and $ -6a_{i}^{\top}x_{\star} \leq u_{i}<0$.}
 Note that $h(u_{i}) = h(0)$ for $u_{i}<0$ and thus
\begin{align*}
  \big[ h(a_{i}^{\top}x_{\star}) - h(u_{i}) \big] \cdot \big[ u_{i} - a_{i}^{\top}x_{\star} \big] &= \big[ h(a_{i}^{\top}x_{\star}) - h(0) \big] \cdot \big[ u_{i} - a_{i}^{\top}x_{\star} \big] \\
  & \overset{\1}{\geq} |h'(6a_{i}^{\top}x_{\star})| \cdot a_{i}^{\top} x_{\star} \cdot \big[ a_{i}^{\top}x_{\star} - u_{i} \big] \\
  & \geq \frac{1}{7} |h'(6a_{i}^{\top}x_{\star})| \cdot \big[ a_{i}^{\top}x_{\star} - u_{i} \big]^{2},
\end{align*}
where in step $\1$ we use the mean value theorem once more, and in the last step we use $u_{i} \in [-6a_{i}^{\top}x_{\star},0]$. This concludes the proof of claim~\eqref{claim:lower-bound-first-part}.

\paragraph{Verifying Eq.~\eqref{explicit-inf-u}} We consider two cases: \\
\textbf{Case 1: $a_{i}^{\top}x_{\star}\alpha/\norm + \beta \widetilde{g}_{i} < -6a_{i}^{\top}x_{\star}$ or $a_{i}^{\top}x_{\star}\alpha/\norm + \beta \widetilde{g}_{i} > 6a_{i}^{\top}x_{\star}$ or $a_{i}^{\top}x_{\star}\leq 0$.} Note that in this case, by letting $u_{i} = a_{i}^{\top}x_{\star}\alpha/\norm + \beta \widetilde{g}_{i}$ and recalling Eq.~\eqref{lower-bound-first-part}, we obtain
\[
  f(a_{i}^{\top}x_{\star},u_{i}) + \gamma \cdot \Big( \alpha a_{i}^{\top}x_{\star}/\norm + \beta\widetilde{g}_{i} - u_{i} \Big)^{2} = 0.
\]
Since $f(a_{i}^{\top}x_{\star},u_{i}) \geq 0$ for all $u_{i} \in \real$, thus $u_{i} = a_{i}^{\top}x_{\star}\alpha/\norm + \beta \widetilde{g}_{i}$ attains the minimum value $0$.
\textbf{Case 2: $-6a_{i}^{\top}x_{\star} \leq a_{i}^{\top}x_{\star}\alpha/\norm + \beta \widetilde{g}_{i} \leq 6a_{i}^{\top}x_{\star}$ and $a_{i}^{\top}x_{\star} > 0$.} Taking the derivative and setting it to zero yields
\[
  |h'(6a_{i}^{\top}x_{\star})| \cdot \big( u_{i} - a_{i}^{\top}x_{\star} \big) \cdot \mathbbm{1} \{ a_{i}^{\top}x_{\star}>0, -6a_{i}^{\top}x_{\star} \leq u_{i} \leq 6a_{i}^{\top}x_{\star}  \} + \gamma \cdot \big(u_{i} - a_{i}^{\top}x_{\star}\alpha/\norm - \beta \widetilde{g}_{i} \big)  = 0 
\]
The above equation has a unique solution
\[
  u_{i} = \frac{|h'(6a_{i}^{\top}x_{\star})| \cdot a_{i}^{\top}x_{\star} + \gamma \cdot(a_{i}^{\top}x_{\star}\alpha/\norm + \beta \widetilde{g}_{i})  }{|h'(6a_{i}^{\top}x_{\star})| + \gamma},
\]
and its corresponding function value is
\[
  f(a_{i}^{\top}x_{\star},u_{i}) + \frac{\gamma}{7} \cdot \Big( \alpha a_{i}^{\top}x_{\star}/\norm + \beta\widetilde{g}_{i} - u_{i} \Big)^{2} = \frac{1}{7}\frac{\gamma \cdot |h'(6a_{i}^{\top}x_{\star})|}{\gamma + |h'(6a_{i}^{\top}x_{\star})| }  \cdot \Big( (\alpha/\norm - 1) a_{i}^{\top} x_{\star} +\beta\widetilde{g}_{i} \Big)^{2}.
\]
It remains to verify the restriction of the variational problem $\mathcal{P}(A)$ to compact sets.

\paragraph{Restriction of the variational problem $\mathcal{P}(A)$ to compact sets}
Note that it is equivalent to consider an alternate definition with
    \[
    \widetilde{\mathcal{P}}(A) := \min_{x \in X} \;\Bigl\{\frac{1}{n} \sum_{i=1}^{n} \bigl[h(a_i^{\top} x) - h(a_i^{\top} x_{\star})\bigr]\cdot \bigl[a_i^{\top}x - a_i^{\top} x_{\star}\bigr] - \nu \| x - x_{\star} \|_2^2\Bigr\}.
    \]
    and show that $\widetilde{\mathcal{P}}(A) \geq 0$.  Moreover, since by assumption $X$ is compact, the minimum is attained.  Next, note that the supremum over the parameter $\lambda \in \mathbb{R}^n$ comes by dualizing the constraint $u = Ax$.  We next show, following the strategy of~\cite[Corollary 5.1]{miolane2021distribution}, that it suffices to maximize over a compact subset of $\mathbb{R}^n$.  In particular, consider the function $G: \mathbb{R}^n \rightarrow \mathbb{R}$ defined as $G(u) = \frac{1}{n} \sum_{i=1}^{n} \bigl[h(u_i) - h(a_i^{\top} x_{\star})\bigr]\cdot \bigl[u_i - a_i^{\top} x_{\star}\bigr]$, noting that $G$ is $C$-Lipschitz, where $C$ depends only on $\{s_j, \mu_j\}_{j=1}^{W}$.  Now, let $\epsilon > 0$ and note that by tightness of the Gaussian measure, it holds that there exists a constant $M > 0$ depending only on $\epsilon$ such that the event
    \[
    \mathcal{A} := \Bigl\{ \| A \|_{\mathsf{op}} \leq M\Bigr\}.
    \]
    satisfies $\mathbb{P}(\mathcal{A}) \geq 1 - \epsilon$.  Further, on the event $\mathcal{A}$, it holds that
    \[
    \widetilde{\mathcal{P}}(A) = \min_{x \in X, u \in \mathbb{B}_2(C M)}\;\; \Bigl\{ G(u) \quad \text{ s.t. } \quad Ax = u\Bigr\}.
    \]
    It thus follows from the Lipschitz nature of $G$ that, for any $t > 0$, 
    \[
    \max_{\| \lambda \|_2 = t} \bigl\{ G(u) + \langle \lambda, Ax - u\rangle \bigr\} \geq G(Ax) + (t - C) \| u - Ax \|_2.
    \]
    Hence, it suffices to maximize $\lambda$ over the set $\mathbb{B}_2(C)$ so that on the event $\mathcal{A}$,
    \[
    \widetilde{\mathcal{P}}(A) = \min_{x \in X, u \in \mathbb{B}_2(C M)} \max_{\lambda \in \mathbb{B}_2(C)}\;\; \Bigl\{ G(u) + \langle \lambda, Ax - u \rangle\Bigr\}.
    \]
    Consequently, by the CGMT, we have
    \[
    \mathbb{P}(\widetilde{\mathcal{P}}(A) \leq \tau) \leq 2 \mathbb{P}(\mathcal{Q}(Ax_{\star}, g, \widetilde{g}) \leq \tau) + \epsilon \quad \text{ for all } \quad \tau \in \mathbb{R}. 
    \]
    Taking $\epsilon \downarrow 0$ yields the desired result.

    This completes the proof.
\qed

\subsection{Proof of Lemma~\ref{lemma-gamma-property}} \label{sec:proof-lemma-gamma-property}
To reduce the notational burden, we define $\ell_{\alpha,\beta}:\real \mapsto \real$ as
\[
  \ell_{\alpha,\beta}(\gamma) =  \frac{1}{n} \sum_{i=1}^{n} \mathcal{T}_{i}(\alpha,\beta,\gamma)  - \frac{\gamma \cdot \beta^{2}}{7n} \cdot \omega(X,x_{\star})^{2},
\]
where $\mathcal{T}_{i}(\alpha,\beta,\gamma)$ is defined in Eq.~\eqref{def-T-alpha-beta-gamma}. Computing the gradient (i.e. derivative) with respect to $\gamma$ yields
\begin{align*}
  \nabla \ell_{\alpha,\beta}(\gamma) &= \frac{1}{7n} \sum_{i=1}^{n} \frac{|h'(6a_{i}^{\top}x_{\star})|^{2}  \big( (\alpha/\norm-1)a_{i}^{\top}x_{\star} + \beta \widetilde{g}_{i} \big)^{2} }{(\gamma + |h'(6a_{i}^{\top}x_{\star})| )^{2}} \times \\
    & \qquad \qquad \mathbbm{1}\Big\{-6a_{i}^{\top}x_{\star} \leq \beta \widetilde{g}_{i} + \frac{\alpha}{\norm} a_{i}^{\top}x_{\star} \leq 6a_{i}^{\top}x_{\star}, a_{i}^{\top}x_{\star}>0 \Big\}  - \frac{\beta^2 \omega(x_{\star},X)^{2}}{7n}
\end{align*}

\paragraph{Proof of finiteness} 
We first prove that $\overline{\gamma}(\alpha,\beta)<\infty$. Note that 
\[
    \lim_{\gamma \rightarrow +\infty } \nabla \ell_{\alpha,\beta}(\gamma) = - \frac{\beta^2 \omega(x_{\star},X)^{2}}{7n} \overset{\1}{\leq} - \frac{\beta^2 \overline{\omega}(x_{\star},X)^{2}/4}{7n} < 0,
\]
where step $\1$ holds with probability at least $1-2\exp\big( - \overline{\omega}(x_{\star},X)^{2}/8 \big)$ by applying inequality~\eqref{Expectation-bound-delta-eta} in Lemma~\ref{lemma:concentration-delta-eta-iota}, and in the last step we use $\beta>0$ and $\overline{\omega}(x_{\star},X) >0$. Since $\ell_{\alpha,\beta}(\gamma)$ is a concave function of $\gamma$, we obtain that with probability 
at least $1-2\exp\big( - \overline{\omega}(x_{\star},X)^{2}/8 \big)$, the scalar $\overline{\gamma}(\alpha,\beta)$ is finite uniformly for all $(\alpha,\beta)$ satisfying Eq.~\eqref{assump-of-alpha-beta}. 

\paragraph{Proof that $\overline{\gamma}(\alpha,\beta)\geq \gamma_{\star}$} To show this, we will show that $\nabla \ell_{\alpha,\beta}(\gamma_{\star})>0$ with the desired high-probability. Note that since $\nabla \ell_{\alpha,\beta}(\gamma)$ is non-increasing over $\gamma$, this implies $\nabla \ell_{\alpha,\beta}(\gamma) \geq \nabla \ell_{\alpha,\beta}(\gamma_{\star})>0$ for all $0 \leq \gamma \leq \gamma_{\star}$. This further implies $\ell_{\alpha,\beta}(\gamma)$ is strictly increasing when $0 \leq \gamma \leq \gamma_{\star}$.
Further by definition
\[
    \overline{\gamma}(\alpha,\beta) = \arg \sup_{\gamma>0} \; \ell_{\alpha,\beta}(\gamma).
\]
We thus conclude that $\overline{\gamma}(\alpha,\beta) \geq \gamma_{\star}$.

We start with obtaining a lower bound on $\nabla \ell_{\alpha,\beta}(\gamma_{\star})$. We begin with a bound on the indicator terms in this expression. 
Substituting inequality~\eqref{ineq-indicator-lower-bounf} into the definition of the gradient, we have
\begin{align}\label{ineq-lower-bound-derivative-ell}
\hspace{-0.5cm}
  &\nabla \ell_{\alpha,\beta}(\gamma_{\star}) \\
  &\quad \geq \frac{1}{7n} \sum_{i=1}^{n} \frac{|h'(6a_{i}^{\top}x_{\star})|^{2}  \big( (\alpha/\norm-1)a_{i}^{\top}x_{\star} + \beta \widetilde{g}_{i} \big)^{2} }{(\gamma_{\star} + |h'(6a_{i}^{\top}x_{\star})| )^{2}}  \mathbbm{1}\Big\{ -\frac{ a_{i}^{\top}x_{\star}}{4\norm} \leq \widetilde{g}_{i} \leq \frac{a_{i}^{\top}x_{\star}}{\norm}, a_{i}^{\top}x_{\star}>0 \Big\} - \frac{\beta^2 \omega(x_{\star},X)^{2}}{7n} \nonumber \\
  &\quad = \frac{(\alpha-\norm)^{2}}{7} \frac{1}{n} \sum_{i=1}^{n} \delta_{i} + \frac{\beta^{2}}{7} \frac{1}{n} \sum_{i=1}^{n} \eta_{i} - \frac{2(\alpha - \norm) \beta}{7} \frac{1}{n} \sum_{i=1}^{n} \iota_{i} - \frac{\beta^2 \omega(x_{\star},X)^{2}}{7n}, 
\end{align} 
where in the last step we use the shorthands
\begin{subequations}
\begin{align}
    \label{def-delta}
    \delta_{i} &= \frac{|h'(6a_{i}^{\top}x_{\star})|^{2}  (a_{i}^{\top}x_{\star})^{2}/\norm^{2} }{(\gamma_{\star} + |h'(6a_{i}^{\top}x_{\star})| )^{2}} \cdot \mathbbm{1}\Big\{ -\frac{ a_{i}^{\top}x_{\star}}{4\norm} \leq \widetilde{g}_{i} \leq \frac{a_{i}^{\top}x_{\star}}{\norm}, a_{i}^{\top}x_{\star}>0 \Big\},\\
    \label{def-eta}
  \eta_{i} &= \frac{|h'(6a_{i}^{\top}x_{\star})|^{2} \cdot \widetilde{g}_{i}^{2}  }{(\gamma_{\star} + |h'(6a_{i}^{\top}x_{\star})| )^{2}} \cdot \mathbbm{1}\Big\{ -\frac{ a_{i}^{\top}x_{\star}}{4\norm} \leq \widetilde{g}_{i} \leq \frac{a_{i}^{\top}x_{\star}}{\norm}, a_{i}^{\top}x_{\star}>0 \Big\}, \\
    \label{def-iota}
  \iota_{i} & = \frac{|h'(6a_{i}^{\top}x_{\star})|^{2} \cdot \widetilde{g}_{i} a_{i}^{\top}x_{\star}/\norm }{(\gamma_{\star} + |h'(6a_{i}^{\top}x_{\star})| )^{2}} \cdot \mathbbm{1}\Big\{ -\frac{ a_{i}^{\top}x_{\star}}{4\norm} \leq \widetilde{g}_{i} \leq \frac{a_{i}^{\top}x_{\star}}{\norm}, a_{i}^{\top}x_{\star}>0 \Big\}.
\end{align}
\end{subequations}
We next invoke the following auxiliary lemma to control the random variables in the RHS of inequality~\eqref{ineq-lower-bound-derivative-ell}.
\begin{lemma}\label{lemma:concentration-delta-eta-iota}
    Let $\{\delta_{i},\eta_{i},\iota_{i}\}_{i=1}^{n}$ be defined in Eq.~\eqref{def-delta}, Eq.~\eqref{def-eta}, and Eq.~\eqref{def-iota}, respectively. Let $\omega(x_{\star},X)$ and $\overline{\omega}(x_{\star},X)$ be defined in Eq.~\eqref{Random-Gaussian-width} and Eq.~\eqref{def-Gaussian-width}, respectively. Then the following holds for all $\epsilon \in (0,1)$ and for all $t>0$:
    \begin{subequations}\label{claim-concentation-delta-eta-iota}
        \begin{align}
      \label{claim-concentation-delta}
      &\Pr\bigg\{ \Big| \frac{1}{n} \sum_{i=1}^{n} \delta_{i} - \EE\{\delta_{i}\} \Big| \geq 8\log(2/\epsilon)/n + \sqrt{64 \EE\{ \delta_{i}\} \log(2/\epsilon)/n} \bigg\} \leq \epsilon,\\
      \label{claim-concentration-eta}
      &\Pr\bigg\{ \Big| \frac{1}{n} \sum_{i=1}^{n} \eta_{i} - \EE\{\eta_{i}\} \Big| \geq 8\log(2/\epsilon)/n + \sqrt{64 \EE\{ \eta_{i}\} \log(2/\epsilon)/n} \bigg\} \leq \epsilon, \\
      \label{claim-concentration-iota}
      &   \Pr\bigg\{ \Big| \frac{1}{n} \sum_{i=1}^{n} \iota_{i} \Big| \geq 8\log(2/\epsilon)/n + \sqrt{ 64\EE\{\delta_{i}\} \log(2/\epsilon)/n }  \bigg\} \leq \epsilon, \\
            \label{Expectation-bound-delta-eta}
            & \EE[\delta_{i}] \wedge \EE[\eta_{i}] \geq  \frac{ \Constant \overline{\omega}(x_{\star},X)^{2}}{n}  \quad \text{and} \quad \Pr\Big\{ |\omega(x_{\star},X) - \overline{\omega}(x_{\star},X)| \geq t \Big\} \leq 2\exp\big(-t^{2}/2 \big).
        \end{align}
    \end{subequations}
\end{lemma}
We provide the proof of Lemma~\ref{lemma:concentration-delta-eta-iota} in Section~\ref{sec:pf-lemma-concentration-delta-eta-iota}. By setting $\epsilon = 2\exp\big(- \overline{\omega}(x_{\star},X)^{2}/64 \big)$ and applying inequality~\eqref{claim-concentation-delta-eta-iota} and the union bound, we obtain that with probability at least $1-6\exp\big(- \overline{\omega}(x_{\star},X)^{2}/64 \big)$, we have
\begin{align*}
    \frac{1}{n} \sum_{i=1}^{n} \delta_{i} &\geq \EE\{\delta_{i}\} - \frac{\overline{\omega}(x_{\star},X)^{2}}{8n} - \sqrt{\EE\{\delta_{i}\}} \sqrt{\frac{\overline{\omega}(x_{\star},X)^{2}}{n}}, \\
    \frac{1}{n} \sum_{i=1}^{n} \eta_{i} &\geq \EE\{\eta_{i}\} - \frac{\overline{\omega}(x_{\star},X)^{2}}{8n} - \sqrt{\EE\{\eta_{i}\}} \sqrt{\frac{\overline{\omega}(x_{\star},X)^{2}}{n}}, \\
    \bigg| \frac{1}{n} \sum_{i=1}^{n} \iota_{i} \bigg| &\leq \frac{\overline{\omega}(x_{\star},X)^{2}}{8n} + \sqrt{\EE\{\delta_{i}\}} \sqrt{\frac{\overline{\omega}(x_{\star},X)^{2}}{n}}.
\end{align*}
Substituting the inequalities in the display above into inequality~\eqref{ineq-lower-bound-derivative-ell} yields
\begin{align*}
    7\nabla \ell_{\alpha,\beta}(\gamma_{\star}) &\geq (\alpha - \norm)^{2} \cdot \bigg( \EE\{\delta_{i}\} - \frac{\overline{\omega}(x_{\star},X)^{2}}{8n} - \sqrt{\EE\{\delta_{i}\}} \sqrt{\frac{\overline{\omega}(x_{\star},X)^{2}}{n}}  \bigg) \\
    & \quad + \beta^{2} \cdot \bigg( \EE\{\eta_{i}\} - \frac{\overline{\omega}(x_{\star},X)^{2}}{8n} - \sqrt{\EE\{\eta_{i}\}} \sqrt{\frac{\overline{\omega}(x_{\star},X)^{2}}{n}}  \bigg) \\
    & \quad -2|\alpha-\norm|\beta \cdot \bigg( \frac{\overline{\omega}(x_{\star},X)^{2}}{8n} + \sqrt{\EE\{\delta_{i}\}} \sqrt{\frac{\overline{\omega}(x_{\star},X)^{2}}{n}} \bigg) - \beta^{2}\frac{\omega(x_{\star},X)^{2}}{n}.
\end{align*}
Note that by the AM-GM inequality, we obtain $2|\alpha-\norm|\beta \leq (\alpha-\norm)^{2} + \beta^{2}$ and 
\[
    2|\alpha-\norm|\beta \sqrt{\EE\{\delta_{i}\}} \sqrt{\frac{\overline{\omega}(x_{\star},X)^{2}}{n}} \leq 2\beta^{2} \frac{\overline{\omega}(x_{\star},X)^{2}}{n} + \frac{(\alpha-\norm)^{2}}{2} \EE\{\delta_{i}\}.
\]
Substituting above, we have
\begin{align*}  
    7\nabla \ell_{\alpha,\beta}(\gamma_{\star}) &\geq (\alpha - \norm)^{2} \cdot \bigg( \frac{1}{2} \EE\{\delta_{i}\} - \frac{\overline{\omega}(x_{\star},X)^{2}}{4n} - \sqrt{\EE\{\delta_{i}\}} \sqrt{\frac{\overline{\omega}(x_{\star},X)^{2}}{n}}  \bigg) \\
    & \quad + \beta^{2} \cdot \bigg(  \EE\{\eta_{i}\} - \frac{(2+1/8)\overline{\omega}(x_{\star},X)^{2}}{n} - \sqrt{\EE\{\eta_{i}\}} \sqrt{\frac{\overline{\omega}(x_{\star},X)^{2}}{n}}  \bigg) - \beta^{2}\frac{\omega(x_{\star},X)^{2}}{n}.
\end{align*}
Continuing, using inequality~\eqref{Expectation-bound-delta-eta} yields
\begin{align*}
    &\frac{1}{2} \EE\{\delta_{i}\} - \frac{\overline{\omega}(x_{\star},X)^{2}}{4n} - \sqrt{\EE\{\delta_{i}\}} \sqrt{\frac{\overline{\omega}(x_{\star},X)^{2}}{n}} \\
    & = 0.25 \EE\{\delta_{i}\} - \frac{\overline{\omega}(x_{\star},X)^{2}}{4n} + 0.25\EE\{\delta_{i}\} - \sqrt{\EE\{\delta_{i}\}} \sqrt{\frac{\overline{\omega}(x_{\star},X)^{2}}{n}} \geq \frac{3\overline{\omega}(x_{\star},X)^{2}}{n},
\end{align*}
and similar calculation yields
\[
    \EE\{\eta_{i}\} - \frac{(2+1/8)\overline{\omega}(x_{\star},X)^{2}}{n} - \sqrt{\EE\{\eta_{i}\}} \sqrt{\frac{\overline{\omega}(x_{\star},X)^{2}}{n}} \geq \frac{3\overline{\omega}(x_{\star},X)^{2}}{n}.
\]
Putting the pieces together yields
\begin{align*}
    7\nabla \ell_{\alpha,\beta}(\gamma_{\star}) &\geq \big( (\alpha - \norm)^{2} + \beta^{2} \big) \cdot \frac{3\overline{\omega} (x_{\star},X)^{2}}{n}  -  \beta^{2}\frac{\omega(x_{\star},X)^{2}}{n} \\
    & \overset{\1}{\geq} \big( (\alpha - \norm)^{2} + \beta^{2} \big) \cdot \frac{(3-9/4)\overline{\omega} (x_{\star},X)^{2}}{n} > 0,
\end{align*}
where in step $\1$ we use inequality~\eqref{Expectation-bound-delta-eta} so that 
\[
    \omega(x_{\star},X) \leq (3/2) \overline{\omega}(x_{\star},X) \quad \text{with probability} \geq 1- 2\exp\big(- \overline{\omega}(x_{\star},X)^{2} / 8 \big).
\]
In conclusion, we obtain that $\nabla \ell_{\alpha,\beta}(\gamma_{\star})>0$ with probability at least $1-8\exp\big(- \overline{\omega}(x_{\star},X)^{2} / 64 \big)$, as desired.
\qed

\subsubsection{Proof of Lemma~\ref{lemma:concentration-delta-eta-iota}}\label{sec:pf-lemma-concentration-delta-eta-iota}
We prove each part separately.
\paragraph{Proof of inequality~\eqref{claim-concentation-delta}}
To reduce the notational burden, we define a function $\psi:\real \mapsto \real$
\begin{align}\label{function-psi}
  \psi(t) = \frac{ |h'(6\norm t)|^{2} }{(\gamma_{\star} + |h'(6\norm t)| )^{2}}, \quad \text{for all } t\in \real.
\end{align}
We first claim that $\delta_i$ satisfies the Bernstein condition (see, e.g.,~\cite[Eq. 2.16]{wainwright2019high})
\begin{align}\label{claim-moment-bound-delta}
  \big| \EE\big\{ (\delta_{i} - \EE[\delta_{i}])^{k}  \big\} \big| \leq \frac{1}{2} \cdot k! \cdot 4^{k-2} \cdot 16 \mathbb{E}[\delta_i] \quad \text{ for all } \quad k = 2, 3, \ldots.
\end{align}
deferring its proof to the end of this section. Consequently, by ~\cite[Proposition 2.3]{wainwright2019high}, we obtain that for $a = 32\EE\{\delta_{i}\}/n$ and $b = 4/n$, 
\begin{align*}
  \EE \Big\{ e^{\lambda \frac{1}{n}\sum_{i=1}^{n}(\delta_{i} - \EE\{ \delta_{i} \} ) } \Big\} = \prod_{i=1}^{n} \EE\Big\{ e^{ \frac{\lambda}{n}(\delta_{i} - \EE\{ \delta_{i} \} ) } \Big\} \leq 
  \exp\bigg\{ \frac{ a \lambda^{2}/2 }{1-|\lambda| b} \bigg\} \quad \text{for all } \lambda \in (-1/b,1/b),
\end{align*}
and
\begin{align*}
  \Pr\Big\{ \frac{1}{n}  \sum_{i=1}^{n} \delta_{i} - \EE\{\delta_{i}\}  \geq t \Big\}  \leq  e^{-\frac{t^{2}/2}{a+bt} }.
\end{align*}
Thus, letting $t = 2b\log(2/\epsilon) + \sqrt{2a\log(2/\epsilon)}$, we obtain that for $\epsilon \in (0,1]$,
\[
  \Pr\Big\{ \frac{1}{n} \sum_{i=1}^{n} \delta_{i} - \EE\{\delta_{i}\} \geq 2b\log(2/\epsilon) + \sqrt{2a\log(2/\epsilon)} \Big\} \leq \epsilon/2.
\]
Following the same steps yields the lower tail bound 
\[
  \Pr\Big\{ \frac{1}{n} \sum_{i=1}^{n} \delta_{i} - \EE\{\delta_{i}\} \leq -2b\log(2/\epsilon) - \sqrt{2a\log(2/\epsilon)} \Big\} \leq \epsilon/2.
\]
Applying the union bound thus yields
\[
  \Pr\Big\{ \Big| \frac{1}{n} \sum_{i=1}^{n} \delta_{i} - \EE\{\delta_{i}\} \Big| \geq 2b\log(2/\epsilon) + \sqrt{2a\log(2/\epsilon)} \Big\} \leq \epsilon.
\]
The desired result follows by noting $a = 32\EE\{\delta_{i}\}/n$ and $b = 4/n$.  It remains to prove the inequality~\eqref{claim-moment-bound-delta}.

\medskip
\noindent \underline{Establishing the Bernstein condition~\eqref{claim-moment-bound-delta}.} 
Let $\overline{\delta}_{i}=\delta_{i}-\EE\{\delta_i\}$.  Applying Jensen's inequality to the convex function $t \mapsto t^k$, we find that
\begin{align*}
  |\overline{\delta}_{i}|^{k} \leq \big( |\delta_{i}| + \EE\{|\delta_{i}|\} \big)^{k} \leq 2^{k-1} \big( |\delta_{i}|^{k} + \EE\{|\delta_{i}|\}^{k}  \big).
\end{align*}
Consequently,
\[
  \EE\big\{ |\overline{\delta}_{i}|^{k} \big\} \leq 2^{k-1}\big( \EE\big\{ |\delta_{i}|^{k} \big\} + \EE\{|\delta_{i}|\}^{k}  \big) \leq 2^{k} \EE\big\{ |\delta_{i}|^{k} \big\}.
\] 
Letting $G,\widetilde{G} \overset{\mathsf{i.i.d.}}{\sim} \NORMAL(0,1)$, we compute
\begin{align}\label{aux-eq1-delta}
  \EE\big\{ |\delta_{i}|^{k} \big\} &= \EE\big\{ \psi(G)^{k} G^{2k} \cdot \mathbbm{1}\{ -G/4 \leq  \widetilde{G} \leq G/4, G>0 \} \big\} \nonumber \\
  &\overset{\1}{\leq} \EE\big\{ \psi(G) G^{2k} \cdot \mathbbm{1}\{ -G/4 \leq  \widetilde{G} \leq G/4, G>0 \} \big\} \nonumber \\
  & = \EE\bigg\{ \psi(G) G^{2k} \int_{-G/4}^{G/4} \phi(t) \mathrm{d}t \cdot \mathbbm{1}\{G>0\} \bigg\},
\end{align}
where step $\1$ follows from the sandwich relation $0 \leq \psi(t) \leq 1$ for all $t\in \real$ and in the final step we have used the shorthand $\phi(t) = \exp\{-t^{2}/2\} / (\sqrt{2\pi})$. Integration by parts yields
\begin{align}\label{aux-eq2-delta}
  &\EE\bigg\{ \psi(G) G^{2k} \int_{-G/4}^{G/4} \phi(t) \cdot \mathbbm{1}\{G>0\} \mathrm{d}t \bigg\} 
  = (2k-1)\EE\bigg\{  \psi(G) G^{2k-2} \int_{-G/4}^{G/4} \phi(t) \mathrm{d}t \cdot \mathbbm{1}\{G>0\}  \bigg\} \nonumber \\
  &\qquad +\EE\bigg\{ \psi'(G)G^{2k-1} \int_{-G/4}^{G/4} \phi(t) \mathrm{d}t \cdot \mathbbm{1}\{G>0\} \bigg\} + \EE\bigg\{ \psi(G) G^{2k-1}  \frac{\phi(G/4)}{2} \cdot \mathbbm{1}\{G>0\} \bigg\}.
\end{align}
We next bound the last two terms in the equality display above. Note that $\psi'(t) \leq 0$ for all $t>0$ since $\psi(t)$~\eqref{function-psi} is non-increasing for $t>0$, and thus the second term is non-positive, i.e.,
\[
  \EE\bigg\{ \psi'(G)G^{2k-1} \int_{-G/4}^{G/4} \phi(t) \mathrm{d}t \cdot \mathbbm{1}\{G>0\} \bigg\} \leq 0.
\]
For the third term, note that
\begin{align*}
  G \frac{ \phi(G/4)}{2} \cdot \mathbbm{1}\{G>0\} \leq \int_{-G/4}^{G/4} \phi(t)\mathrm{d}t \cdot \mathbbm{1}\{G>0\},
\end{align*}
and thus we can bound the third term as
\[
  \EE\bigg\{ \psi(G) G^{2k-1} \frac{ \phi(G/4)}{2} \cdot \mathbbm{1}\{G>0\} \bigg\} \leq \EE\bigg\{ \psi(G) G^{2k-2} \int_{-G/4}^{G/4} \phi(t)\mathrm{d}t \cdot \mathbbm{1}\{G>0\}  \bigg\}.
\]
Substituting the bounds of the second and third term into Eq.~\eqref{aux-eq2-delta} yields that for $k\geq 1$
\begin{align*}
\EE\bigg\{ \psi(G) G^{2k} \int_{-G/4}^{G/4} \phi(t) \cdot \mathbbm{1}\{G>0\} \mathrm{d}t \bigg\}  \leq 2k \cdot \EE\bigg\{ \psi(G) G^{2k-2} \int_{-G/4}^{G/4} \phi(t)\mathrm{d}t \cdot \mathbbm{1}\{G>0\}  \bigg\}.
\end{align*}
Note that the above inequality holds for all natural numbers $k\geq 1$. Iteratively applying the inequality in the display above yields, for $k\geq 1$,
\begin{align} \label{ineq-reduce-order-G}
\EE\bigg\{ \psi(G) G^{2k} \int_{-G/4}^{G/4} \phi(t) \cdot \mathbbm{1}\{G>0\} \mathrm{d}t \bigg\}  \leq 2^{k-1} k! \cdot \EE\bigg\{ \psi(G) G^{2} \int_{-G/4}^{G/4} \phi(t)\mathrm{d}t \cdot \mathbbm{1}\{G>0\}  \bigg\}.
\end{align}
Consequently, recalling inequality~\eqref{aux-eq1-delta}, we obtain the upper bound
\begin{align*}
  \EE\big\{ |\delta_{i}|^{k} \big\} \leq 2^{k-1} k! \cdot \EE\bigg\{ \psi(G) G^{2} \int_{-G/4}^{G/4} \phi(t)\mathrm{d}t \cdot \mathbbm{1}\{G>0\}  \bigg\} = 2^{k-1} k! \cdot \EE\{ \delta_{i}\},
\end{align*}
where in the last step we use the definition of $\delta_{i}$~\eqref{def-delta}.  Putting the pieces together yields the desired result.

\paragraph{Proof of inequality~\eqref{claim-concentration-eta}} We first claim that the random variable $\eta_{i} - \EE\{\eta_{i}\}$ satisfies the following Bernstein condition, deferring its proof to the end,
\begin{align}\label{claim-moment-bound-eta}
  \big| \EE\big\{ (\eta_{i} - \EE[\eta_{i}])^{k}  \big\} \big| \leq \frac{1}{2} \cdot k! \cdot 4^{k-2} \cdot 16 \mathbb{E}[\eta_i] \quad \text{ for all } \quad k = 2, 3, 4, \ldots
\end{align}
The rest of proof is identical to the proof of inequality~\eqref{claim-concentation-delta}, so we omit it. 

It remains to verify inequality~\eqref{claim-moment-bound-eta}. Let $\overline{\eta}_{i} = \eta_{i} - \EE\{\eta_{i}\}$. Following identical steps as in the proof of inequality~\eqref{claim-moment-bound-delta}, we obtain
\begin{align}\label{aux-eq0-eta}
  \EE\big\{ \lvert \overline{\eta}_i \rvert^{k} \big\} \leq 2^k \EE\bigl\{ \lvert \eta_i \rvert^k \bigr\}.
\end{align}
Continuing, by definition of $\eta_{i}$, we obtain that for $G,\widetilde{G} \overset{\mathsf{i.i.d.}}{\sim} \NORMAL(0,1)$
\begin{align}\label{aux-eq1-eta}
  \EE\big\{ |\eta_{i}|^{k} \big\} &= \EE\bigg\{ \psi(G)^{k}\widetilde{G}^{2k} \cdot \mathbbm{1}\big\{ -G/4\leq  \widetilde{G} \leq G/4, G>0 \big\} \bigg\} \nonumber \\
  & \leq \EE\bigg\{ \psi(G)\widetilde{G}^{2k} \cdot \mathbbm{1}\big\{ -G/4\leq  \widetilde{G} \leq G/4, G>0 \big\} \bigg\},
\end{align}
where in the last step we use $0 \leq \psi(t) \leq 1$ for all $t>0$. Conditioning on $G$ and for $G>0$, we obtain 
\begin{align*}
  \EE\Big\{ \widetilde{G}^{2k} \cdot \mathbbm{1}\big\{ -G/4\leq  \widetilde{G} \leq G/4 \big\}  \Big\} &= \int_{-G/4}^{G/4} t^{2k} \phi(t) \mathrm{d}t \\
  &\overset{\1}{=} (2k-1) \cdot \int_{-G/4}^{G/4} t^{2k-2}\phi(t) \mathrm{d}t - t^{2k-1}\phi(t) \Big\vert_{-G/4}^{G/4} \\
  & \overset{\2}{\leq} (2k-1) \cdot \int_{-G/4}^{G/4} t^{2k-2}\phi(t) \mathrm{d}t \\
  & = (2k-1) \cdot \EE\Big\{ \widetilde{G}^{2k-2} \cdot \mathbbm{1}\big\{ -G/4\leq  \widetilde{G} \leq G/4 \big\}  \Big\},
\end{align*}
where in step $\1$ we use integration by parts and in step $\2$ we use 
$
  t^{2k-1}\phi(t) \Big\vert_{-G/4}^{G/4} \geq 0.
$ 
The inequality in the display holds for any positive integer $k$ and thus applying it sequentially yields that, conditionally on $G$ and $G>0$,
\begin{align}\label{ineq-reduce-order-tilde-G}
  \EE\Big\{ \widetilde{G}^{2k} \cdot \mathbbm{1}\big\{ -G/4\leq \widetilde{G} \leq G/4 \big\}  
    &\Big\} \leq \prod_{i=1}^{k}(2i-1) \cdot \EE\Big\{ \widetilde{G}^{2} \cdot \mathbbm{1}\big\{ -G/4\leq  \widetilde{G} \leq G/4 \big\}  \Big\} \nonumber \\
  &\leq 2^{k}k! \cdot \EE\Big\{ \widetilde{G}^{2} \cdot \mathbbm{1}\big\{ -G/4 \leq \widetilde{G} \leq G/4 \big\}  \Big\}.
\end{align}
Substituting the inequality in the display into inequality~\eqref{aux-eq1-eta}, we obtain 
\[
  \EE\big\{ |\eta_{i}|^{k} \big\} \leq 2^{k}k! \cdot  \EE\bigg\{ \psi(G)\widetilde{G}^{2} \cdot \mathbbm{1}\big\{ -G/4 \leq  \widetilde{G} \leq G/4, G>0 \big\} \bigg\} = 2^{k}k! \cdot \EE\{\eta_{i}\}
\]
The desired result follows by substituting the inequality in the display into inequality~\eqref{aux-eq0-eta}.

\paragraph{Proof of inequality~\eqref{claim-concentration-iota}} We again show that $\iota_i$ satisfies a Bernstein condition.  First, note that for odd $k$, 
\[
  \EE\{ \iota_{i}^{k} \} = \EE\Big\{ \psi(G)^{k} \widetilde{G}^{k} G^{k} \mathbbm{1}\big\{ -G/4 \leq  \widetilde{G} \leq G/4, G>0 \big\}  \Big\} = 0,
\]
as $\widetilde{G}^{k} \mathbbm{1}\big\{ -G/4 \leq  \widetilde{G} \leq G/4, G>0 \big\}$ is an odd function of $\widetilde{G}$ and $\widetilde{G}, G \overset{\mathsf{i.i.d.}}{\sim} \NORMAL(0,1)$.   Continuing, using the fact $0 \leq \psi(t) \leq 1$ for all $t\in \real$, we obtain
\begin{align*}
  \EE\big\{ \iota_{i}^{2k} \big\} &\leq \EE\Big\{ \psi(G) G^{2k} \widetilde{G}^{2k} \mathbbm{1}\{G>0\} \mathbbm{1}\big\{  - G/4 \leq  \widetilde{G} \leq  G/4 \big\}    \Big\} 
    \\ & \overset{\1}{\leq} 2^{k} k! \cdot \EE\Big\{ \psi(G) G^{2k}  \mathbbm{1}\{G>0\} \mathbbm{1}\big\{  - G/4 \leq \widetilde{G} \leq  G/4 \big\}    \Big\} 
    \\ & \overset{\2}{\leq}  (2^{k} k!)^{2} \cdot \EE\Big\{ \psi(G) G^{2}  \mathbbm{1}\{G>0\} \mathbbm{1}\big\{  - G/4 \leq  \widetilde{G} \leq  G/4 \big\}  \Big\} = (2^{k} k!)^{2} \cdot \EE\{\delta_{i}\},
\end{align*}
where in step $\1$ we use inequality~\eqref{ineq-reduce-order-tilde-G} so that conditionally on $G$ and $G>0$,
\[
    \EE\Big\{ \widetilde{G}^{2k} \mathbbm{1}\big\{  - G/4\leq \widetilde{G} \leq  G/4 \big\}   \Big\} \leq 2^{k} k! \cdot 
    \EE\Big\{ \mathbbm{1}\big\{  - G/4 \leq  \widetilde{G} \leq  G/4 \big\}   \Big\},
\]
and in step $\2$ we use inequality~\eqref{ineq-reduce-order-G}.  We thus have
\[
\EE\bigl\{\iota_i^{2k} \bigr\} \leq \frac{1}{2} \cdot (2k)! \cdot 4^{2k - 2} \cdot 16 \mathbb{E}\{\delta_i\}, \quad \text{ for all } \quad k = 2, 3, 4, \ldots .
\]
Following the identical steps in the proof of inequality~\eqref{claim-concentation-delta}, we obtain that for all $\epsilon \in (0,1)$,
\[
    \Pr\bigg\{ \Big| \frac{1}{n} \sum_{i=1}^{n} \iota_{i} \Big| \geq 8\log(2/\epsilon)/n + \sqrt{ 64\EE\{\delta_{i}\} \log(2/\epsilon)/n }  \bigg\} \leq \epsilon.
\]

\paragraph{Proof of inequality~\eqref{Expectation-bound-delta-eta}} We first lower bound $\EE\{\delta_{i}\}$. 
Expanding the definition of $\delta_{i}$~\eqref{def-delta} yields
\begin{align*}
    \EE\{\delta_{i}\} = \EE\bigg\{ \frac{|h'(6\norm G)|^{2} G^{2} \mathbbm{1}\{G>0\}}{ \big( |h'(6\norm G)|^{2} + \gamma_{\star}  \big)^{2} }
    \cdot \int_{- G/ 4}^{ G/4} \frac{e^{-t^{2}/2}}{\sqrt{2\pi}}  \mathrm{d}t
    \bigg\}, \quad \text{where } G,\widetilde{G} \overset{\mathsf{i.i.d.}}{\sim} \NORMAL(0,1).
\end{align*}
Further note that for $G>0$,
\[
    \int_{-G/4}^{G/4} \frac{t^{2}e^{-t^{2}/2}}{\sqrt{2\pi}} \mathrm{d}t  \leq \Big(\frac{G}{4}\Big)^{2} \int_{- G/ 4}^{ G/4} \frac{e^{-t^{2}/2}}{\sqrt{2\pi}}  \mathrm{d}t.
\]
Hence, invoking the definition of $\gamma_{\star}$ as in Eq.~\eqref{fixed-point}, we find that
\[
    \EE\{ \delta_{i} \} \geq 16 \EE\bigg\{ \frac{|h'(6\norm G)|^{2}  \mathbbm{1}\{G>0\}}{ \big( |h'(6\norm G)|^{2} + \gamma_{\star}  \big)^{2} }
    \cdot \int_{-G/4}^{G/4} \frac{t^{2}e^{-t^{2}/2}}{\sqrt{2\pi}} \mathrm{d}t 
    \bigg\} \geq \frac{ \Constant \overline{\omega}(x_{\star},X)^{2}}{n}.
\]
 We next turn to  lower bounding $\EE\{\eta_{i}\}$. By definition of $\eta_{i}$~\eqref{def-eta},
\[
    \EE\{\eta_{i}\} = \EE\bigg\{ \frac{|h'(6\norm G)|^{2} \widetilde{G}^{2} \mathbbm{1}\{G>0\}}{ \big( |h'(6\norm G)|^{2} + \gamma_{\star}  \big)^{2} }
    \cdot \mathbbm{1} \big\{  - G/4 \leq  \widetilde{G} \leq  G/4 \big\}
    \bigg\}.
\]
Conditioning on $G$ in tandem with the event $G>0$, we have
\begin{align*}
    \EE\Big\{ \widetilde{G}^{2}\mathbbm{1} \big\{  - G/4 \leq  \widetilde{G} \leq  G/4 \big\} \Big\} = \int_{- G/4}^{  G/4}  \frac{t^{2} e^{-t^{2}/2}  }{\sqrt{2\pi}} \mathrm{d} t.
\end{align*}
Putting the two pieces together yields
\[
    \EE\{\eta_{i}\} \geq \EE\bigg\{ \frac{|h'(6\norm G)|^{2} \mathbbm{1}\{G>0\}}{ \big( |h'(6\norm G)|^{2} + \gamma_{\star}  \big)^{2} }
    \cdot \int_{-G/4}^{ G/4}  \frac{t^{2} e^{-t^{2}/2}  }{\sqrt{2\pi}} \mathrm{d} t
    \bigg\} \geq \frac{\Constant \overline{\omega}(x_{\star},X)^{2}}{n},
\]
where the last step follows from the definition of $\gamma_{\star}$, i.e., Eq.~\eqref{fixed-point}. We next turn to prove the concentration of $\omega(x_{\star},X)$~\eqref{Random-Gaussian-width}. To reduce the notational burden, we let $X^{\perp} = \{ P_{x_{\star}}^{\perp}x : x \in X \}$ and define $\varPsi:\real^{d} \mapsto \real$ as $\varPsi(g) = \sup_{v \in X^{\perp}} \langle v/\|v\|_{2}, g\rangle$. Note that by definition~\eqref{Random-Gaussian-width}, we have $\omega(x_{\star},X) = \varPsi(g)$ for $g \sim \NORMAL(0,I_{d})$. It is equivalent to prove concentration of the random variable $\varPsi(g)$ for $g \sim \NORMAL(0,I_{d})$. Note that for all $g_{1},g_{2}$, we obtain
\[
    \varPsi(g_{1}) - \varPsi(g_{2}) = \sup_{v \in X^{\perp}} \Bigl \langle \frac{v}{\|v\|_{2}}, g_{1} \Bigr\rangle - \sup_{v \in X^{\perp}} \Bigl\langle \frac{v}{\|v\|_{2}}, g_{2}\Bigr\rangle  \leq \sup_{v \in X^{\perp}} \Bigl\langle \frac{v}{\|v\|_{2}}, g_{1} - g_{2}\Bigr\rangle  \leq \| g_{1} - g_{2} \|_{2}.
\]
Similar arguments show $\varPsi(g_{2}) - \varPsi(g_{1}) \leq \| g_{1} - g_{2} \|_{2}$ for all $g_{1},g_{2} \in \real^{d}$. Therefore, $\varPsi(\cdot)$ is a $1$-Lipschitz function, so that applying~\cite[e.g.,][Theorem 2.26]{wainwright2019high} yields
\[
    \Pr\Big\{ | \varPsi(g) - \EE\{ \varPsi(g) \}  | \geq t \Big\} \leq 2\exp(-t^{2}/2) \quad  \text{for all } t>0.
\]
The desired result follows by noting $\omega(x_{\star},X) = \varPsi(g)$ and $\overline{\omega}(x_{\star},X) = \EE\{ \varPsi(g) \}$ for $g \sim \NORMAL(0,I_{d})$.
\qed

\subsection{Proof of Lemma~\ref{lemma:concentration-theta-xi}} \label{sec:pf-concentration-theta-xi} 
Note that by definition of $\{\theta_{i},\vartheta_{i},\xi_{i}\}$ in Eq.~\eqref{def-theta-xi},
\[
    |\theta_{i}| \leq \sum_{j=1}^{\N} s_{j}\coeff_{j} \frac{(a_{i}^{\top}x_{\star})^{2}}{\norm^{2}}, \quad |\vartheta_{i}| \leq \sum_{j=1}^{\N} s_{j}\coeff_{j} \widetilde{g}_{i}^{2},\quad |\xi_{i}| \leq \sum_{j=1}^{\N} s_{j}\coeff_{j} \frac{a_{i}^{\top}x_{\star}}{\norm}\widetilde{g}_{i}.
\]
Since $a_{i} \sim \NORMAL(0,I_{d})$ and $\widetilde{g}_{i} \sim \NORMAL(0,1)$, it follows that $\theta_{i}$, $\vartheta_{i}$, and $\xi_{i}$ are sub-exponential random variables, i.e., there exists a universal constant $C>0$ such that
\[
    \| \theta_{i} \|_{\psi_{1}} \; \vee \; \| \vartheta_{i} \|_{\psi_{1}} \; \vee \; \| \xi_{i} \|_{\psi_{1}} \leq C \sum_{j=1}^{\N} s_{j} \coeff_{j}.
\]
Applying Bernstein’s inequality~\cite[Theorem 2.8.1]{vershynin2018high}, we deduce that there exists a universal constant $C'>0$ such that with probability at least $1-6n^{-10}$,
\begin{align*}
    \bigg| \frac{1}{n} \sum_{i=1}^{n} \theta_{i} - \EE\{ \theta_{i} \} \bigg|\; \vee \;
    \bigg| \frac{1}{n} \sum_{i=1}^{n} \vartheta_{i} - \EE\{ \vartheta_{i} \} \bigg| \; \vee \; \bigg|  \frac{1}{n} \sum_{i=1}^{n} \xi_{i} - \EE\{ \xi_{i} \}  \bigg| \leq  C' \sum_{j=1}^{\N}s_{j}\coeff_{j}  \sqrt{\frac{\log(n)}{n}}.
\end{align*}
We claim that there exists a universal and positive constant $c$ such that,
\begin{align*}
    \EE\{ \theta_{i} \} \geq   c \sum_{j=1}^{\N} \frac{s_{j} \coeff_{j}}{ \big(\coeff_{j} \norm\big)^{4} \vee 1 },\quad  
     \EE\{ \vartheta_{i} \} \geq   c \sum_{j=1}^{\N} \frac{s_{j} \coeff_{j}}{ \big(\coeff_{j} \norm\big)^{4} \vee 1 }, \quad \EE\{\xi_{i}\} = 0,
\end{align*}
deferring its proof to the end of this section. Combining the inequalities in the display with condition~\eqref{assump-sample-size-thm} and adjusting constants, we obtain that the deviation $\sqrt{\log(n)/n}$ is of smaller order term than both $\EE\{\theta_{i}\}$ and $\EE\{\vartheta_{i}\}$. We thus conclude that with probability at least $1-6n^{-10}$,
\[
    \frac{1}{n} \sum_{i=1}^{n} \theta_{i} \; \wedge \; \frac{1}{n} \sum_{i=1}^{n} \vartheta_{i}  \gtrsim \sum_{j=1}^{\N} \frac{s_{j} \coeff_{j}}{ \big(\coeff_{j} \norm\big)^{4} \vee 1 } \quad \text{and} \quad \bigg| \frac{1}{n} \sum_{i=1}^{n} \xi_{i} \bigg| \lesssim  \sum_{j=1}^{\N}s_{j} \coeff_{j} \cdot  \sqrt{\frac{\log(n)}{ n}}.
\]
This concludes the main proof. It remains to establish the claim on $\EE\{\theta_{i}\}$, $\EE\{\vartheta_{i}\}$, and $\EE\{\xi_{i}\}$.

\paragraph{Lower bound on $\EE\{\theta_{i}\}$} By definition~\eqref{def-theta-xi}, we obtain that for $G, \widetilde{G} \overset{\mathsf{i.i.d.}}{\sim} \NORMAL(0,1)$,
\begin{align*}
    \EE\{\theta_{i}\} &= \EE\Big\{ |h'(6\norm G)| G^{2} \mathbbm{1}\big\{ -G/4 \leq \widetilde{G} \leq G/4, G>0 \big\} \Big\} \\
    & = \EE\bigg\{|h'(6\norm G)|G^{2} \mathbbm{1}\{G>0\} \int_{-G/4}^{G/4} \frac{e^{-t^{2}/2}}{\sqrt{2\pi}} \mathrm{d}t  \bigg\} \\
    & \geq \EE\bigg\{ |h'(6\norm G)|G^{2} \mathbbm{1}\{G>0\} \frac{G}{2}\frac{e^{-G^{2}/32}}{\sqrt{2\pi}}   \bigg\}.
\end{align*}
Note that by Eq.~\eqref{function-h-plus}, $|h'(t)| = \sum_{j=1}^{\N} s_{j} \coeff_{j} \exp(-\coeff_{j} t)$ for all $t>0$. Thus, we obtain
\begin{align} \label{ineq-lower-bound-E-theta}
    \EE\{\theta_{i}\} &\geq \frac{1}{2\sqrt{2\pi}} \sum_{j=1}^{\N} s_{j} \coeff_{j} \EE\Big\{ \exp\big(-6\coeff_{j}\norm G - G^{2}/32 \big) \mathbbm{1}\{G>0\} G^{3} \Big\} \nonumber \\
    & = \frac{1}{4\pi} \sum_{j=1}^{\N} s_{j} \coeff_{j} \int_{0}^{+\infty} e^{-6\coeff_{j}\norm t - t^{2}/32 } \cdot  t^{3} \cdot e^{-t^{2}/2}   \mathrm{d}t  \nonumber \\
    & \geq \frac{1}{4\pi} \sum_{j=1}^{\N} s_{j} \coeff_{j} \int_{0}^{+\infty} e^{-6\coeff_{j}\norm t - t^{2} } \cdot  t^{3}    \mathrm{d}t.
\end{align}
Note that if $\coeff_{j} \norm \leq 1$, then
\[
    \int_{0}^{+\infty} e^{-6\coeff_{j}\norm t - t^{2} } \cdot  t^{3}    \mathrm{d}t \geq 
    \int_{0}^{+\infty} e^{-6 t - t^{2} } \cdot  t^{3}    \mathrm{d}t := c_{0},
\]
where $c_{0}$ is a universal and positive constant. If $\coeff_{j} \norm \geq 1$, then
\begin{align*}
    \int_{0}^{+\infty} e^{-6\coeff_{j}\norm t - t^{2} } \cdot  t^{3}  \mathrm{d}t &\geq \int_{0}^{(3\coeff_{j} \norm)^{-1}} e^{-6\coeff_{j}\norm t - t^{2} } \cdot  t^{3}    \mathrm{d}t \\
    & \overset{\1}{\geq} e^{-3} \int_{0}^{(3\coeff_{j} \norm)^{-1}} t^{3} \mathrm{d}t \gtrsim \frac{1}{\coeff_{j}^{4} \norm^{4}},
\end{align*}
where in step $\1$ we use 
\[
    -6\coeff_{j}\norm t - t^{2} \geq -2 - (3\coeff_{j} \norm)^{-2} \geq -3, \quad \text{for all } 0 \leq t \leq (3\coeff_{j} \norm)^{-1} \text{ and } \coeff_{j} \norm \geq 1.
\]
Putting the two lower bounds together yields that there exists a universal and positive constant $c$ such that
\begin{align}\label{aux-ineq-int-cubic}
    \int_{0}^{+\infty} e^{-6\coeff_{j}\norm t - t^{2} } \cdot  t^{3}  \mathrm{d}t \geq  \frac{c}{\big(\coeff_{j}^{4} \norm^{4}\big) \vee 1}.
\end{align}
Substituting the inequality in the display above into inequality~\eqref{ineq-lower-bound-E-theta} yields
\begin{align*}
    \EE\{\theta_{i}\} \geq \frac{c}{4\pi} \sum_{j=1}^{\N} \frac{s_{j} \coeff_{j}}{ \big(\coeff_{j} \norm\big)^{4} \vee 1  },
\end{align*}
which concludes the proof.

\paragraph{Lower bound on $\EE\{\vartheta_{i}\}$} By definition~\eqref{def-theta-xi}, we obtain that for $G, \widetilde{G} \overset{\mathsf{i.i.d.}}{\sim} \NORMAL(0,1)$,
\begin{align*}
    \EE\{\varphi_{i}\} &= \EE\Big\{ |h'(6\norm G)| \widetilde{G}^{2} \mathbbm{1}\big\{ -G/4 \leq \widetilde{G} \leq G/4, G>0 \big\} \Big\} \\
    & = \EE\bigg\{|h'(6\norm G)| \mathbbm{1}\{G>0\} \int_{-G/4}^{G/4} \frac{t^{2}e^{-t^{2}/2}}{\sqrt{2\pi}} \mathrm{d}t  \bigg\} \\
    & \gtrsim \EE\bigg\{ |h'(6\norm G)| \mathbbm{1}\{G>0\} e^{-G^{2}/32} G^{3}   \bigg\},
\end{align*}
where in the last step we use for $G>0$
\[
    \int_{-G/4}^{G/4} \frac{t^{2}e^{-t^{2}/2}}{\sqrt{2\pi}} \mathrm{d}t \gtrsim e^{-G^{2}/32} \int_{-G/4}^{G/4} t^{2} \mathrm{d}t \gtrsim e^{-G^{2}/32} G^{3}.
\]
Consequently, we obtain 
\begin{align*} 
    \EE\{\vartheta_{i}\} &\gtrsim \sum_{j=1}^{\N} s_{j} \coeff_{j} \EE\bigg\{ \exp\Big(-6\coeff_{j}\norm G - G^{2}/32 \Big) \mathbbm{1}\{G>0\} G^{3} \bigg\}  \\
    & = \sum_{j=1}^{\N} s_{j} \coeff_{j} \int_{0}^{+\infty} e^{-6\coeff_{j}\norm t - t^{2}/32 } \cdot  t^{3} \cdot e^{-t^{2}/2}   \mathrm{d}t   \\
    & \geq  \sum_{j=1}^{\N} s_{j} \coeff_{j} \int_{0}^{+\infty} e^{-6\coeff_{j}\norm t - t^{2} } \cdot  t^{3}    \mathrm{d}t  \gtrsim \sum_{j=1}^{\N} \frac{ s_{j} \coeff_{j} }{ (\coeff_{j} \norm)^{4} \vee 1 },
\end{align*}
where in the last step we use inequality~\eqref{aux-ineq-int-cubic}. This concludes the proof.

\paragraph{Proof that $\EE\{\xi_{i}\}=0$} By definition~\eqref{def-theta-xi}, we obtain
\begin{align*}
    \EE\{\xi_{i}\} = \EE\bigg\{ |h'(6\norm G)| G \widetilde{G} \cdot \mathbbm{1}\Big\{ G>0, -G/4 \leq \widetilde{G} \leq G/4 \Big\} \bigg\}.
\end{align*}
Since for any $G>0$, $\widetilde{G} \cdot \mathbbm{1}\big\{G>0, -G/4 \leq  \widetilde{G} \leq G/4 \big\} $ is an odd function of $\widetilde{G}$ and $G,\widetilde{G} \overset{\mathsf{i.i.d.}}{\sim} \NORMAL(0,1)$, then by condition on $G$, 
\[
    \EE\Big\{ \widetilde{G} \cdot \mathbbm{1}\big\{G>0,  -G/4 \leq  \widetilde{G} \leq G/4 \big\} \Big\} = 0.
\]
We thus obtain $\EE\{\xi_{i}\} = 0$.
\qed

\end{document}